\documentclass[prx,twocolumn,superscriptaddress]{revtex4-2}
\usepackage{amsmath,amssymb,bm,mathrsfs,graphicx, braket, times,color}
\usepackage[colorlinks=true,citecolor=green!70!black]{hyperref}
\usepackage{soul}
\usepackage{xcolor}
\usepackage{colortbl}
\usepackage[makeroom]{cancel}
\usepackage{array,multirow,graphicx}
\usepackage{tikz}
\usetikzlibrary{shadings}
\usepackage{txfonts}  
\usepackage{txfontsb} 

\makeatletter 
\renewcommand{\section}{\@startsection{section}{1}{0mm}
  {-\baselineskip}{0.5\baselineskip}{\bf\leftline}}
\makeatother
\begin{document}

\title{Micro-scale opto-thermo-mechanical actuation in the dry adhesive regime}

\author{Weiwei Tang}
\thanks{These authors contributed equally to this work.}
\affiliation{Key Laboratory of 3D Micro/Nano Fabrication and Characterization of Zhejiang Province, School of Engineering, Westlake University, 18 Shilongshan Road, Hangzhou 310024, Zhejiang Province, China}
\affiliation{Institute of Advanced Technology, Westlake Institute for Advanced Study, 18 Shilongshan Road, Hangzhou 310024, Zhejiang Province, China}

\author{Wei Lv}
\thanks{These authors contributed equally to this work.}
\affiliation{Key Laboratory of 3D Micro/Nano Fabrication and Characterization of Zhejiang Province, School of Engineering, Westlake University, 18 Shilongshan Road, Hangzhou 310024, Zhejiang Province, China}
\affiliation{Institute of Advanced Technology, Westlake Institute for Advanced Study, 18 Shilongshan Road, Hangzhou 310024, Zhejiang Province, China}

\author{Jinsheng~Lu}
\thanks{These authors contributed equally to this work.}
\affiliation{State Key Laboratory of Modern Optical Instrumentation, College of Optical Science
and Engineering, Zhejiang University, Hangzhou 310027, China}

\author{Fengjiang~Liu}
\affiliation{Key Laboratory of 3D Micro/Nano Fabrication and Characterization of Zhejiang Province, School of Engineering, Westlake University, 18 Shilongshan Road, Hangzhou 310024, Zhejiang Province, China}
\affiliation{Institute of Advanced Technology, Westlake Institute for Advanced Study, 18 Shilongshan Road, Hangzhou 310024, Zhejiang Province, China}

\author{Jiyong~Wang}
\affiliation{Key Laboratory of 3D Micro/Nano Fabrication and Characterization of Zhejiang Province, School of Engineering, Westlake University, 18 Shilongshan Road, Hangzhou 310024, Zhejiang Province, China}
\affiliation{Institute of Advanced Technology, Westlake Institute for Advanced Study, 18 Shilongshan Road, Hangzhou 310024, Zhejiang Province, China}

\author{Wei Yan}
\thanks{To whom correspondence should be addressed; wyanzju@gmail.com}
\affiliation{Key Laboratory of 3D Micro/Nano Fabrication and Characterization of Zhejiang Province, School of Engineering, Westlake University, 18 Shilongshan Road, Hangzhou 310024, Zhejiang Province, China}
\affiliation{Institute of Advanced Technology, Westlake Institute for Advanced Study, 18 Shilongshan Road, Hangzhou 310024, Zhejiang Province, China}

\author{Min Qiu}
\thanks{To whom correspondence should be addressed; qiumin@westlake.edu.cn }
\affiliation{Key Laboratory of 3D Micro/Nano Fabrication and Characterization of Zhejiang Province, School of Engineering, Westlake University, 18 Shilongshan Road, Hangzhou 310024, Zhejiang Province, China}
\affiliation{Institute of Advanced Technology, Westlake Institute for Advanced Study, 18 Shilongshan Road, Hangzhou 310024, Zhejiang Province, China}

\begin{abstract}

Realizing optical manipulation of microscopic objects is crucial in the research fields of life science, condensed matter physics, and physical chemistry.
In non-liquid environments, this task is commonly regarded as difficult due to strong adhesive surface force ($\sim\mu\rm N$) attached to solid interfaces that makes tiny optical driven force ($\sim\rm pN$) insignificant.
Here, by recognizing the microscopic interaction mechanism between friction force---the parallel component of surface force on a contact surface---and thermoelastic waves induced by pulsed optical absorption, we establish a general principle enabling the actuation of micro-objects on dry frictional surfaces based on the opto-thermo-mechanical effects.
Theoretically, we predict that nanosecond pulsed optical absorption with mW-scale peak power is sufficient to tame $\mu\rm N$-scale friction force.
Experimentally, we demonstrate the two-dimensional spiral motion of gold plates on micro-fibers driven by nanosecond laser pulses, and reveal the rules of motion control.
Our results pave the way for future development of micro-scale actuators in non-liquid environments.
\end{abstract}
\maketitle

\section*{Introduction}
In his landmark lecture “{\it There is plenty of room at the bottom}" in 1959, Richard Feynman envisioned a fundamental scaling challenge in the coming era of nanotechnology: macroscopic designing rules for mechanical devices shall become invalid at micro-scales where adhesive surface force plays a dominant role in the mechanical response of devices as a result of increased surface-to-volume ratio~\cite{Feynman:1960}.
To bypass this challenge, the micro-scale mechanical devices, especially {\it actuators}---indispensable in various applications, such as optical communications~\cite{Neukermans:2001,Wu:2006,Kan:2015,Haffner:2019}, optical displays~\cite{VanKessel:1998} and molecular cargo~\cite{Wang:2012,Tottori:2012}---, nowadays generally exploit membrane architecture to mitigate undesirable surface effects and employ scale-invariant electrostatic force~\cite{Judy:2001,DelRio:2005}. Alternatively, they are limited in low adhesive environments (e.g., immersed in liquids) using tiny optical force ($\sim\rm pN$)~\cite{Ashkin:1970,Grier:2003,MacDonald:2003,Neale:2005,Gao:2017,Tanaka:2020,Ren:2020}, photoelectric force ($\sim\rm pN$)~\cite{Shvedov:2010,Shvedov:2014,Ren:2020} or taking advantage of bio-inspired actuation strategy~\cite{Palagi:2018}.
Realizing micro-scale actuators that can freely walk on two-dimensional dry (non-liquid) surfaces directly against strong resistance from the parallel component of surface force, i.e., {\it friction force} ($\sim\,\mu\rm N$), is challenging~\cite{Kendall:1994}.

In this regard, a few recent studies have pointed out a new direction: the excitations of elastic waves in micro-objects could facilitate their locomotion on dry surfaces~\cite{Lu:2017,Lu:2019,Linghu:2021}, as an extension of macroscopic surface elastic-wave motors~\cite{Kurosawa:1998,Shigematsu:2003}.
{In these pioneering studies, elastic waves are excited by temperature changes in micro-objects through optical absorption based on fiber-based systems.
Essentially, this is different from another scheme widely adopted in micro-fludics, wherein elastic waves are generated by the electronic excitations with piezoelectric substrates and are used to drive the locomotion of cells, droplets, particles, etc. for lab-on-a-chip technologies~\cite{Destgeer:2015}.
Besides, the findings in these studies have demonstrated motions of specific-shaped gold plates either along the azimuthal~\cite{Lu:2019} or axial directions~\cite{Lu:2017,Linghu:2021} of the micro-fibers, but not both.}

However, these existing results severely lack a convincing theory---that takes into account of the fundamental role of friction force at micro-scales---to inspire new designs, and experimentally only access to the one dimensional motion.
Here we resolve all these limitations both theoretically and experimentally.
A theory that takes microscopic interactions between friction force and thermally excited elastic waves into account---and is supplemented with an illustrative case study---is established. It features a predictive equation for the threshold optical power needed to overcome friction resistance. The double role of friction force in both hindering and promoting actuation is explicitly revealed.
Experimentally, we observe two-dimensional spiral motion of gold plates {with various shapes (including triangle, rectangle, square, hexagon and circle)} on micro-fibers driven by a nanosecond pulsed laser, and demonstrate, validate the general rules of motion control for gold plates.
Further, we provide the direct observation of the threshold optical power of motion, thereby validating the proposed theory.
These results advantageously provide a solid framework for understanding, interpreting, modelling the actuation dynamics involving friction force at micro-scales and for further inspiring new designs of miniature actuators based on similar principles.

\begin{figure*}[ht!]
\includegraphics[width=0.9\textwidth]{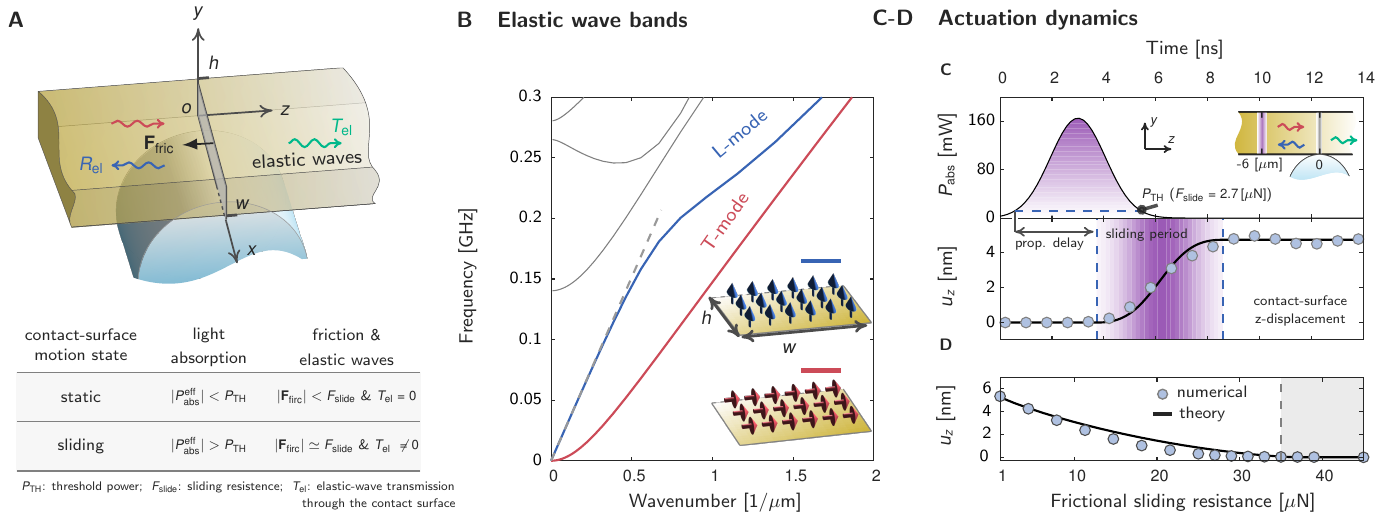}
\caption{{\bf Principle of opto-thermo-mechanical actuation overcoming $\mu \rm N$-scale friction force at micro-scales}.
{\bf A}. Driving locomotion of a rectangular plate on a frictional surface through excitations of elastic waves by pulsed optical absorption. The contact surface is a line-shaped region rendered by the surface curvature of the substrate.
Table: relationships between motion states of the plate, (instantaneous) effective absorbed light power, friction force and elastic waves.
{\bf B}. Band structure of elastic waveguide modes in a gold plate (width, $w=4\,\mu\rm m$; height, $h=60\,\rm nm$). Insets: modal profiles of fundamental longitudinal ($L$)- and transverse ($T$)-like modes at static frequency (arrows specify directions of elastic oscillations).
{\bf C}. Sliding displacement of the contact surface of the gold plate (same as in {\bf B}; lower panel) in the $z$-direction driven by a nanosecond optical pulse (upper panel) with frictional sliding resistance $F_{\rm slide}=2.7\, \mu\rm N$. The absorbed optical power distributes uniformly in the $x-y$ plane, has a Gaussian width of $1\,\mu \rm m$ in the $z$-direction and localizes on the left side of the contact surface (sketched in the inset of the upper panel).
{\bf D}. Stabilized sliding displacement as a function of sliding resistance, $F_{\rm slide}$.
In {\bf C}--{\bf D}, the numerical results are obtained with the COMSOL Multiphysics (circles) and agree with the theoretical predictions [solid lines; computed with Eq.~(2.18) in the Supplementary Material]. The elastic attenuation is ignored.
}
\label{fig::principle}
\end{figure*}

\section*{Results}

\noindent {\bf Principle: Interplay between Friction Force and Elastic Waves }
\label{sec::principle}


%
\noindent We start with a case study to pedagogically interpret the principle of the proposed micro-scale opto-thermo-mechanical actuation scheme and to clarify the key role of elastic waves---excited by mild (mW-scale) optical absorption---in overcoming friction force.
As shown in Fig.~\ref{fig::principle}A, a gold rectangular plate is placed on a curved substrate. The contact surface of two structures is a line-shaped region, resembling the later demonstrated experiment.
Illuminated by laser pulses, the plate absorbs light and then converts it to heat that induces lattice oscillations, i.e., elastic waves~\cite{Landau:2012}.
The excited elastic waves intend to drive the plate to move, which, however, is resisted by the friction force on the contact surface~\cite{Bowden:1950}.
Thereupon, {\it the elementary physical process here lies in the interaction between incident elastic waves---that propagate toward the contact surface from one side---and the friction force} (sketched in Fig.~\ref{fig::principle}A), which we shall discuss below.
%



%
We first examine the excited elastic waves in the plate by representing them with waveguide modes.
Figure~\ref{fig::principle}B shows the band structure of these modes (propagating along the $z$-direction) supported by a rectangular plate with cross-sectional width $w=4\,\mu\rm m$ and thickness $h=60\,\rm nm$.
Among a variety of modes, we highlight the fundamental longitudinal ($L$)- and transverse ($T$)-like modes (see Sec. 1 in the Supplementary Material for the derivations of their dispersion relations at low frequencies), which show no cutoff, and oscillate parallel and perpendicular to their propagation directions, respectively (see the modal profiles in the insets of Fig.~\ref{fig::principle}B).
These modes are important since they can be dominantly excited, e.g., by nanosecond laser pulses (as in our experiment), and moreover their static-frequency (Fourier) contributions determine the stabilized locomotion state of the plate along the longitudinal ($z-$) and transverse ($x-$) directions, respectively.

From a microscopic wave picture, the friction force behaves as a fence resisting the elastic waves from passing through the contact surface (see the table in Fig.~\ref{fig::principle}A, and Fig. S5 in the Supplementary Material for numerical illustrations).
In the static regime, the static friction force is large enough to totally reflect back the incident elastic waves and thus nullify mechanical oscillations on the contact surface (that is, the incident and reflected elastic waves cancel each other out perfectly), thereby preserving the contact surface in the still state.
On the contrary, in the dynamic regime, the elastic waves are so strong that even the maximum allowable static friction force---the so-called sliding resistance force denoted by $F_{\rm silde}$---cannot nullify them on the contact surface. The plate thus slides.

Apparently, the static-to-sliding transition occurs at a critical point when the induced elastic waves due to the sliding resistance $F_{\rm silde}$ perfectly cancel with the incident elastic waves due to optical absorption.
In view of this, we introduce the {\it sliding threshold power} $P_{\rm TH}$---the minimum of the magnitude of the {\it instantaneous} effective absorbed optical power, $P_{\rm abs}^{\rm eff}$, required to overcome the sliding resistance---to quantify the power cost of this transition.
Here, the effective absorbed optical power is defined by $P_{\rm abs}^{\rm eff}\equiv P_{\rm abs}- P_{\rm leak}$, i.e., the absorbed optical power $P_{\rm abs}$ minus the leaked power $P_{\rm leak}$.
For simplicity of analysis, we assume that the optical absorption---that induces the incident elastic waves---uniformly distributes along the oscillation direction of the $T$-modes ($x$-direction), so that only the sliding along the longitudinal direction ($z$-direction) relating to the $L$-modes is activated.
%
%
%

%
The sliding threshold power is then interpreted as the magnitude of the effective absorbed optical power that together with the sliding resistance results in zero amplitude of the $L$-modes on the contact surface, and it is derived as (see Sec. 2 in the Supplementary Material)
\begin{align}
P_{\rm TH}\simeq \frac{c_pF_{\rm silde}}{\alpha_{\rm th}v_{\rm 3D}^{\rm L}}e^{t_0/\tau_{\rm ac}}.
\label{eq::PTH}
\end{align}
Here, $c_p$ and $\alpha_{\rm th}$ denote the specific heat capacity and the coefficient of linear thermal expansion of gold, respectively. $v_{\rm 3D}^{\rm L}\equiv\sqrt{E/\rho}\simeq 2000\, \rm m/s$ ($E$ and $\rho$ denoting Young's modulus and density of gold, respectively) is the velocity of the $L$-modes at low frequencies, wherein a linear dispersion relation emerges [see the dashed line in Fig.~\ref{fig::principle}B and Eq.~(1.6a) in the Supplementary Material]. $t_0$ is the time for elastic waves traveling from the center of the absorbed optical power to the contact surface and $\tau_{\rm ac}$ is the life time of the elastic waves~\cite{Kor:1972,Ruijgrok:2012}.\\

\noindent {\bf Implications of Threshold Power}

\noindent We now discuss the dependence of the threshold power $P_{\rm TH}$ on the dimension of the plate. First, the term of the sliding resistance, $F_{\rm silde}$, renders $P_{\rm TH}$ a linear dependence on the width of the plate along the $x$-axis, i.e., $w$. Specially, $F_{\rm silde}$ linearly scales with the area of the contact surface~\cite{Bowden:1950}, which---in our case study---is proportional to $w$. Second, we observe that the threshold power increases with the propagation time of the elastic waves $t_0$, which indirectly associates with the width of the plate along the $z$-axis (that is, the larger of the width, potentially the longer $t_0$). Finally, we note that the thickness dependence is absent in our theory, which holds considering that the thickness of the plate is much smaller than the elastic wavelength of interest (see Sec. 2 in the Supplementary Material).

Equation~\eqref{eq::PTH} signifies the feasibility of the actuation of the plate by using practical optical power.
Specifically, $P_{\rm TH}\simeq  F_{\rm silde} \times 2\, {\rm mW/\mu N}$ for gold (neglecting elastic attenuation; see Table I in the Supplementary Material for thermal and mechanical parameters of gold), implying that $\rm m W$-scale effective absorbed optical power is sufficient to overcome $\mu {\rm N}$-scale friction force.
Such mild instantaneous absorbtion power can be readily realized with micro-/nano- structures made of gold---and also a large variety of other lossy materials---with broadband losses, and by additionally exploiting resonance effects (e.g., plasmonic excitations and Mie's resonances)~\cite{Maier:2007,Hao:2010}.
More importantly, regarding laser sources, Eq.~\eqref{eq::PTH} guides that {\it pulsed lasers are the ideal choice for the proposed actuation scheme}, featuring two unique advantages: (i) high peak power facilitating the effective absorbed instantaneous power to surpass $P_{\rm TH}$ and (ii) low single-pulse energy and relatively long repetition period benefiting thermal cooling.

Equation~\eqref{eq::PTH} also implies that continuous wave (CW) lasers are deficient for the proposed actuation scheme.
This is because that, with the CW lasers, the plate will rise to constant temperature with $P_{\rm abs}=P_{\rm leak}$, which leads to a zero value of the effective absorbed power, $P_{\rm abs}^{\rm eff}=0$.
Accordingly, the threshold power $P_{\rm TH}$ is failed to be overcome.

\begin{figure*}[ht!]
\includegraphics[width=0.9\textwidth]{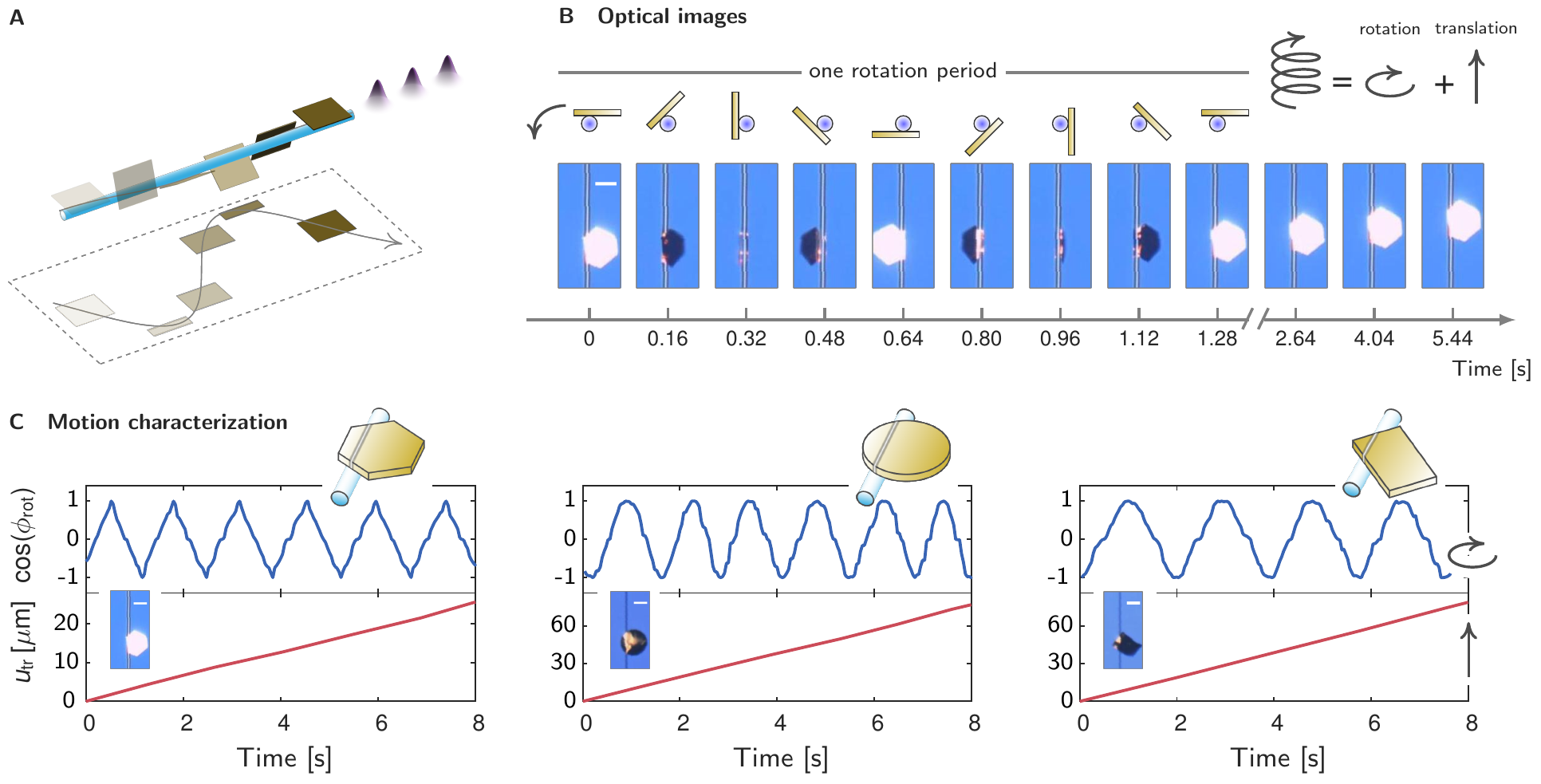}
\caption{{\bf Systematic measurement of spiral motion of gold plates around micro-fibers in ambient conditions}.
{\bf A.} Sketch of the observed spiral motion.
{\bf B.} Temporal sequencing of optical images of a hexagonal gold plate spirally moving around a micro-fiber.
The fiber has a diameter of $2\,\mu\rm m$, and the side length and the thickness of the plate are $27.72\,\mu\rm m$ and 30 nm, respectively.
{\bf C.} Cosine of rotation angle $\phi_{\rm rot}$  (upper panels), translation displacement (lower panels) as functions of time for gold plates with hexagonal, circular and rectangular base shapes. All scale bars represent $15\,\mu\rm m$. The used super-continuum laser pulses have 6.8--mW average power, 3--ns temporal width and 6.13--kHz repetition rate.
}
\label{fig::sprial}
\end{figure*}

Employing pulsed lasers, the plate slides within the period when the magnitude of the effective absorbed optical power is above $P_{\rm TH}$.
To straightforwardly reveal this actuation dynamics, we perform numerical simulations by considering that a gold rectangular plate---with width $w=4\,\mu\rm m$ and height $h=60\,\rm nm$---(same as in Fig.~\ref{fig::principle}B) is driven by a nanosecond pulsed absorbed optical power.
The plate is placed on a curved cylindrical surface, and the friction resistance between them is $F_{\rm silde}=2.7\,\mu\rm N$. Consequently, the threshold power predicted with Eq.~\eqref{eq::PTH} is about $P_{\rm TH}\simeq 5.4\,\rm mW$. We set the peak power of the optical absorption to be 160 mW (see the upper panel of Fig.~\ref{fig::principle}C for its temporal profile), which exceeds $P_{\rm TH}$, and additionally ensures that maximum rising temperature of the plate (about $600^{\rm o}{\rm C}$) is well below the melting temperature of gold.
Besides, the absorbed optical power, which localizes in a line-shaped region (see the caption of Fig.~\ref{fig::principle} for more details), is set to be $6\,\mu\rm m$ away from the contact surface in the negative the $z$ direction, from where the thermally excited elastic waves impinge on the contact surface.

As is shown in the lower panel of Fig.~\ref{fig::principle}C, the sliding takes place when
$|P_{\rm abs}^{\rm eff}(t-t_0)|>P_{\rm TH}$, which can be simplified to $P_{\rm abs}(t-t_0)>P_{\rm TH}$ considering that the leaked power $P_{\rm leak}$ is negligible (see Fig. S10 in the Supplementary Material).
In the post-sliding period, the absorbed optical power rapidly decreases. Concurrently, the slow cooling process begins
with $|P_{\rm abs}^{\rm eff}(t-t_0)|<P_{\rm TH}$.
Accordingly, the contact surface ceases the sliding (see Fig.~\ref{fig::principle}C), while preserving the previously accumulated sliding distance thanks to the static friction force that prevents the contact surface from returning to its original position.
Thus, in the whole process, the friction force plays a double role: mitigating the actuation in the sliding period, while
preserving the actuation in the post-sliding period.
Moreover, as shown in Fig.~\ref{fig::principle}D, the stabilized sliding distance decreases as the sliding resistance $F_{\rm silde}$ increases.
Notably, the sliding completely ceases when $F_{\rm silde}$ exceeds a critical value ($\sim 35\,\mu\rm N$), where $P_{\rm TH}$ approximately equals the absorbed peak power.\\

\noindent{\bf Spiral Motion of Gold Plates}


\begin{figure*}[htp!]
\includegraphics[width=0.9\textwidth]{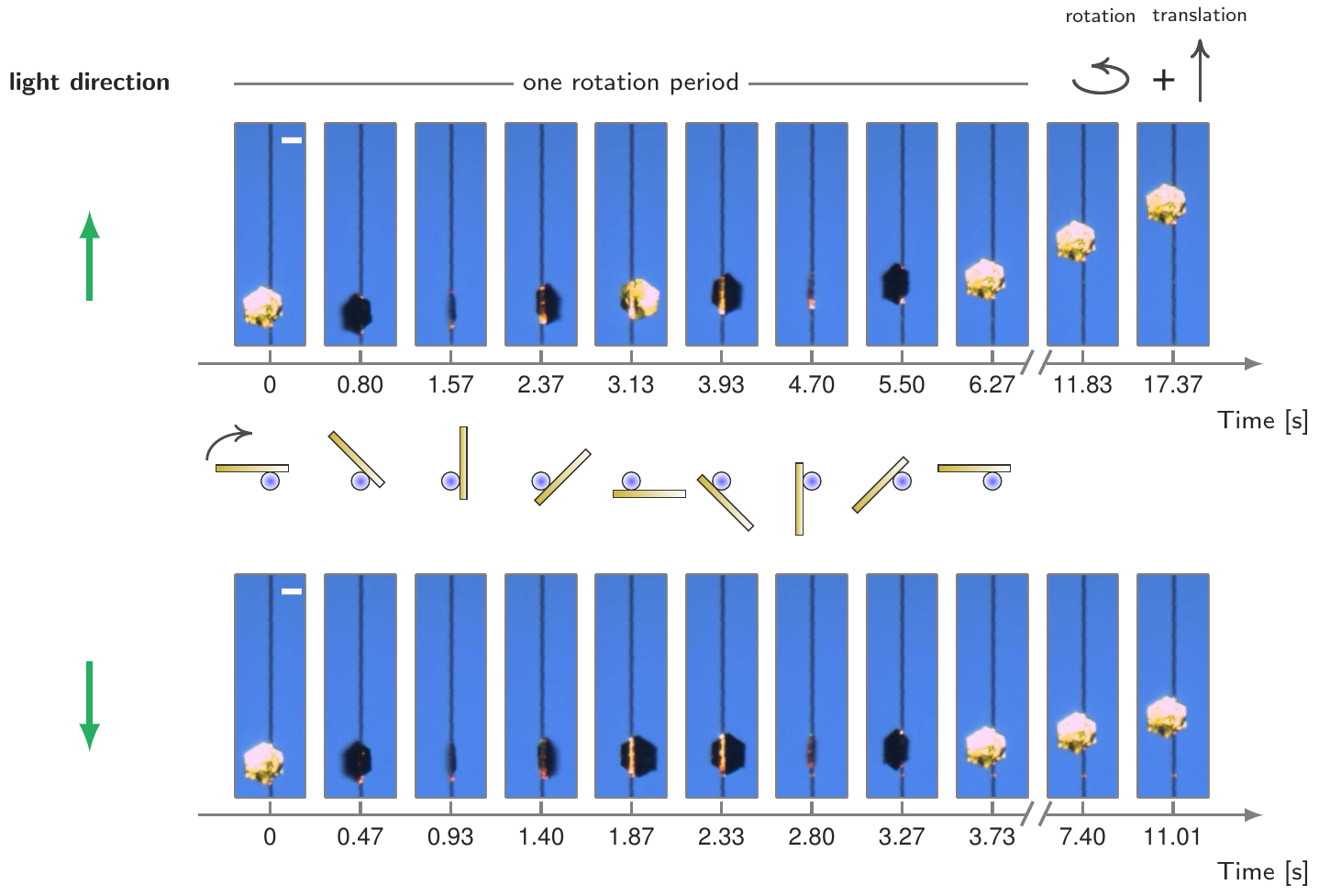}
\caption{{\bf Sequential optical images of a hexagonal gold plates moving around micro-fibers spirally in ambient conditions, showing that both rotation and translation directions are independent of incident light directions}. All scale bars represent $15\,\mu\rm m$. The laser source has the same parameters as in Fig.~\ref{fig::sprial}.}
\label{fig:::motion_dir}
\end{figure*}

\noindent Experimentally, in ambient conditions, we set up a gold-plate and micro-fiber coupled system (Fig.~\ref{fig::sprial}A)---which hosts two-dimensional locomotion of gold plates on curved surfaces (that is, rotation and translation in the azimuthal and axial directions of the micro-fibers, respectively)---to concretize the proposed theoretical concept.
We fabricated gold plates of various base shapes, including triangle, rectangle, square, hexagon and circle (see Fig. S14 in the Supplementary Material), to evidence that the proposed scheme works insensitive of structure geometry.
The plates have thicknesses varying between 30 nm and 60 nm and their base dimensions are typically about a few tens of micro-meters.
The micro-fibers with diameter about a few micro-meters (1.5 $\mu$m-4 $\mu$m) were tapered from standard optical fibers using the flame-heated drawing technique.
The plates were placed on top of the micro-fibers. They stick together by adhesive surface force of the order of $\mu\rm N$, and, accordingly support friction force of the same order~\cite{Kendall:1994,Lu:2019}.
To surpass the sliding threshold power endowed by the friction force, while mitigating undesirable heating effects, we used an ns-pulsed super-continuum light source (wavelength range, 450 nm to 2400 nm; see Fig. S9A in the Supplementary Material for the laser spectrum) to drive the plates.
As the laser pulses propagate inside the micro-fibers, their evanescent electric fields leak out and interact with the plates, which are then converted into heat by Ohmic losses of gold (see Fig. S9C for measured absorption spectrum in the Supplementary Material).
The electromagnetic simulations (see Fig. S8 in the Supplementary Material) demonstrate that the fiber modes with a dominant electric field component perpendicular to the plate surface can be more efficiently absorbed by the gold plates due to plasmonic effects than those without~\cite{Maier:2007}. Moreover, the absorbed optical power localizes within a narrow line-shaped subwavelength region centralized at the touching line between the plate and the micro-fiber as a result of plasmonic confinements.

The nanosecond pulsed optical absorption---spatially overlapping with the contact surface---excites the fundamental elastic $T$- and $L$-modes (together with
other less important high-order modes). These modes bounce back and forth inside the plates, rendering the locomotion of the plates two degrees of freedom: rotation ($L$-modes) and translation ($T$-modes).
As the elastic waves pass through the contact surface, they interact with the friction force following the elementary process revealed in the pedagogical model, and drive the plates to slide.
The combination of rotation and translation freedoms results in the spiral motion of the plates, as shown in Fig.~\ref{fig::sprial}B with the recorded temporal sequential optical images for a hexagonal plate. This observation generalizes the previously observed single motion freedom either rotation~\cite{Lu:2019} or translation~\cite{Lu:2017,Linghu:2021}.
Specifically, the plate translates towards the upper axial direction of the micro-fiber with a speed of 3.4 $\mu\rm m/s$, and, at the mean time, it rotates anti-clockwise (as viewed from the top side of the optical images) with a speed of 46 rpm (characterized in the leftmost panel of Fig.~\ref{fig::sprial}C).
The measured translation and rotation speeds increase approximately linearly with the repetition rate of the laser pulses (see Fig.~S17 in the Supplementary Material and Supplementary Movie 1), evidencing that the spiral motion of the plates is driven by individual single pulses.
Besides, the spiral motion was observed for gold plates with various geometries (see Fig.~\ref{fig::sprial}C, Fig.~S15 in the Supplementary Material, and Supplementary Movie 2), and with side length even up to two hundred micro-meters.
The gold plates can move stably on the micro-fibers
over a distance of several centimeters.
We also performed experiments in a vacuum chamber with the same observations on the spiral motion of the gold plates (see Fig. S19 in the Supplementary Material), which thus indicates that air molecules here play a negligible role.
\\

\noindent{\bf Motion Control}
\label{Sec:MC}

\begin{figure*}[ht!]
\includegraphics[width=0.9\textwidth]{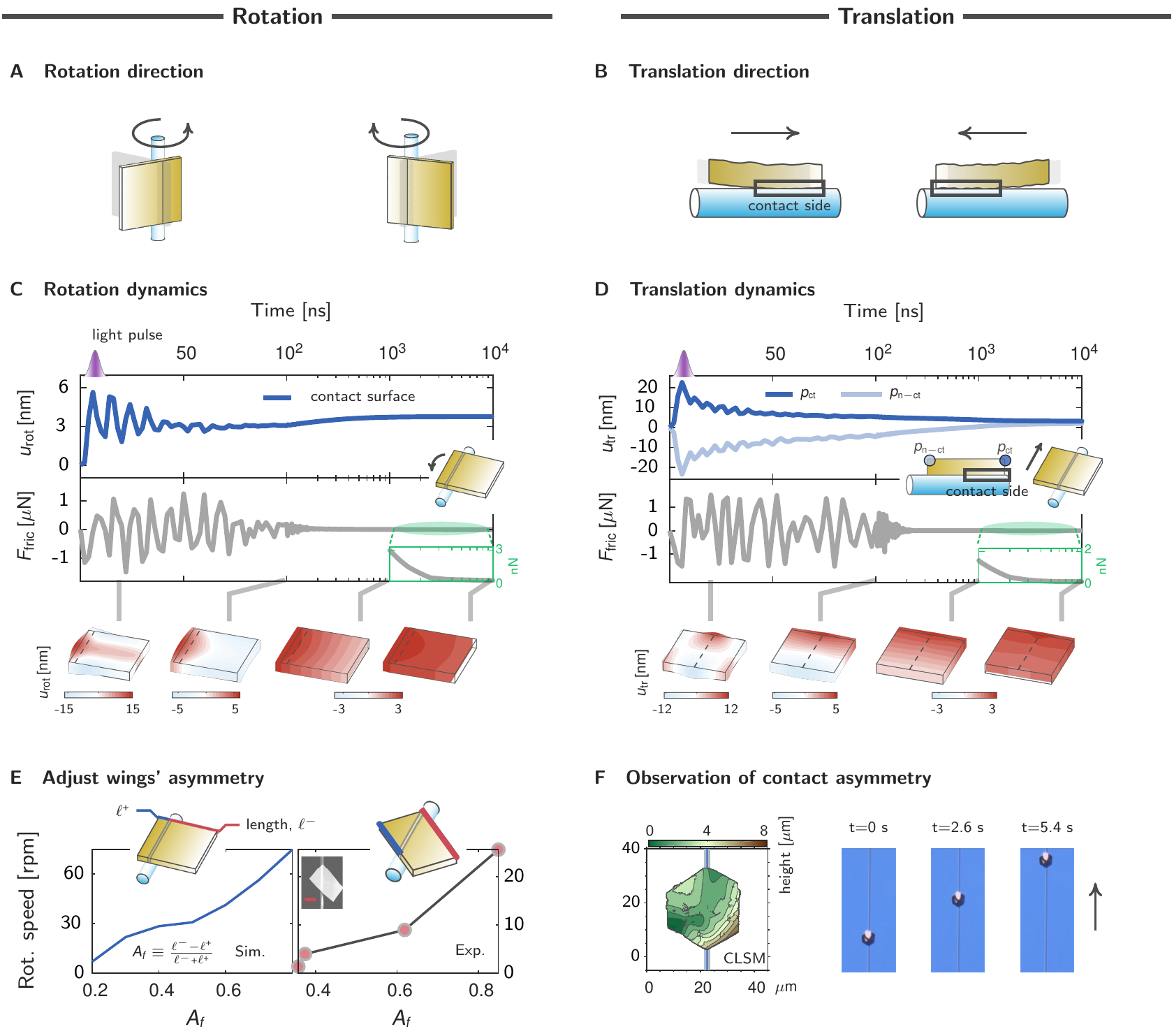}
\caption{{\bf Mechanism of spiral motion of gold plates.}
The spiral motion consists of two degrees of freedom---rotation ({\bf A, C, E}) and translation ({\bf B, D, F}).
{\bf A-B.} Sketch of rotation ({\bf A}) and translation ({\bf B}) directions of gold plates and of their dependencies on wings' and contact-surface's asymmetries, respectively.
{\bf C-D.} Simulation results of (1) sliding displacements (top panels) and (2) friction forces (middle panels) as functions of time for square gold plates driven by a single laser pulse, and of (3) displacement profiles at different time (lower panels).
The square plates have 30-nm thickness and 10-$\mu\rm m$ side length; the frictional sliding resistance is $1.5\,\mu\rm N$; the life time of the elastic waves is 20 ns.
The fiber has a diameter of $1.8\,\mu \rm m$.
The pulsed optical absorption has a 100-mW peak power and a 3-ns width, and its spatial distribution is specified in Eq. (5.1) in the Supplementary Material.
Moreover, in the rotation ({\bf C}) case, the two wings of the plate have unequal lengths, 8-$\mu\rm m$ and 2-$\mu\rm m$, wherein the translation freedom is locked due to the absence of the contact asymmetry. In the translation ({\bf D}) case, the two wings are of equal length to lock the rotation freedom, and the contact surface situates in the half side of the plate.
The temperature distributions at different time in the plate are provided in Figs. S11B and S12B in the Supplementary Material.
{\bf E.} Simulated (left) and measured (right) rotation speeds as functions of wings' asymmetry factor $A_f$. The repetition rate of laser pulses is 2.04-kHz. Scale bar: 4 $\mu\rm m$.
{\bf F.} Confocal laser scanning microscopy image (left) and temporal sequencing of optical images (right) of a hexagonal gold plate on a micro-fiber, evidencing the direct correlation between the contact asymmetry and the translation direction.
}
\label{fig::mechansim}
\end{figure*}

\noindent As an essential step towards realizing practical actuators, a complete control of motion direction is indispensable.
For actuators driven by optical force, their control could be implemented by changing the propagation directions of the incident light.
However, the same scheme fails here, wherein the dominant force is the friction force.
Instead, we observed that the motion directions of the gold plates are independent of the light directions, as shown in Fig.~\ref{fig:::motion_dir} and  Supplementary Movie 3.
This observation has been confirmed with similar tests involving 40 different gold plates~\footnote{Among 40 tests, the independence of the motion directions on the light directions was observed for 38 gold plates.
Only for 2 gold plates, their translation directions changed with the light directions.
However, in such 2 exceptional examples, the positions of the plates on the micro-fibers were noticeably changed while we reversed the light directions.
In view of this, we exclude these two exceptions and conclude that the motion directions of the gold plates are independent of the light directions.
}.
Furthermore, we shall demonstrate below that both rotation and translation directions can be manipulated by adjusting how the gold plates are placed on and contact with the micro-fibers.

{\it Rotation}---The dimension asymmetry in two wings of the gold plates---demarcated by the micro-fibers---is the key enabling the rotation of the plates (Fig.~\ref{fig::mechansim}A), which has been firstly observed by Lu et al.~\cite{Lu:2019} and is confirmed here for the spiral motion.
As shown in Fig.~\ref{fig::sprial}B and Fig. S15 in the Supplementary Material, the rotation direction points from the long wing to the short wing (that is, the long wing seems to push forward the short wing to rotate).
In the presence of the symmetry in two wings, the rotation stops.
The simulation and experimental results in Fig.~\ref{fig::mechansim}E confirm this, showing that, the rotation speed monotonically decreases as the asymmetry factor $A_f$ (see the inset for the definition of $A_f$) decreases (i.e., two wings become more symmetric)
The rotation mechanism can be qualitatively comprehended with the insights from the pedagogically model of Fig.~\ref{fig::principle} that considers a specific case of rectangular plates,
and by additionally taking multiple reflections of the $L$-modes due to finite size effects of the gold plates into account.
For a more precise quantitative prediction, one should refer to numerical simulations as will be discussed with Fig.~\ref{fig::mechansim}.
Briefly, the optical-absorption hotspot centralized around the contact surface generates the $L-$modes, which propagate towards the long- and short-wing sides, and, for convenience, are termed as “long-wing (LW)" and “short-wing (SW)" waves, respectively. These initial LW and SW waves carry the dominant elastic displacements towards the long-wing and short-wing sides, respectively.
As these LW and SW waves reach the ends of the plates, their propagation directions reverse and turn towards the contact surface.
Next, when these waves return back to the contact surface, the contact surface rotates as a result of the elementary interaction between an incident elastic waves and the friction force (see Fig.~\ref{fig::principle}).
This elementary interaction continuously happens as the $L$-modes bounce back and forth inside the plates, manifesting in the temporal oscillations of the rotation displacement (see the upper panel in Fig.~\ref{fig::mechansim}C); and the sliding gradually weakens due to elastic attenuation, as indicated in Eq.~\eqref{eq::PTH} that the power threshold increases with the propagation time $t_0$ of the reflected elastic waves.
Notably, albeit the reversion of the propagation direction due to the reflection, the LW and SW waves retain the same directions of the elastic oscillations, i.e., pointing towards the long- and short-wing sides, respectively.
This is due to that the reflection adds negligible phase changes to the $L$-modes (see Fig.~S7 in the Supplementary Material).
Consequently, the LW and SW waves consistently drive the plates to rotate towards the long- and short-wing sides oppositely.
The net rotation points towards the short-wing side since the SW waves are less attenuated thanks to their shorter traveling distance. This process is captured with the simulated profiles of the elastic displacement in the rotation direction at different time, as shown in the lower panel of Fig.~\ref{fig::mechansim}C.
In this process, the friction force plays the same double role as in Fig.~\ref{fig::principle}.
First, as the contact surface initially rotates towards the short-wing side, the friction force on average points towards the long-wing side resisting the rotation (see the middle panel in Fig.~\ref{fig::mechansim}C).
Then, as time increases and the thermal cooling begins, the friction force plays an opposite role: it instead points towards the short-wing side against thermal contraction (see the inset in the middle panel of Fig.~\ref{fig::mechansim}C), thereby preventing the contact surface from returning to its original position.
The finishing of this cooling process takes time of about $10^4$ ns (see Fig. S10B in the Supplementary Material).

{\it Translation}---The translation of the gold plates essentially requires the breaking of reflection symmetry in the axial direction of the micro-fibers.
Otherwise, if the reflection symmetry is respected, there is no directional bias to trigger the translation.
Here, the asymmetry in the contact surface is identified to be the key enabling the translation, which is consistent with our observation that the translation direction is independent of light direction.
This novel translation mechanism marks the essential difference between our scheme and another in a recent publication~\cite{Linghu:2021}---that proposes the asymmetry in the spatial distribution of optical absorption as the key factor and observed the dependence of the translation direction on light direction (see Sec. 6-B2 in the Supplementary Material for more discussions).

\begin{figure}[ht!]
\includegraphics[width=8cm]{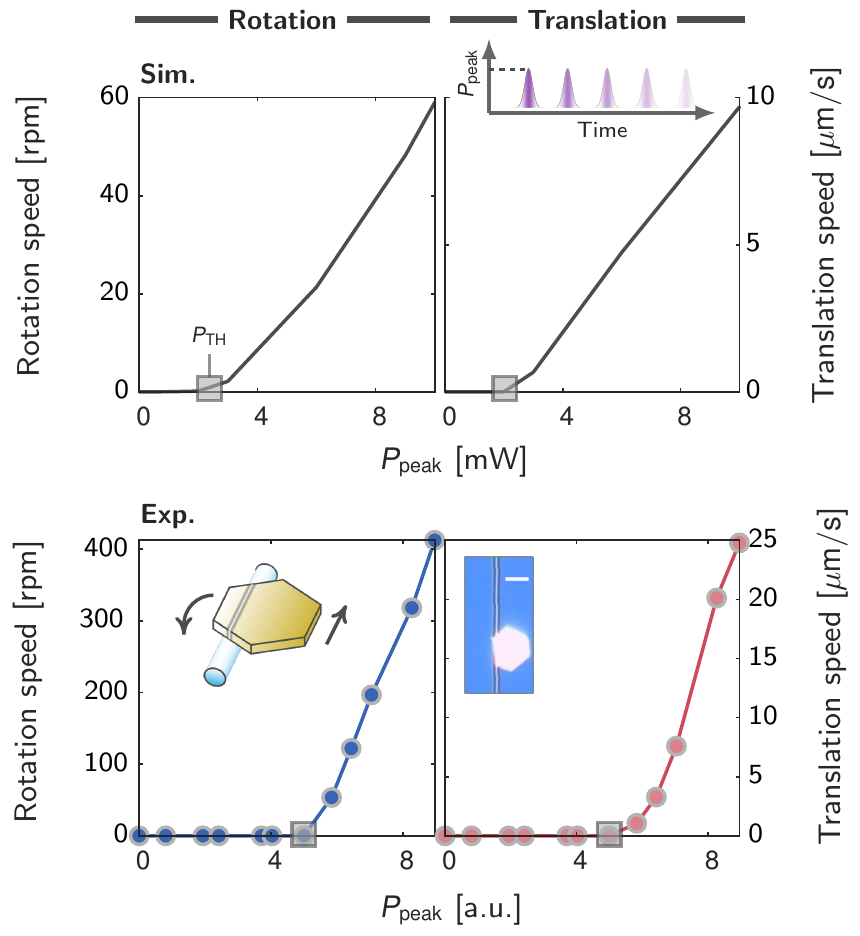}
\caption{ {\bf Observation of sliding threshold power $P_{\rm TH}$ in spiral motion of gold plates.} The numerical results of rotation and translation speeds as functions of absorbed peak power (top panel) for rectangular gold plates [same as those studied in Fig.~\ref{fig::mechansim}{\bf C} (rotation) and {\bf D} (translation), respectively] feature the mW-scale sliding power threshold $P_{\rm TH}$, which qualitatively agree with the measurement results (lower panel) for a hexangular plate (same as in Fig.~\ref{fig::sprial}{\bf B}). The numerical simulations adopt the same parameters of the pulsed absorbed optical power, the micro-fiber, and the frictional sliding resistance as in Fig.~\ref{fig::mechansim}{\bf C}-{\bf D}. The used super-continuum laser pulses have 6.8--mW average power, 3--ns temporal width and 17.50--kHz repetition rate. The scale bar in the inset optical image is 15 $\mu\rm m$.
}
\label{fig::power}
\end{figure}

The origin of the contact asymmetry is the surface curvature of the plates.
Interestingly, the surface curvature is {\it unintentionally} introduced during the process of transferring the plates onto the micro-fibers.
The plates are initially placed on glass substrates.
In order to peal the plates off the substrates and then drag them to the micro-fibers, the tapered fibers are used as levers to tilt the plates up, wherein the plates are inevitably bent (see Fig. S20 in the Supplementary Material).
Notably, the surface curvature can even be as large as a few micro-meters, as shown by confocal-laser-scanning-microscopy (CLSM) images in Fig.~\ref{fig::mechansim}F and Fig. S21 in the Supplementary Material.

As a result of the existing surface curvature, the real contact surface potentially only occupies parts of the ideal flat surface (in the absence of the surface curvature), thereby dividing the latter into the non-contact and contact sides, respectively (see Figs.~\ref{fig::mechansim}B and F).
The temporal sequencing of optical images for a hexagonal plate in Fig.~\ref{fig::mechansim}F shows that the translation direction points from the non-contact side to the contact side.
We numerically reproduce the same observation in Fig.~\ref{fig::mechansim}D.
Specifically, the upper panel of Fig.~\ref{fig::mechansim}D contrasts the temporal evolution of the translation displacements of two points on the non-contact ($P_{\rm n-ct}$) and contact ($P_{\rm ct}$) sides, respectively, and shows that the stabilized translation direction points from the $P_{\rm n-ct}$ point to the $P_{\rm ct}$ point.

The dependence of the translation direction on the contact asymmetry can be appreciated by analyzing the friction force (see the middle panel in Fig.~\ref{fig::mechansim}D).
First, in the initial sliding period, the thermally-excited elastic waves expand the two ends of the plate in opposite directions, and, the friction force on-average pointing towards the $P_{\rm n-ct}$ point resists the thermal expansion of the contact-side.
Next, as the thermal cooling starts, the friction force instead, pointing towards the $P_{\rm ct}$ point, preserves the previously accumulated sliding distance of the contact surface, and, accordingly, drags the non-contact side toward the contact side (see the lower panel in Fig.~\ref{fig::mechansim}D).
Consequently, the net motion direction is towards the contact side, which physically relates with the excitations of the $T$-modes whose elastic oscillation direction is parallel to the axial direction of the micro-fiber (see the inset in Fig.~\ref{fig::principle}B).
Note that, in the cooling period, the contact side does not stay in place after the initial quick expansion, and instead slowly reduces to a smaller value.
This is attributed to that the excited elastic waves bounce back and forth inside in the plate, resulting in a homogenized distribution of the elastic waves, and, accordingly, making the translation displacements of the two sides approach each other.
With these insights, we underline that the translation direction can be controlled by adjusting the contact asymmetry.
In view of this, to adjust the translation direction, one can either rotate the plates with a angle of $\pi$ or change the relative positions between the plates and the micro-fibers (see Fig.~S18 in the Supplementary Material and Supplementary Movie 4), so that the spatial configurations of the non-contact and contact sides could change.

\section*{Discussion}

The experimental results presented here demonstrate two-dimensional locomotion of micro-scale objects on dry surfaces in the presence of $\mu\rm N$-scale friction force by exploiting thermally-excited elastic waves.
The theoretical analysis---reinforced by the experimental results---reveals the elementary interaction dynamics between elastic waves and friction force, highlighting the double role of friction force in both hindering and promoting actuation at different stages.
The feasibility of the proposed scheme manifests in mild mW-scale threshold optical absorption instantaneous power $P_{\rm TH}$ required to enable actuation.

Experimentally, we observed the threshold power $P_{\rm TH}$ by monitoring the power dependence of the rotation and translation speeds (see Fig.~\ref{fig::power} and Supplementary Movie 5; noting the qualitative agreement between the theoretical and experimental results).
This result---together with the predicted linear relation between $P_{\rm TH}$ and friction force [Eq.~\eqref{eq::PTH}]---suggests that the proposed actuation system can also be used to probe, analyze friction force at micro-scales by monitoring changes of $P_{\rm TH}$, and further to investigate friction laws.
Regarding this, a qualitative determination of the friction resistance requires a precise measurement of $P_{\rm TH}$. This requirement favors the use of a narrow-band pulsed laser, with which $P_{\rm TH}$ could be more precisely measured, e.g., using a power meter that generally has a limited bandwidth.

The demonstrated locomotion is driven by single laser pulses in a step wise fashion. The step actuation distance can be as small as a few nanometers and even approach sub-nanometers.
The motion speed increases with the pulse repetition rate and the pulse power.
In addition, we discover that the motion direction is controllable by mechanically adjusting the relative positions and contact configurations between the gold plates and the microfibers.
Further, to improve this controllability---particularly relating with the contact configurations---, one practical direction is to fabricate micro-plates with predefined surface curvatures, e.g., with electron-beam lithography methods.
Besides, it is also meaningful to explore new designs for the realization of the motion control with a full-optical way, i.e., by tuning the propagation direction, wavelength, and fiber modes of the incident laser pulses (see Sec. 6-B2 in the Supplementary Material).

We envision the proposed actuation scheme can in principle find practical applications in various fields that require to precisely manipulate micro-objects in non-liquid environments. For instance, integrating our technique with an on-chip waveguide coupled network, one can in principle achieve optical modulation by adjusting positions of a gold plate on top of the waveguide to control waveguide transmission via tuning coupling between nearby waveguides.
Moreover, it can also be used for transporting dielectric particles attached to the surface of a gold plate along a micro-fiber/nano-wire, which is essential in lab-on-a-chip technologies, e.g., for life-science applications.

\section*{Materials and methods}

\noindent{\bf Experiments}

\noindent The micrometer-sized hexagonal gold plates are synthesized in large quantities by introducing aniline ($\rm C_6H_7N$) to a heated ethylene glycol (EG) solution of hydrogen tetrachloroaurate ($\rm HAuC_{l4}·4H_2O$), following the recipe reported in Ref.~\cite{Guo:2006}. The typical side length and thickness of synthesized gold plates are in the order of several tens of micrometers and of nanometers, respectively. The gold plates in the shape of triangle, circle, rectangle and square are fabricated by electron beam lithography. The process was carried out on a $\rm SiO_2$ substrate spin-coated at 5000 r.p.m. with poly (methyl methacrylate) (MicroChem 950 PMMA A4) electronic resist. An electron beam lithography machine (ELPHY Quantum, Raith, Germany) equipped with a pattern generator was used for direct writing the designed micro-structures. Then the exposed resist was developed in a conventional solution of methyl isobutyl ketone-isopropyl alcohol (MIBK-IPA) (1:3) for 60 s. Using electron beam evaporation, 60-nm Au was deposited on the sample. Finally, a gap was formed at the interface between the exposed and unexposed regime, making it easy to transfer towards the microfiber by a tapered fiber.

The microfibers are fabricated using flame-heated drawing technique from standard optical fibers~\cite{Tong:2013}. The fabrication procedures can be described as follows. A hydrogen flame is used for heating the fiber. Then, the fiber is stretched and elongated gradually with the reduced diameter until the desired diameter of the microfiber is reached, under a optimized pulling force exerted on the standard fiber at both ends.

The used pulsed supercontinuum light (SuperK Compact, NKT photonics) has  a wavelength range of 450 nm-2400 nm, a pulse duration of 2.6 ns, and tunable repetition rates ranging from 230 Hz to 23 kHz. The variable optical attenuator is used for controlling the output attenuation. Then, the pulsed light with tunable single pules energy is first lens-coupled into a standard silica fiber (SMF-28, Corning) and next delivered into a microfiber. The movements of the gold plates on microfibers in ambient conditions are monitored using a 10X microscope objective and a CCD camera.\\

\noindent{\bf Simulations}

\noindent Numerical simulations, in Fig. 1C and D, Fig. 4C and D, are performed using the commercial finite-element simulation software, COMSOL Multiphysics. The coupled heat-mechanical simulations are performed by coupling the COMSOL modules of ‘‘Heat Transfer in Solids" and ‘‘Solid Mechanics". Friction force is introduced through the augmented Lagrangian method using the contact node under the ‘‘Solid Mechanics" module. The COMSOL files can be requested by contacting W.Y..
\\
\\

\noindent{\bf Acknowledgements} \\
\noindent This project was supported by the National Key Research and Development Program of China (2017YFA0205700), the National Natural Science Foundation of China (61927820, 61905200, 12004313), and the China Postdoctoral Science Foundation (2020M671809).
\\
\\

\noindent{\bf Author contributions} \\
W. Tang, W. Lv and J. Lu performed the experiments with help from F. Liu and J. Wang. W. Yan performed the theoretical analysis and numerical simulations with W. Tang. M. Qiu supervised the project.
All the authors contributed to the final manuscript.
\\
\\

\noindent{\bf Conflict of interest} \\
The authors declare no competing interests.
\bibliography{ms}

\begin{thebibliography}{36}%
\makeatletter
\providecommand \@ifxundefined [1]{%
 \@ifx{#1\undefined}
}%
\providecommand \@ifnum [1]{%
 \ifnum #1\expandafter \@firstoftwo
 \else \expandafter \@secondoftwo
 \fi
}%
\providecommand \@ifx [1]{%
 \ifx #1\expandafter \@firstoftwo
 \else \expandafter \@secondoftwo
 \fi
}%
\providecommand \natexlab [1]{#1}%
\providecommand \enquote  [1]{``#1''}%
\providecommand \bibnamefont  [1]{#1}%
\providecommand \bibfnamefont [1]{#1}%
\providecommand \citenamefont [1]{#1}%
\providecommand \href@noop [0]{\@secondoftwo}%
\providecommand \href [0]{\begingroup \@sanitize@url \@href}%
\providecommand \@href[1]{\@@startlink{#1}\@@href}%
\providecommand \@@href[1]{\endgroup#1\@@endlink}%
\providecommand \@sanitize@url [0]{\catcode `\\12\catcode `\$12\catcode
  `\&12\catcode `\#12\catcode `\^12\catcode `\_12\catcode `\%12\relax}%
\providecommand \@@startlink[1]{}%
\providecommand \@@endlink[0]{}%
\providecommand \url  [0]{\begingroup\@sanitize@url \@url }%
\providecommand \@url [1]{\endgroup\@href {#1}{\urlprefix }}%
\providecommand \urlprefix  [0]{URL }%
\providecommand \Eprint [0]{\href }%
\providecommand \doibase [0]{https://doi.org/}%
\providecommand \selectlanguage [0]{\@gobble}%
\providecommand \bibinfo  [0]{\@secondoftwo}%
\providecommand \bibfield  [0]{\@secondoftwo}%
\providecommand \translation [1]{[#1]}%
\providecommand \BibitemOpen [0]{}%
\providecommand \bibitemStop [0]{}%
\providecommand \bibitemNoStop [0]{.\EOS\space}%
\providecommand \EOS [0]{\spacefactor3000\relax}%
\providecommand \BibitemShut  [1]{\csname bibitem#1\endcsname}%
\let\auto@bib@innerbib\@empty
\bibitem [{\citenamefont {Feynman}(1960)}]{Feynman:1960}%
  \BibitemOpen
  \bibfield  {author} {\bibinfo {author} {\bibfnamefont {R.~P.}\ \bibnamefont
  {Feynman}},\ }\bibfield  {title} {\bibinfo {title} {{\it There's Plenty of
  Room at the Bottom}},\ }\href {http://calteches.library.caltech.edu/1976}
  {\bibfield  {journal} {\bibinfo  {journal} {Engineering and Science}\
  }\textbf {\bibinfo {volume} {23}},\ \bibinfo {pages} {22} (\bibinfo {year}
  {1960})}\BibitemShut {NoStop}%
\bibitem [{\citenamefont {Neukermans}\ and\ \citenamefont
  {Ramaswami}(2001)}]{Neukermans:2001}%
  \BibitemOpen
  \bibfield  {author} {\bibinfo {author} {\bibfnamefont {A.}~\bibnamefont
  {Neukermans}}\ and\ \bibinfo {author} {\bibfnamefont {R.}~\bibnamefont
  {Ramaswami}},\ }\bibfield  {title} {\bibinfo {title} {{\it MEMS technology
  for optical networking applications}},\ }\href
  {https://doi.org/10.1109/35.894378} {\bibfield  {journal} {\bibinfo
  {journal} {IEEE Commun. Mag.}\ }\textbf {\bibinfo {volume} {39}},\ \bibinfo
  {pages} {62} (\bibinfo {year} {2001})}\BibitemShut {NoStop}%
\bibitem [{\citenamefont {Wu}\ \emph {et~al.}(2006)\citenamefont {Wu},
  \citenamefont {Solgaard},\ and\ \citenamefont {Ford}}]{Wu:2006}%
  \BibitemOpen
  \bibfield  {author} {\bibinfo {author} {\bibfnamefont {M.~C.}\ \bibnamefont
  {Wu}}, \bibinfo {author} {\bibfnamefont {O.}~\bibnamefont {Solgaard}},\ and\
  \bibinfo {author} {\bibfnamefont {J.~E.}\ \bibnamefont {Ford}},\ }\bibfield
  {title} {\bibinfo {title} {{\it Optical MEMS for Lightwave Communication}},\
  }\href {https://doi.org/10.1109/JLT.2006.886405} {\bibfield  {journal}
  {\bibinfo  {journal} {J. Lightwave Technol.}\ }\textbf {\bibinfo {volume}
  {24}},\ \bibinfo {pages} {4433} (\bibinfo {year} {2006})}\BibitemShut
  {NoStop}%
\bibitem [{\citenamefont {Kan}\ \emph {et~al.}(2015)\citenamefont {Kan},
  \citenamefont {Isozaki}, \citenamefont {Kanda}, \citenamefont {Nemoto},
  \citenamefont {Konishi}, \citenamefont {Takahashi}, \citenamefont
  {Kuwata-Gonokami}, \citenamefont {Matsumoto},\ and\ \citenamefont
  {Shimoyama}}]{Kan:2015}%
  \BibitemOpen
  \bibfield  {author} {\bibinfo {author} {\bibfnamefont {T.}~\bibnamefont
  {Kan}}, \bibinfo {author} {\bibfnamefont {A.}~\bibnamefont {Isozaki}},
  \bibinfo {author} {\bibfnamefont {N.}~\bibnamefont {Kanda}}, \bibinfo
  {author} {\bibfnamefont {N.}~\bibnamefont {Nemoto}}, \bibinfo {author}
  {\bibfnamefont {K.}~\bibnamefont {Konishi}}, \bibinfo {author} {\bibfnamefont
  {H.}~\bibnamefont {Takahashi}}, \bibinfo {author} {\bibfnamefont
  {M.}~\bibnamefont {Kuwata-Gonokami}}, \bibinfo {author} {\bibfnamefont
  {K.}~\bibnamefont {Matsumoto}},\ and\ \bibinfo {author} {\bibfnamefont
  {I.}~\bibnamefont {Shimoyama}},\ }\bibfield  {title} {\bibinfo {title} {{\it
  Enantiomeric switching of chiral metamaterial for terahertz polarization
  modulation employing vertically deformable MEMS spirals}},\ }\href
  {https://doi.org/10.1038/ncomms9422} {\bibfield  {journal} {\bibinfo
  {journal} {Nat. Commun.}\ }\textbf {\bibinfo {volume} {6}},\ \bibinfo {pages}
  {1} (\bibinfo {year} {2015})}\BibitemShut {NoStop}%
\bibitem [{\citenamefont {Haffner}\ \emph {et~al.}(2019)\citenamefont
  {Haffner}, \citenamefont {Joerg}, \citenamefont {Doderer}, \citenamefont
  {Mayor}, \citenamefont {Chelladurai}, \citenamefont {Fedoryshyn},
  \citenamefont {Roman}, \citenamefont {Mazur}, \citenamefont {Burla},
  \citenamefont {Lezec}, \citenamefont {Aksyuk},\ and\ \citenamefont
  {Leuthold}}]{Haffner:2019}%
  \BibitemOpen
  \bibfield  {author} {\bibinfo {author} {\bibfnamefont {C.}~\bibnamefont
  {Haffner}}, \bibinfo {author} {\bibfnamefont {A.}~\bibnamefont {Joerg}},
  \bibinfo {author} {\bibfnamefont {M.}~\bibnamefont {Doderer}}, \bibinfo
  {author} {\bibfnamefont {F.}~\bibnamefont {Mayor}}, \bibinfo {author}
  {\bibfnamefont {D.}~\bibnamefont {Chelladurai}}, \bibinfo {author}
  {\bibfnamefont {Y.}~\bibnamefont {Fedoryshyn}}, \bibinfo {author}
  {\bibfnamefont {C.~I.}\ \bibnamefont {Roman}}, \bibinfo {author}
  {\bibfnamefont {M.}~\bibnamefont {Mazur}}, \bibinfo {author} {\bibfnamefont
  {M.}~\bibnamefont {Burla}}, \bibinfo {author} {\bibfnamefont {H.~J.}\
  \bibnamefont {Lezec}}, \bibinfo {author} {\bibfnamefont {V.~A.}\ \bibnamefont
  {Aksyuk}},\ and\ \bibinfo {author} {\bibfnamefont {J.}~\bibnamefont
  {Leuthold}},\ }\bibfield  {title} {\bibinfo {title} {{\it
  Nano{\textendash}opto-electro-mechanical switches operated at CMOS-level
  voltages}},\ }\href {https://doi.org/10.1126/science.aay8645} {\bibfield
  {journal} {\bibinfo  {journal} {Science}\ }\textbf {\bibinfo {volume}
  {366}},\ \bibinfo {pages} {860} (\bibinfo {year} {2019})}\BibitemShut
  {NoStop}%
\bibitem [{\citenamefont {Van~Kessel}\ \emph {et~al.}(1998)\citenamefont
  {Van~Kessel}, \citenamefont {Hornbeck}, \citenamefont {Meier},\ and\
  \citenamefont {Douglass}}]{VanKessel:1998}%
  \BibitemOpen
  \bibfield  {author} {\bibinfo {author} {\bibfnamefont {P.~F.}\ \bibnamefont
  {Van~Kessel}}, \bibinfo {author} {\bibfnamefont {L.~J.}\ \bibnamefont
  {Hornbeck}}, \bibinfo {author} {\bibfnamefont {R.~E.}\ \bibnamefont
  {Meier}},\ and\ \bibinfo {author} {\bibfnamefont {M.~R.}\ \bibnamefont
  {Douglass}},\ }\bibfield  {title} {\bibinfo {title} {{\it A MEMS-based
  projection display}},\ }\href {https://doi.org/10.1109/5.704274} {\bibfield
  {journal} {\bibinfo  {journal} {Proc. IEEE}\ }\textbf {\bibinfo {volume}
  {86}},\ \bibinfo {pages} {1687} (\bibinfo {year} {1998})}\BibitemShut
  {NoStop}%
\bibitem [{\citenamefont {Wang}(2012)}]{Wang:2012}%
  \BibitemOpen
  \bibfield  {author} {\bibinfo {author} {\bibfnamefont {J.}~\bibnamefont
  {Wang}},\ }\bibfield  {title} {\bibinfo {title} {{\it Cargo-towing synthetic
  nanomachines: Towards active transport in microchip devices}},\ }\href
  {https://doi.org/10.1039/C2LC00003B} {\bibfield  {journal} {\bibinfo
  {journal} {Lab Chip}\ }\textbf {\bibinfo {volume} {12}},\ \bibinfo {pages}
  {1944} (\bibinfo {year} {2012})}\BibitemShut {NoStop}%
\bibitem [{\citenamefont {Tottori}\ \emph {et~al.}(2012)\citenamefont
  {Tottori}, \citenamefont {Zhang}, \citenamefont {Qiu}, \citenamefont
  {Krawczyk}, \citenamefont {Franco-Obreg{\ifmmode\acute{o}\else\'{o}\fi}n},\
  and\ \citenamefont {Nelson}}]{Tottori:2012}%
  \BibitemOpen
  \bibfield  {author} {\bibinfo {author} {\bibfnamefont {S.}~\bibnamefont
  {Tottori}}, \bibinfo {author} {\bibfnamefont {L.}~\bibnamefont {Zhang}},
  \bibinfo {author} {\bibfnamefont {F.}~\bibnamefont {Qiu}}, \bibinfo {author}
  {\bibfnamefont {K.~K.}\ \bibnamefont {Krawczyk}}, \bibinfo {author}
  {\bibfnamefont {A.}~\bibnamefont
  {Franco-Obreg{\ifmmode\acute{o}\else\'{o}\fi}n}},\ and\ \bibinfo {author}
  {\bibfnamefont {B.~J.}\ \bibnamefont {Nelson}},\ }\bibfield  {title}
  {\bibinfo {title} {{\it Magnetic Helical Micromachines: Fabrication,
  Controlled Swimming, and Cargo Transport}},\ }\href
  {https://doi.org/10.1002/adma.201103818} {\bibfield  {journal} {\bibinfo
  {journal} {Adv. Mater.}\ }\textbf {\bibinfo {volume} {24}},\ \bibinfo {pages}
  {811} (\bibinfo {year} {2012})}\BibitemShut {NoStop}%
\bibitem [{\citenamefont {Judy}(2001)}]{Judy:2001}%
  \BibitemOpen
  \bibfield  {author} {\bibinfo {author} {\bibfnamefont {J.~W.}\ \bibnamefont
  {Judy}},\ }\bibfield  {title} {\bibinfo {title} {{\it Microelectromechanical
  systems (MEMS):}},\ }\href {https://doi.org/10.1088/0964-1726/10/6/301}
  {\bibfield  {journal} {\bibinfo  {journal} {Smart Mater. Struct.}\ }\textbf
  {\bibinfo {volume} {10}},\ \bibinfo {pages} {1115} (\bibinfo {year}
  {2001})}\BibitemShut {NoStop}%
\bibitem [{\citenamefont {DelRio}\ \emph {et~al.}(2005)\citenamefont {DelRio},
  \citenamefont {de~Boer}, \citenamefont {Knapp}, \citenamefont {Reedy},
  \citenamefont {Clews},\ and\ \citenamefont {Dunn}}]{DelRio:2005}%
  \BibitemOpen
  \bibfield  {author} {\bibinfo {author} {\bibfnamefont {F.~W.}\ \bibnamefont
  {DelRio}}, \bibinfo {author} {\bibfnamefont {M.~P.}\ \bibnamefont {de~Boer}},
  \bibinfo {author} {\bibfnamefont {J.~A.}\ \bibnamefont {Knapp}}, \bibinfo
  {author} {\bibfnamefont {E.~D.}\ \bibnamefont {Reedy}}, \bibinfo {author}
  {\bibfnamefont {P.~J.}\ \bibnamefont {Clews}},\ and\ \bibinfo {author}
  {\bibfnamefont {M.~L.}\ \bibnamefont {Dunn}},\ }\bibfield  {title} {\bibinfo
  {title} {{\it The role of van der Waals forces in adhesion of micromachined
  surfaces}},\ }\href {https://doi.org/10.1038/nmat1431} {\bibfield  {journal}
  {\bibinfo  {journal} {Nat. Mater.}\ }\textbf {\bibinfo {volume} {4}},\
  \bibinfo {pages} {629} (\bibinfo {year} {2005})}\BibitemShut {NoStop}%
\bibitem [{\citenamefont {Ashkin}(1970)}]{Ashkin:1970}%
  \BibitemOpen
  \bibfield  {author} {\bibinfo {author} {\bibfnamefont {A.}~\bibnamefont
  {Ashkin}},\ }\bibfield  {title} {\bibinfo {title} {{\it Acceleration and
  Trapping of Particles by Radiation Pressure}},\ }\href
  {https://doi.org/10.1103/PhysRevLett.24.156} {\bibfield  {journal} {\bibinfo
  {journal} {Phys. Rev. Lett.}\ }\textbf {\bibinfo {volume} {24}},\ \bibinfo
  {pages} {156} (\bibinfo {year} {1970})}\BibitemShut {NoStop}%
\bibitem [{\citenamefont {Grier}(2003)}]{Grier:2003}%
  \BibitemOpen
  \bibfield  {author} {\bibinfo {author} {\bibfnamefont {D.~G.}\ \bibnamefont
  {Grier}},\ }\bibfield  {title} {\bibinfo {title} {{\it A revolution in
  optical manipulation}},\ }\href {https://doi.org/10.1038/nature01935}
  {\bibfield  {journal} {\bibinfo  {journal} {Nature}\ }\textbf {\bibinfo
  {volume} {424}},\ \bibinfo {pages} {810} (\bibinfo {year}
  {2003})}\BibitemShut {NoStop}%
\bibitem [{\citenamefont {MacDonald}\ \emph {et~al.}(2003)\citenamefont
  {MacDonald}, \citenamefont {Spalding},\ and\ \citenamefont
  {Dholakia}}]{MacDonald:2003}%
  \BibitemOpen
  \bibfield  {author} {\bibinfo {author} {\bibfnamefont {M.~P.}\ \bibnamefont
  {MacDonald}}, \bibinfo {author} {\bibfnamefont {G.~C.}\ \bibnamefont
  {Spalding}},\ and\ \bibinfo {author} {\bibfnamefont {K.}~\bibnamefont
  {Dholakia}},\ }\bibfield  {title} {\bibinfo {title} {{\it Microfluidic
  sorting in an optical lattice}},\ }\href
  {https://doi.org/10.1038/nature02144} {\bibfield  {journal} {\bibinfo
  {journal} {Nature}\ }\textbf {\bibinfo {volume} {426}},\ \bibinfo {pages}
  {421} (\bibinfo {year} {2003})}\BibitemShut {NoStop}%
\bibitem [{\citenamefont {Neale}\ \emph {et~al.}(2005)\citenamefont {Neale},
  \citenamefont {MacDonald}, \citenamefont {Dholakia},\ and\ \citenamefont
  {Krauss}}]{Neale:2005}%
  \BibitemOpen
  \bibfield  {author} {\bibinfo {author} {\bibfnamefont {S.~L.}\ \bibnamefont
  {Neale}}, \bibinfo {author} {\bibfnamefont {M.~P.}\ \bibnamefont
  {MacDonald}}, \bibinfo {author} {\bibfnamefont {K.}~\bibnamefont
  {Dholakia}},\ and\ \bibinfo {author} {\bibfnamefont {T.~F.}\ \bibnamefont
  {Krauss}},\ }\bibfield  {title} {\bibinfo {title} {{\it All-optical control
  of microfluidic components using form birefringence}},\ }\href
  {https://doi.org/10.1038/nmat1411} {\bibfield  {journal} {\bibinfo  {journal}
  {Nat. Mater.}\ }\textbf {\bibinfo {volume} {4}},\ \bibinfo {pages} {530}
  (\bibinfo {year} {2005})}\BibitemShut {NoStop}%
\bibitem [{\citenamefont {Gao}\ \emph {et~al.}(2017)\citenamefont {Gao},
  \citenamefont {Ding}, \citenamefont {Nieto-Vesperinas}, \citenamefont {Ding},
  \citenamefont {Rahman}, \citenamefont {Zhang}, \citenamefont {Lim},\ and\
  \citenamefont {Qiu}}]{Gao:2017}%
  \BibitemOpen
  \bibfield  {author} {\bibinfo {author} {\bibfnamefont {D.}~\bibnamefont
  {Gao}}, \bibinfo {author} {\bibfnamefont {W.}~\bibnamefont {Ding}}, \bibinfo
  {author} {\bibfnamefont {M.}~\bibnamefont {Nieto-Vesperinas}}, \bibinfo
  {author} {\bibfnamefont {X.}~\bibnamefont {Ding}}, \bibinfo {author}
  {\bibfnamefont {M.}~\bibnamefont {Rahman}}, \bibinfo {author} {\bibfnamefont
  {T.}~\bibnamefont {Zhang}}, \bibinfo {author} {\bibfnamefont
  {C.}~\bibnamefont {Lim}},\ and\ \bibinfo {author} {\bibfnamefont {C.-W.}\
  \bibnamefont {Qiu}},\ }\bibfield  {title} {\bibinfo {title} {{\it Optical
  manipulation from the microscale to the nanoscale: fundamentals, advances and
  prospects}},\ }\href {https://doi.org/10.1038/lsa.2017.39} {\bibfield
  {journal} {\bibinfo  {journal} {Light Sci. Appl.}\ }\textbf {\bibinfo
  {volume} {6}},\ \bibinfo {pages} {e17039} (\bibinfo {year}
  {2017})}\BibitemShut {NoStop}%
\bibitem [{\citenamefont {Tanaka}\ \emph {et~al.}(2020)\citenamefont {Tanaka},
  \citenamefont {Albella}, \citenamefont {Rahmani}, \citenamefont {Giannini},
  \citenamefont {Maier},\ and\ \citenamefont {Shimura}}]{Tanaka:2020}%
  \BibitemOpen
  \bibfield  {author} {\bibinfo {author} {\bibfnamefont {Y.~Y.}\ \bibnamefont
  {Tanaka}}, \bibinfo {author} {\bibfnamefont {P.}~\bibnamefont {Albella}},
  \bibinfo {author} {\bibfnamefont {M.}~\bibnamefont {Rahmani}}, \bibinfo
  {author} {\bibfnamefont {V.}~\bibnamefont {Giannini}}, \bibinfo {author}
  {\bibfnamefont {S.~A.}\ \bibnamefont {Maier}},\ and\ \bibinfo {author}
  {\bibfnamefont {T.}~\bibnamefont {Shimura}},\ }\bibfield  {title} {\bibinfo
  {title} {{\it Plasmonic linear nanomotor using lateral optical forces}},\
  }\href {https://doi.org/10.1126/sciadv.abc3726} {\bibfield  {journal}
  {\bibinfo  {journal} {Sci. Adv.}\ }\textbf {\bibinfo {volume} {6}},\ \bibinfo
  {pages} {eabc3726} (\bibinfo {year} {2020})}\BibitemShut {NoStop}%
\bibitem [{\citenamefont {Ren}\ \emph {et~al.}(2020)\citenamefont {Ren},
  \citenamefont {Zeng}, \citenamefont {Zhou}, \citenamefont {Kong},
  \citenamefont {Mao}, \citenamefont {Qiu}, \citenamefont {Tsia},\ and\
  \citenamefont {Wong}}]{Ren:2020}%
  \BibitemOpen
  \bibfield  {author} {\bibinfo {author} {\bibfnamefont {Y.-X.}\ \bibnamefont
  {Ren}}, \bibinfo {author} {\bibfnamefont {X.}~\bibnamefont {Zeng}}, \bibinfo
  {author} {\bibfnamefont {L.-M.}\ \bibnamefont {Zhou}}, \bibinfo {author}
  {\bibfnamefont {C.}~\bibnamefont {Kong}}, \bibinfo {author} {\bibfnamefont
  {H.}~\bibnamefont {Mao}}, \bibinfo {author} {\bibfnamefont {C.-W.}\
  \bibnamefont {Qiu}}, \bibinfo {author} {\bibfnamefont {K.~K.}\ \bibnamefont
  {Tsia}},\ and\ \bibinfo {author} {\bibfnamefont {K.~K.~Y.}\ \bibnamefont
  {Wong}},\ }\bibfield  {title} {\bibinfo {title} {{\it Photonic Nanojet
  Mediated Backaction of Dielectric Microparticles}},\ }\href
  {https://doi.org/10.1021/acsphotonics.0c00242} {\bibfield  {journal}
  {\bibinfo  {journal} {ACS Photonics}\ }\textbf {\bibinfo {volume} {7}},\
  \bibinfo {pages} {1483} (\bibinfo {year} {2020})}\BibitemShut {NoStop}%
\bibitem [{\citenamefont {Shvedov}\ \emph {et~al.}(2010)\citenamefont
  {Shvedov}, \citenamefont {Rode}, \citenamefont {Izdebskaya}, \citenamefont
  {Desyatnikov}, \citenamefont {Krolikowski},\ and\ \citenamefont
  {Kivshar}}]{Shvedov:2010}%
  \BibitemOpen
  \bibfield  {author} {\bibinfo {author} {\bibfnamefont {V.~G.}\ \bibnamefont
  {Shvedov}}, \bibinfo {author} {\bibfnamefont {A.~V.}\ \bibnamefont {Rode}},
  \bibinfo {author} {\bibfnamefont {Y.~V.}\ \bibnamefont {Izdebskaya}},
  \bibinfo {author} {\bibfnamefont {A.~S.}\ \bibnamefont {Desyatnikov}},
  \bibinfo {author} {\bibfnamefont {W.}~\bibnamefont {Krolikowski}},\ and\
  \bibinfo {author} {\bibfnamefont {Y.~S.}\ \bibnamefont {Kivshar}},\
  }\bibfield  {title} {\bibinfo {title} {{\it Giant Optical Manipulation}},\
  }\href {https://doi.org/10.1103/PhysRevLett.105.118103} {\bibfield  {journal}
  {\bibinfo  {journal} {Phys. Rev. Lett.}\ }\textbf {\bibinfo {volume} {105}},\
  \bibinfo {pages} {118103} (\bibinfo {year} {2010})}\BibitemShut {NoStop}%
\bibitem [{\citenamefont {Shvedov}\ \emph {et~al.}(2014)\citenamefont
  {Shvedov}, \citenamefont {Davoyan}, \citenamefont {Hnatovsky}, \citenamefont
  {Engheta},\ and\ \citenamefont {Krolikowski}}]{Shvedov:2014}%
  \BibitemOpen
  \bibfield  {author} {\bibinfo {author} {\bibfnamefont {V.}~\bibnamefont
  {Shvedov}}, \bibinfo {author} {\bibfnamefont {A.~R.}\ \bibnamefont
  {Davoyan}}, \bibinfo {author} {\bibfnamefont {C.}~\bibnamefont {Hnatovsky}},
  \bibinfo {author} {\bibfnamefont {N.}~\bibnamefont {Engheta}},\ and\ \bibinfo
  {author} {\bibfnamefont {W.}~\bibnamefont {Krolikowski}},\ }\bibfield
  {title} {\bibinfo {title} {{\it A long-range polarization-controlled optical
  tractor beam}},\ }\href {https://doi.org/10.1038/nphoton.2014.242} {\bibfield
   {journal} {\bibinfo  {journal} {Nat. Photonics}\ }\textbf {\bibinfo {volume}
  {8}},\ \bibinfo {pages} {846} (\bibinfo {year} {2014})}\BibitemShut {NoStop}%
\bibitem [{\citenamefont {Palagi}\ and\ \citenamefont
  {Fischer}(2018)}]{Palagi:2018}%
  \BibitemOpen
  \bibfield  {author} {\bibinfo {author} {\bibfnamefont {S.}~\bibnamefont
  {Palagi}}\ and\ \bibinfo {author} {\bibfnamefont {P.}~\bibnamefont
  {Fischer}},\ }\bibfield  {title} {\bibinfo {title} {{\it Bioinspired
  microrobots}},\ }\href {https://doi.org/10.1038/s41578-018-0016-9} {\bibfield
   {journal} {\bibinfo  {journal} {Nat. Rev. Mater.}\ }\textbf {\bibinfo
  {volume} {3}},\ \bibinfo {pages} {113} (\bibinfo {year} {2018})}\BibitemShut
  {NoStop}%
\bibitem [{\citenamefont {Kendall}(1994)}]{Kendall:1994}%
  \BibitemOpen
  \bibfield  {author} {\bibinfo {author} {\bibfnamefont {K.}~\bibnamefont
  {Kendall}},\ }\bibfield  {title} {\bibinfo {title} {{\it Adhesion: molecules
  and mechanics}},\ }\href {https://doi.org/10.1126/science.263.5154.1720}
  {\bibfield  {journal} {\bibinfo  {journal} {Science}\ }\textbf {\bibinfo
  {volume} {263}},\ \bibinfo {pages} {1720} (\bibinfo {year}
  {1994})}\BibitemShut {NoStop}%
\bibitem [{\citenamefont {Lu}\ \emph {et~al.}(2017)\citenamefont {Lu},
  \citenamefont {Yang}, \citenamefont {Zhou}, \citenamefont {Yang},
  \citenamefont {Luo}, \citenamefont {Li},\ and\ \citenamefont
  {Qiu}}]{Lu:2017}%
  \BibitemOpen
  \bibfield  {author} {\bibinfo {author} {\bibfnamefont {J.}~\bibnamefont
  {Lu}}, \bibinfo {author} {\bibfnamefont {H.}~\bibnamefont {Yang}}, \bibinfo
  {author} {\bibfnamefont {L.}~\bibnamefont {Zhou}}, \bibinfo {author}
  {\bibfnamefont {Y.}~\bibnamefont {Yang}}, \bibinfo {author} {\bibfnamefont
  {S.}~\bibnamefont {Luo}}, \bibinfo {author} {\bibfnamefont {Q.}~\bibnamefont
  {Li}},\ and\ \bibinfo {author} {\bibfnamefont {M.}~\bibnamefont {Qiu}},\
  }\bibfield  {title} {\bibinfo {title} {{\it Light-Induced Pulling and Pushing
  by the Synergic Effect of Optical Force and Photophoretic Force}},\ }\href
  {https://doi.org/10.1103/PhysRevLett.118.043601} {\bibfield  {journal}
  {\bibinfo  {journal} {Phys. Rev. Lett.}\ }\textbf {\bibinfo {volume} {118}},\
  \bibinfo {pages} {043601} (\bibinfo {year} {2017})}\BibitemShut {NoStop}%
\bibitem [{\citenamefont {Lu}\ \emph {et~al.}(2019)\citenamefont {Lu},
  \citenamefont {Li}, \citenamefont {Qiu}, \citenamefont {Hong}, \citenamefont
  {Ghosh},\ and\ \citenamefont {Qiu}}]{Lu:2019}%
  \BibitemOpen
  \bibfield  {author} {\bibinfo {author} {\bibfnamefont {J.}~\bibnamefont
  {Lu}}, \bibinfo {author} {\bibfnamefont {Q.}~\bibnamefont {Li}}, \bibinfo
  {author} {\bibfnamefont {C.-W.}\ \bibnamefont {Qiu}}, \bibinfo {author}
  {\bibfnamefont {Y.}~\bibnamefont {Hong}}, \bibinfo {author} {\bibfnamefont
  {P.}~\bibnamefont {Ghosh}},\ and\ \bibinfo {author} {\bibfnamefont
  {M.}~\bibnamefont {Qiu}},\ }\bibfield  {title} {\bibinfo {title} {{\it
  Nanoscale Lamb wave{\textendash}driven motors in nonliquid environments}},\
  }\href {https://doi.org/10.1126/sciadv.aau8271} {\bibfield  {journal}
  {\bibinfo  {journal} {Sci. Adv.}\ }\textbf {\bibinfo {volume} {5}},\ \bibinfo
  {pages} {eaau8271} (\bibinfo {year} {2019})}\BibitemShut {NoStop}%
\bibitem [{\citenamefont {Linghu}\ \emph {et~al.}(2021)\citenamefont {Linghu},
  \citenamefont {Gu}, \citenamefont {Lu}, \citenamefont {Fang}, \citenamefont
  {Yang}, \citenamefont {Yu}, \citenamefont {Li}, \citenamefont {Zhu},
  \citenamefont {Peng}, \citenamefont {Zhan}, \citenamefont {Zhuang},
  \citenamefont {Gu},\ and\ \citenamefont {Gu}}]{Linghu:2021}%
  \BibitemOpen
  \bibfield  {author} {\bibinfo {author} {\bibfnamefont {S.}~\bibnamefont
  {Linghu}}, \bibinfo {author} {\bibfnamefont {Z.}~\bibnamefont {Gu}}, \bibinfo
  {author} {\bibfnamefont {J.}~\bibnamefont {Lu}}, \bibinfo {author}
  {\bibfnamefont {W.}~\bibnamefont {Fang}}, \bibinfo {author} {\bibfnamefont
  {Z.}~\bibnamefont {Yang}}, \bibinfo {author} {\bibfnamefont {H.}~\bibnamefont
  {Yu}}, \bibinfo {author} {\bibfnamefont {Z.}~\bibnamefont {Li}}, \bibinfo
  {author} {\bibfnamefont {R.}~\bibnamefont {Zhu}}, \bibinfo {author}
  {\bibfnamefont {J.}~\bibnamefont {Peng}}, \bibinfo {author} {\bibfnamefont
  {Q.}~\bibnamefont {Zhan}}, \bibinfo {author} {\bibfnamefont {S.}~\bibnamefont
  {Zhuang}}, \bibinfo {author} {\bibfnamefont {M.}~\bibnamefont {Gu}},\ and\
  \bibinfo {author} {\bibfnamefont {F.}~\bibnamefont {Gu}},\ }\bibfield
  {title} {\bibinfo {title} {{\it Plasmon-driven nanowire actuators for on-chip
  manipulation}},\ }\href {https://doi.org/10.1038/s41467-020-20683-2}
  {\bibfield  {journal} {\bibinfo  {journal} {Nat. Commun.}\ }\textbf {\bibinfo
  {volume} {12}},\ \bibinfo {pages} {385} (\bibinfo {year} {2021})}\BibitemShut
  {NoStop}%
\bibitem [{\citenamefont {Kurosawa}\ \emph {et~al.}(1998)\citenamefont
  {Kurosawa}, \citenamefont {Takahashi},\ and\ \citenamefont
  {Higuchi}}]{Kurosawa:1998}%
  \BibitemOpen
  \bibfield  {author} {\bibinfo {author} {\bibfnamefont {M.~K.}\ \bibnamefont
  {Kurosawa}}, \bibinfo {author} {\bibfnamefont {M.}~\bibnamefont
  {Takahashi}},\ and\ \bibinfo {author} {\bibfnamefont {T.}~\bibnamefont
  {Higuchi}},\ }\bibfield  {title} {\bibinfo {title} {{\it Elastic contact
  conditions to optimize friction drive of surface acoustic wave motor}},\
  }\href {https://doi.org/10.1109/58.726448} {\bibfield  {journal} {\bibinfo
  {journal} {IEEE Trans. Ultrason. Ferroelectr. Freq. Control}\ }\textbf
  {\bibinfo {volume} {45}},\ \bibinfo {pages} {1229} (\bibinfo {year}
  {1998})}\BibitemShut {NoStop}%
\bibitem [{\citenamefont {Shigematsu}\ \emph {et~al.}(2003)\citenamefont
  {Shigematsu}, \citenamefont {Kurosawa},\ and\ \citenamefont
  {Asai}}]{Shigematsu:2003}%
  \BibitemOpen
  \bibfield  {author} {\bibinfo {author} {\bibfnamefont {T.}~\bibnamefont
  {Shigematsu}}, \bibinfo {author} {\bibfnamefont {M.~K.}\ \bibnamefont
  {Kurosawa}},\ and\ \bibinfo {author} {\bibfnamefont {K.}~\bibnamefont
  {Asai}},\ }\bibfield  {title} {\bibinfo {title} {{\it Nanometer stepping
  drives of surface acoustic wave motor}},\ }\href
  {https://doi.org/10.1109/TUFFC.2003.1197960} {\bibfield  {journal} {\bibinfo
  {journal} {IEEE Trans. Ultrason. Ferroelectr. Freq. Control}\ }\textbf
  {\bibinfo {volume} {50}},\ \bibinfo {pages} {376} (\bibinfo {year}
  {2003})}\BibitemShut {NoStop}%
\bibitem [{\citenamefont {Destgeer}\ and\ \citenamefont
  {Sung}(2015)}]{Destgeer:2015}%
  \BibitemOpen
  \bibfield  {author} {\bibinfo {author} {\bibfnamefont {G.}~\bibnamefont
  {Destgeer}}\ and\ \bibinfo {author} {\bibfnamefont {H.~J.}\ \bibnamefont
  {Sung}},\ }\bibfield  {title} {\bibinfo {title} {{\it Recent advances in
  microfluidic actuation and micro-object manipulation via surface acoustic
  waves}},\ }\href {https://doi.org/10.1039/c5lc00265f} {\bibfield  {journal}
  {\bibinfo  {journal} {Lab on a Chip}\ }\textbf {\bibinfo {volume} {15}},\
  \bibinfo {pages} {2722} (\bibinfo {year} {2015})}\BibitemShut {NoStop}%
\bibitem [{\citenamefont {Landau}\ \emph {et~al.}(2012)\citenamefont {Landau},
  \citenamefont {Pitaevskii}, \citenamefont {Kosevich},\ and\ \citenamefont
  {Lifshitz}}]{Landau:2012}%
  \BibitemOpen
  \bibfield  {author} {\bibinfo {author} {\bibfnamefont {L.~D.}\ \bibnamefont
  {Landau}}, \bibinfo {author} {\bibfnamefont {L.~P.}\ \bibnamefont
  {Pitaevskii}}, \bibinfo {author} {\bibfnamefont {A.~M.}\ \bibnamefont
  {Kosevich}},\ and\ \bibinfo {author} {\bibfnamefont {E.~M.}\ \bibnamefont
  {Lifshitz}},\ }\href@noop {} {\emph {\bibinfo {title} {{\it Theory of
  Elasticity}}}}\ (\bibinfo  {publisher} {Butterworth-Heinemann},\ \bibinfo
  {address} {Oxford, England, UK},\ \bibinfo {year} {2012})\BibitemShut
  {NoStop}%
\bibitem [{\citenamefont {Bowden}\ and\ \citenamefont
  {Tabor}(1950)}]{Bowden:1950}%
  \BibitemOpen
  \bibfield  {author} {\bibinfo {author} {\bibfnamefont {F.~P.}\ \bibnamefont
  {Bowden}}\ and\ \bibinfo {author} {\bibfnamefont {D.}~\bibnamefont {Tabor}},\
  }\href@noop {} {\emph {\bibinfo {title} {{\it The friction and lubrication of
  solids}}}}\ (\bibinfo  {publisher} {Oxford Univ. Press},\ \bibinfo {address}
  {New York, USA},\ \bibinfo {year} {1950})\BibitemShut {NoStop}%
\bibitem [{\citenamefont {Kor}\ \emph {et~al.}(1972)\citenamefont {Kor},
  \citenamefont {Tandon},\ and\ \citenamefont {Rai}}]{Kor:1972}%
  \BibitemOpen
  \bibfield  {author} {\bibinfo {author} {\bibfnamefont {S.~K.}\ \bibnamefont
  {Kor}}, \bibinfo {author} {\bibfnamefont {U.~S.}\ \bibnamefont {Tandon}},\
  and\ \bibinfo {author} {\bibfnamefont {G.}~\bibnamefont {Rai}},\ }\bibfield
  {title} {\bibinfo {title} {{\it Ultrasonic Attenuation in Copper, Silver, and
  Gold}},\ }\href {https://doi.org/10.1103/PhysRevB.6.2195} {\bibfield
  {journal} {\bibinfo  {journal} {Phys. Rev. B}\ }\textbf {\bibinfo {volume}
  {6}},\ \bibinfo {pages} {2195} (\bibinfo {year} {1972})}\BibitemShut
  {NoStop}%
\bibitem [{\citenamefont {Ruijgrok}\ \emph {et~al.}(2012)\citenamefont
  {Ruijgrok}, \citenamefont {Zijlstra}, \citenamefont {Tchebotareva},\ and\
  \citenamefont {Orrit}}]{Ruijgrok:2012}%
  \BibitemOpen
  \bibfield  {author} {\bibinfo {author} {\bibfnamefont {P.~V.}\ \bibnamefont
  {Ruijgrok}}, \bibinfo {author} {\bibfnamefont {P.}~\bibnamefont {Zijlstra}},
  \bibinfo {author} {\bibfnamefont {A.~L.}\ \bibnamefont {Tchebotareva}},\ and\
  \bibinfo {author} {\bibfnamefont {M.}~\bibnamefont {Orrit}},\ }\bibfield
  {title} {\bibinfo {title} {{\it Damping of Acoustic Vibrations of Single Gold
  Nanoparticles Optically Trapped in Water}},\ }\href
  {https://doi.org/10.1021/nl204311q} {\bibfield  {journal} {\bibinfo
  {journal} {Nano Lett.}\ }\textbf {\bibinfo {volume} {12}},\ \bibinfo {pages}
  {1063} (\bibinfo {year} {2012})}\BibitemShut {NoStop}%
\bibitem [{\citenamefont {Maier}(2007)}]{Maier:2007}%
  \BibitemOpen
  \bibfield  {author} {\bibinfo {author} {\bibfnamefont {S.~A.}\ \bibnamefont
  {Maier}},\ }\href {https://doi.org/10.1007/978-0-387-37825-1} {\emph
  {\bibinfo {title} {{\it Plasmonics: Fundamentals and Applications}}}}\
  (\bibinfo  {publisher} {Springer US},\ \bibinfo {address} {Springer-Verlag
  US},\ \bibinfo {year} {2007})\BibitemShut {NoStop}%
\bibitem [{\citenamefont {Hao}\ \emph {et~al.}(2010)\citenamefont {Hao},
  \citenamefont {Wang}, \citenamefont {Liu}, \citenamefont {Padilla},
  \citenamefont {Zhou},\ and\ \citenamefont {Qiu}}]{Hao:2010}%
  \BibitemOpen
  \bibfield  {author} {\bibinfo {author} {\bibfnamefont {J.}~\bibnamefont
  {Hao}}, \bibinfo {author} {\bibfnamefont {J.}~\bibnamefont {Wang}}, \bibinfo
  {author} {\bibfnamefont {X.}~\bibnamefont {Liu}}, \bibinfo {author}
  {\bibfnamefont {W.~J.}\ \bibnamefont {Padilla}}, \bibinfo {author}
  {\bibfnamefont {L.}~\bibnamefont {Zhou}},\ and\ \bibinfo {author}
  {\bibfnamefont {M.}~\bibnamefont {Qiu}},\ }\bibfield  {title} {\bibinfo
  {title} {{\it High performance optical absorber based on a plasmonic
  metamaterial}},\ }\href {https://doi.org/10.1063/1.3442904} {\bibfield
  {journal} {\bibinfo  {journal} {Appl. Phys. Lett.}\ }\textbf {\bibinfo
  {volume} {96}},\ \bibinfo {pages} {251104} (\bibinfo {year}
  {2010})}\BibitemShut {NoStop}%
\bibitem [{Note1()}]{Note1}%
  \BibitemOpen
  \bibinfo {note} {Among 40 tests, the independence of the motion directions on
  the light directions was observed for 38 gold plates. Only for 2 gold plates,
  their translation directions changed with the light directions. However, in
  such 2 exceptional examples, the positions of the plates on the micro-fibers
  were noticeably changed while we reversed the light directions. In view of
  this, we exclude these two exceptions and conclude that the motion directions
  of the gold plates are independent of the light directions.}\BibitemShut
  {Stop}%
\bibitem [{\citenamefont {Guo}\ \emph {et~al.}(2006)\citenamefont {Guo},
  \citenamefont {Zhang}, \citenamefont {DuanMu}, \citenamefont {Xu},
  \citenamefont {Xie},\ and\ \citenamefont {Gu}}]{Guo:2006}%
  \BibitemOpen
  \bibfield  {author} {\bibinfo {author} {\bibfnamefont {Z.}~\bibnamefont
  {Guo}}, \bibinfo {author} {\bibfnamefont {Y.}~\bibnamefont {Zhang}}, \bibinfo
  {author} {\bibfnamefont {Y.}~\bibnamefont {DuanMu}}, \bibinfo {author}
  {\bibfnamefont {L.}~\bibnamefont {Xu}}, \bibinfo {author} {\bibfnamefont
  {S.}~\bibnamefont {Xie}},\ and\ \bibinfo {author} {\bibfnamefont
  {N.}~\bibnamefont {Gu}},\ }\bibfield  {title} {\bibinfo {title} {{\it Facile
  synthesis of micrometer-sized gold nanoplates through an aniline-assisted
  route in ethylene glycol solution}},\ }\href
  {https://doi.org/10.1016/j.colsurfa.2005.11.075} {\bibfield  {journal}
  {\bibinfo  {journal} {Colloids Surf. A Physicochem. Eng. Asp.}\ }\textbf
  {\bibinfo {volume} {278}},\ \bibinfo {pages} {33} (\bibinfo {year}
  {2006})}\BibitemShut {NoStop}%
\bibitem [{\citenamefont {Wu}\ and\ \citenamefont {Tong}(2013)}]{Tong:2013}%
  \BibitemOpen
  \bibfield  {author} {\bibinfo {author} {\bibfnamefont {X.}~\bibnamefont
  {Wu}}\ and\ \bibinfo {author} {\bibfnamefont {L.}~\bibnamefont {Tong}},\
  }\bibfield  {title} {\bibinfo {title} {{\it Optical microfibers and
  nanofibers}},\ }\href {https://doi.org/10.1016/j.optcom.2012.07.068}
  {\bibfield  {journal} {\bibinfo  {journal} {Nanophotonics}\ }\textbf
  {\bibinfo {volume} {2}},\ \bibinfo {pages} {407} (\bibinfo {year}
  {2013})}\BibitemShut {NoStop}%
\end{thebibliography}%


\begin{thebibliography}{3}%
\makeatletter
\providecommand \@ifxundefined [1]{%
 \@ifx{#1\undefined}
}%
\providecommand \@ifnum [1]{%
 \ifnum #1\expandafter \@firstoftwo
 \else \expandafter \@secondoftwo
 \fi
}%
\providecommand \@ifx [1]{%
 \ifx #1\expandafter \@firstoftwo
 \else \expandafter \@secondoftwo
 \fi
}%
\providecommand \natexlab [1]{#1}%
\providecommand \enquote  [1]{``#1''}%
\providecommand \bibnamefont  [1]{#1}%
\providecommand \bibfnamefont [1]{#1}%
\providecommand \citenamefont [1]{#1}%
\providecommand \href@noop [0]{\@secondoftwo}%
\providecommand \href [0]{\begingroup \@sanitize@url \@href}%
\providecommand \@href[1]{\@@startlink{#1}\@@href}%
\providecommand \@@href[1]{\endgroup#1\@@endlink}%
\providecommand \@sanitize@url [0]{\catcode `\\12\catcode `\$12\catcode
  `\&12\catcode `\#12\catcode `\^12\catcode `\_12\catcode `\%12\relax}%
\providecommand \@@startlink[1]{}%
\providecommand \@@endlink[0]{}%
\providecommand \url  [0]{\begingroup\@sanitize@url \@url }%
\providecommand \@url [1]{\endgroup\@href {#1}{\urlprefix }}%
\providecommand \urlprefix  [0]{URL }%
\providecommand \Eprint [0]{\href }%
\providecommand \doibase [0]{http://dx.doi.org/}%
\providecommand \selectlanguage [0]{\@gobble}%
\providecommand \bibinfo  [0]{\@secondoftwo}%
\providecommand \bibfield  [0]{\@secondoftwo}%
\providecommand \translation [1]{[#1]}%
\providecommand \BibitemOpen [0]{}%
\providecommand \bibitemStop [0]{}%
\providecommand \bibitemNoStop [0]{.\EOS\space}%
\providecommand \EOS [0]{\spacefactor3000\relax}%
\providecommand \BibitemShut  [1]{\csname bibitem#1\endcsname}%
\let\auto@bib@innerbib\@empty
\bibitem [{\citenamefont {Landau}\ \emph {et~al.}(2012)\citenamefont {Landau},
  \citenamefont {Pitaevskii}, \citenamefont {Kosevich},\ and\ \citenamefont
  {Lifshitz}}]{Landau:2012}%
  \BibitemOpen
  \bibfield  {author} {\bibinfo {author} {\bibfnamefont {L.~D.}\ \bibnamefont
  {Landau}}, \bibinfo {author} {\bibfnamefont {L.~P.}\ \bibnamefont
  {Pitaevskii}}, \bibinfo {author} {\bibfnamefont {A.~M.}\ \bibnamefont
  {Kosevich}}, \ and\ \bibinfo {author} {\bibfnamefont {E.~M.}\ \bibnamefont
  {Lifshitz}},\ }\href
  {https://www.elsevier.com/books/theory-of-elasticity/landau/978-0-08-057069-3}
  {\emph {\bibinfo {title} {{Theory of Elasticity}}}}\ (\bibinfo  {publisher}
  {Butterworth-Heinemann},\ \bibinfo {address} {Oxford, England, UK},\ \bibinfo
  {year} {2012})\BibitemShut {NoStop}%
\bibitem [{\citenamefont {Lu}\ \emph {et~al.}(2017)\citenamefont {Lu},
  \citenamefont {Yang}, \citenamefont {Zhou}, \citenamefont {Yang},
  \citenamefont {Luo}, \citenamefont {Li},\ and\ \citenamefont
  {Qiu}}]{Lu:2017}%
  \BibitemOpen
  \bibfield  {author} {\bibinfo {author} {\bibfnamefont {J.}~\bibnamefont
  {Lu}}, \bibinfo {author} {\bibfnamefont {H.}~\bibnamefont {Yang}}, \bibinfo
  {author} {\bibfnamefont {L.}~\bibnamefont {Zhou}}, \bibinfo {author}
  {\bibfnamefont {Y.}~\bibnamefont {Yang}}, \bibinfo {author} {\bibfnamefont
  {S.}~\bibnamefont {Luo}}, \bibinfo {author} {\bibfnamefont {Q.}~\bibnamefont
  {Li}}, \ and\ \bibinfo {author} {\bibfnamefont {M.}~\bibnamefont {Qiu}},\
  }\href {\doibase 10.1103/PhysRevLett.118.043601} {\bibfield  {journal}
  {\bibinfo  {journal} {Phys. Rev. Lett.}\ }\textbf {\bibinfo {volume} {118}},\
  \bibinfo {pages} {043601} (\bibinfo {year} {2017})}\BibitemShut {NoStop}%
\bibitem [{\citenamefont {Batchelor}(1967)}]{Batchelor:1967}%
  \BibitemOpen
  \bibfield  {author} {\bibinfo {author} {\bibfnamefont {G.}~\bibnamefont
  {Batchelor}},\ }\href@noop {} {\emph {\bibinfo {title} {{An introduction to
  fluid dynamics}}}}\ (\bibinfo  {publisher} {Cambridge University Press},\
  \bibinfo {address} {Cambridge, England, UK},\ \bibinfo {year}
  {1967})\BibitemShut {NoStop}%
\end{thebibliography}%

\end{document}


\title{\color{black!88!white} SUPPLEMENTARY INFORMATION\vskip .5em
Micro-scale opto-thermo-mechanical actuation in the dry adhesive regime}


\author{Weiwei~Tang}
\affiliation{Key Laboratory of 3D Micro/Nano Fabrication and Characterization of Zhejiang Province, School of Engineering, Westlake University, 18 Shilongshan Road, Hangzhou 310024, Zhejiang Province, China}
\affiliation{Institute of Advanced Technology, Westlake Institute for Advanced Study, 18 Shilongshan Road, Hangzhou 310024, Zhejiang Province, China}

\author{Wei~Lv}
\affiliation{Key Laboratory of 3D Micro/Nano Fabrication and Characterization of Zhejiang Province, School of Engineering, Westlake University, 18 Shilongshan Road, Hangzhou 310024, Zhejiang Province, China}
\affiliation{Institute of Advanced Technology, Westlake Institute for Advanced Study, 18 Shilongshan Road, Hangzhou 310024, Zhejiang Province, China}

\author{Jinsheng~Lu}
\affiliation{State Key Laboratory of Modern Optical Instrumentation, College of Optical Science
and Engineering, Zhejiang University, Hangzhou 310027, China}

\author{Fengjiang~Liu}

\affiliation{Key Laboratory of 3D Micro/Nano Fabrication and Characterization of Zhejiang Province, School of Engineering, Westlake University, 18 Shilongshan Road, Hangzhou 310024, Zhejiang Province, China}
\affiliation{Institute of Advanced Technology, Westlake Institute for Advanced Study, 18 Shilongshan Road, Hangzhou 310024, Zhejiang Province, China}

\author{Jiyong~Wang}

\affiliation{Key Laboratory of 3D Micro/Nano Fabrication and Characterization of Zhejiang Province, School of Engineering, Westlake University, 18 Shilongshan Road, Hangzhou 310024, Zhejiang Province, China}
\affiliation{Institute of Advanced Technology, Westlake Institute for Advanced Study, 18 Shilongshan Road, Hangzhou 310024, Zhejiang Province, China}

\author{Wei~Yan}
\affiliation{Key Laboratory of 3D Micro/Nano Fabrication and Characterization of Zhejiang Province, School of Engineering, Westlake University, 18 Shilongshan Road, Hangzhou 310024, Zhejiang Province, China}
\affiliation{Institute of Advanced Technology, Westlake Institute for Advanced Study, 18 Shilongshan Road, Hangzhou 310024, Zhejiang Province, China}
\email{wyanzju@gmail.com}

\author{Min~Qiu}
\affiliation{Key Laboratory of 3D Micro/Nano Fabrication and Characterization of Zhejiang Province, School of Engineering, Westlake University, 18 Shilongshan Road, Hangzhou 310024, Zhejiang Province, China}
\affiliation{Institute of Advanced Technology, Westlake Institute for Advanced Study, 18 Shilongshan Road, Hangzhou 310024, Zhejiang Province, China}
\email{qiumin@westlake.edu.cn }

\maketitle

\color{black!88!white}


\noindent{\small\textbf{\textsf{CONTENTS}}}\\
\onecolumngrid
\begin{spacing}{1}
{
\begingroup 
\let\bfseries\relax
\let\tocdepth\relax

\tableofcontents
\endgroup
}
\end{spacing}


\onecolumngrid

\section{Elastic waveguide modes in gold rectangular plates}
\label{Sec:modes}

In this section, we derive dispersion relations of elastic waveguide modes in gold plates with rectangular cross section, {\bf supplementing Fig. 1B in the main text}.

\begin{figure}[!htp]
\centering
\includegraphics[width=10cm]{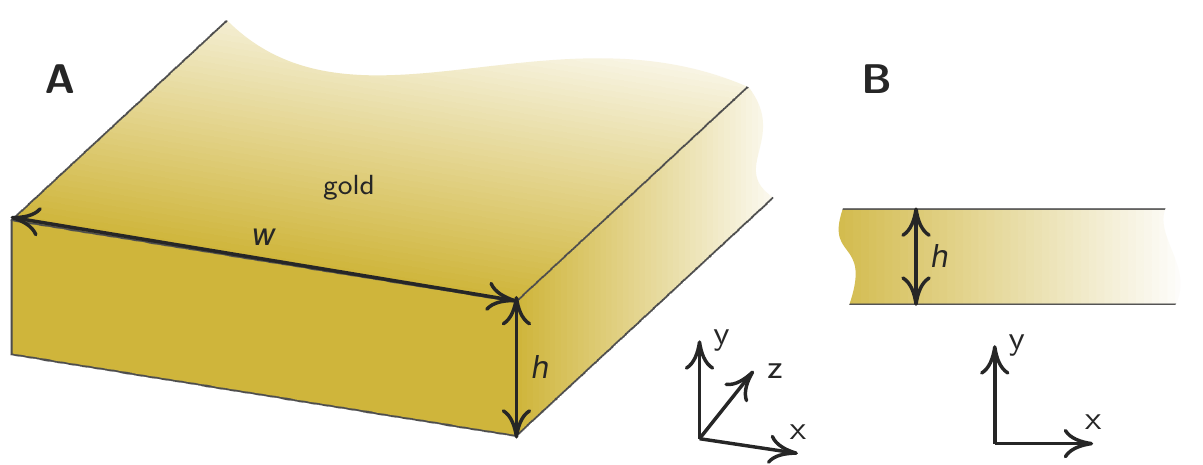}
\caption{
{\bf Sketch of elastic waveguides made of gold.}
{\bf A.} A three-dimensional gold plate with rectangular cross section.
{\bf B.} A two-dimensional gold slab. The elastic waveguide modes of the
2D slab are used as the basis to represent the modes of the 3D plate and to derive the dispersion relations of the latter.
}
\label{Fig:waveguide}
\end{figure}

As shown in Fig.~\ref{Fig:waveguide}A, the cross section (in the $x-y$ plane) of the studied gold plate has geometrical dimensions of width $h$ and thickness $h$, and the plate extends to the infinity in the $z$-direction.
Elastic waveguide modes are eigensolutions of the frequency-domain linear elastic equation, read as~\cite{Landau:2012}
\begin{subequations}
\begin{align}
\nablav\times\nablav\times \uv(\rv;
\omega)-\frac{2(1-\mu)}{1-2\mu}\nablav\nablav\cdot\uv(\rv;
\omega) - \omega^2 \frac{2\rho(1+\mu)}{E}\uv(\rv;
\omega)=0,
\end{align}
and they are constrained by the boundary conditions (BCs)---surface traction vectors (force per unit area) on the plate's outside walls vanish---:
\begin{align}
\hat{\mathbf n}\cdot \bm\sigma (\rv;\omega)  =0,
\label{eq:Bcs}
\end{align}
\end{subequations}
where $\uv \equiv u_x \hat x + u_y \hat y + u_z \hat z $ denotes the elastic displacement vector; $\rho$, $E$ and $\mu$ denote the mass density, Young's modulus and Poison's ratio of gold, respectively; $\sigma$ is the stress tensor.

Note that it is difficult to derive analytic expressions for waveguidng modes of 3D rectangular plates directly.
Nevertheless, since the plates used in our experiments are very thin with $h$ only about tens of nanometers, the long wavelength approximation along the thickness dimension (requiring $h\ll \lambda$, where $\lambda$ denotes elastic wavelength of our interest) can be exploited to simplify derivations.
Moreover, we will firstly start with a simpler case, a two-dimensional (2D) slab (Fig.~\ref{Fig:waveguide}B), and then use the modes of the slab (derived under the long wavelength approximation) as the basis to construct, derive the modes of 3D plates.

\subsection{2D Slab}

A 2D slab, with thickness $h$ extending from $-h/2$ to $h/2$ in the $y$-direction, is sketched in Fig.~\ref{Fig:waveguide}B. The elastic waveguide modes propagate in the $x$-direction and their wavenumbers are denoted by $\beta_{\rm 2D}$. The displacement vectors of the modes, denoted by $\uv_{\rm 2D}\equiv u_{{\rm 2D} ; x} \hat x + u_{{\rm 2D} ; y} \hat y + u_{{\rm 2D} ; z} \hat z  $, are represented by linear combinations of transverse and longitudinal elastic plane waves in bulk gold:
\begin{subequations}
\begin{align}
u_{{\rm 2D};x}  & =         [T_1 k_y^{\rm T}\cos(k_y^{\rm T} y) + T_2 k_y^{\rm T}            \sin(k_y^{\rm T} y)
       +   L_1  \beta      \cos(k_y^{\rm L} y)  + L_2 \beta k_y^{\rm L}  \sin(k_y^{\rm L} y)]\exp(i\beta_{\rm 2D} x),\\
u_{{\rm 2D};y}  & =       [-i T_1 \beta    \sin(k_y^{\rm T} y) + i T_2 \beta              \cos (k_y^{\rm T} y)        +  i L_1 k_y^{\rm L} \sin (k_y^{\rm L} y)  - i L_2 k_y^{\rm L} \cos (k_y^{\rm L} y)]\exp(i\beta_{\rm 2D} x),\\
u_{{\rm 2D};z}  & =         [T_3 k_y^{\rm T}\cos(k_y^{\rm T} y) + T_4 k_y^{\rm T}            \sin(k_y^{\rm T} y) ]
      \exp(i\beta_{\rm 2D} x),
\label{eq:u2D}
\end{align}
with
\begin{align}
\underbrace{\left(k_y^{\rm T} \right) ^2+ \beta_{\rm 2D}^2=\omega^2\frac{2\rho(1+\mu)}{E}}_{\text{dispersion relation of transverse plane waves}},\quad
\underbrace{\left(k_y^{\rm L} \right) ^2+ \beta_{\rm 2D}^2=\omega^2\frac{\rho(1+\mu)(1-2\mu)}{E(1-\mu)}}_{\text{dispersion relation of longitudinal plane waves}},
\label{eq:PW}
\end{align}
where $T_{1,2,3,4}$ and $L_{1,2}$ are the (unknown) modal coefficients of the transverse and longitudinal plane waves, respectively.
\label{eq:u-express}
\end{subequations}

We solve the dispersion relations of the elastic waveguide modes of the 2D slab (i.e., $\beta_{\rm 2D}$ as a function of $\omega$) by plugging Eqs.~\eqref{eq:u-express} into the BCs, Eq.~\eqref{eq:Bcs}. Notably, as a result of the reflection symmetry of the slab about the $x$-axis, the waveguide modes can be partitioned into four sets, relating to the different terms in Eqs.~\eqref{eq:u-express} associated with $\left\{ T_1, L_1 \right\}$, $\left\{ T_3 \right\}$, $\left\{ T_2, L_2 \right\}$ and $\left\{ T_4 \right\}$, respectively. Among the four decoupled solutions, the former two deserve more attentions, because they lead to fundamental modes showing no cutoff and more importantly because their linear combinations give the essential modes of 3D plates (i.e., fundamental $L$- and $T$-like modes, highlighted in Fig. 1B in the main text) that are relevant to our experimental observations. Thereupon, we here below focus on the two sets of the modes concerning $\left\{ T_1, L_1 \right\}$ and $\left\{ T_3 \right\}$, respectively. Employing the long wavelength approximation~\footnote{Approximate that $\cos(k_y^{\rm T, L} y) \simeq 1$ and $\sin(k_y^{\rm T,L} y) \simeq k_y^{\rm T, L} y$.}, we derive that:
\begin{subequations}
\begin{align}
\text{longitudinal-like slab modes:}   \quad   &\beta_{\rm 2D}^{\rm L} \simeq \omega\sqrt{\frac{\rho(1+\mu)(1-\mu)}{E}},\nonumber\\
                                      & \uv_{\rm 2D}^{\rm L} \simeq N e^{i\beta_{\rm 2D}^{\rm L}x}\hat x - iN\frac{\beta_{\rm slab}^{\rm L} \mu}{1-\mu}y e^{i \beta_{\rm 2D}^{\rm L}x}\hat y,\\
\text{transverse slab modes:}    \quad  & \beta_{\rm 2D}^{\rm T}= \omega\sqrt{\frac{2\rho(1+\mu)}{E}},\nonumber
                                    \\
                                    &  \uv_{\rm 2D}^{\rm T} = N e^{i \beta_{\rm 2D}^{\rm T} x } \hat z.
\end{align}
Here, the two modes are labeled by the superscripts “L" and “T", implying that their longitudinal (L) and transverse (T) characters, respectively. The longitudinal-like slab modes have dominant longitudinal component and negligible transverse component, while the transverse slab modes are the same as transverse plane waves in bulk gold.
\label{eq:modes2D}
\end{subequations}

\begin{figure}[!htp]
\centering
\includegraphics[width=10cm]{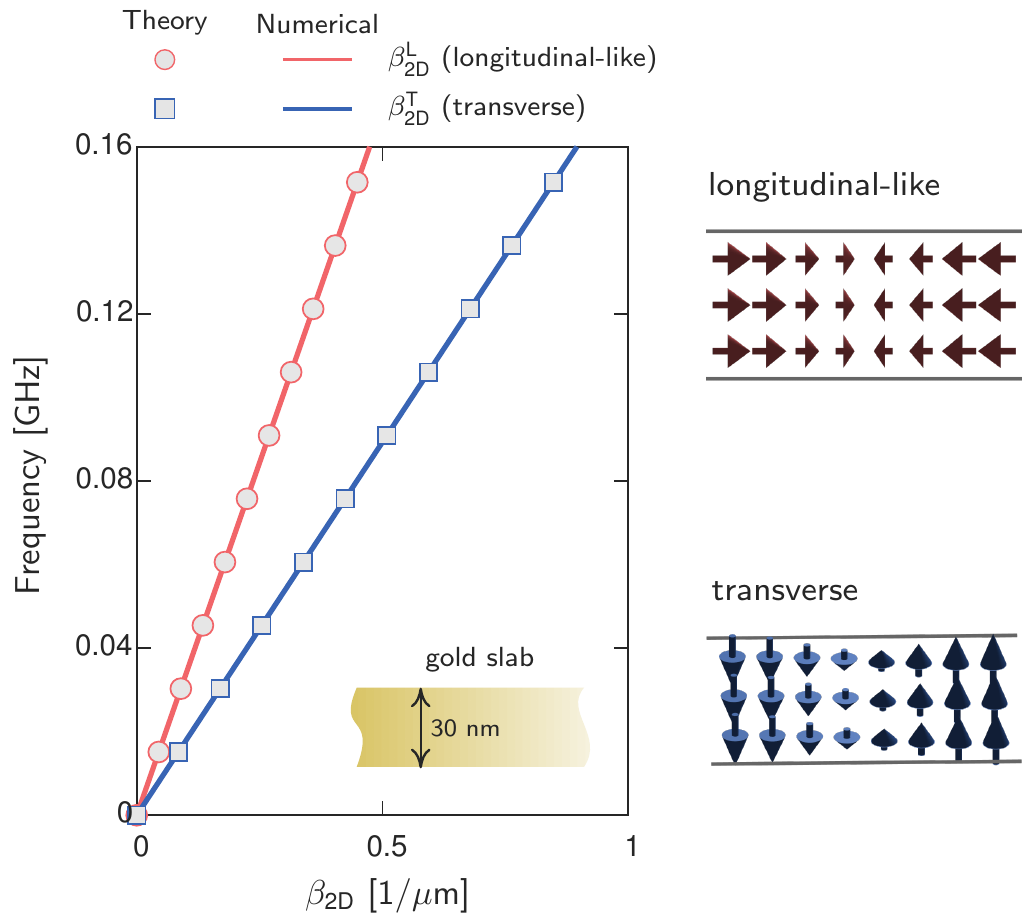}
\caption{{\bf Validation of Eqs.~\eqref{eq:modes2D} for predicting dispersion relations of fundamental elastic waveguide modes of a 2D gold slab with thickness 30 nm}. Right: modal profiles of longitudinal-like and transverse slab modes at the static frequency with arrows specifying oscillation directions of elastic waves.
}
\label{Fig:2D_slab}
\end{figure}

We validate the predictive accuracy of Eqs.~\eqref{eq:modes2D} by examining the dispersion relations of the waveguide modes of a gold slab with $h=30\,\rm nm$.
As shown in Fig.~\ref{Fig:2D_slab}, the theoretical predictions (markers) from Eqs.~\eqref{eq:modes2D} and the fully-numerical results (solid lines) obtained with the COMSOL Multiphysics show a remarkable agreement.

\subsection{3D rectangular plate}
\label{Sec:3DRWG}

We derive the dispersion relations of the waveguide modes in a 3D rectangular plate by expanding the displacement vector $\uv_{\scriptscriptstyle \rm 3D}$ with $\uv_{\rm 2D}^{\rm L,T}$ derived in Eqs.~\eqref{eq:modes2D}:
\begin{subequations}
\begin{align}
u_{{\rm 3D};x}  & = \left[T_1\frac{\beta_{\rm 3D}} {\beta_{\rm 2D}^{\rm T}}  \sin(k_{x}^{\rm T} x)-
L_1\frac{k_{x}^{\rm L}}{\beta_{\rm 2D}^{\rm L}} \sin(k_x^{\rm L} x)
-T_2 \frac{\beta_{\rm 3D}} {\beta_{\rm 2D}^{\rm T}} \cos(k_x^{\rm T} x)+L_2\frac{k_x^{\rm L}}{\beta_{\rm 2D}^{\rm L}} \cos(k_x^{\rm L} x)
\right]
\exp(i\beta_{ \rm 3D}z),
\\
u_{{ \rm 3D};y}  & = \left[L_1\frac{\mu}{1-\mu} \beta_{\rm 2D}^{\rm L} y \cos(k_x^{\rm L} x)
+L_2\frac{\mu}{1-\mu} \beta_{\rm 2D}^{\rm L} y \sin(k_x^{\rm L} x)\right]
\exp(i\beta_{\rm 3D}z),\\
u_{{ \rm 3D};z}  & = \left[i T_1\frac{k_x^{\rm T}}{\beta_{\rm 2D}^{\rm T}}\cos(k_x^{\rm T} x) + i
L_1 \frac{\beta_{ \rm 3D}} {\beta_{\rm 2D}^{\rm L}} \cos(k_x^{\rm L} x)
+ iT_2\frac{k_x^{\rm T}}{\beta_{\rm 2D}^{\rm T}}\sin(k_x^{\rm T} x) +i
L_2 \frac{\beta_{\rm 3D}} {\beta_{\rm 2D}^{\rm L}} \sin(k_x^{\rm L} x) \right]\exp(i\beta_{\rm 3D}z),
\end{align}
\label{eq:u3D}
\end{subequations}
where $\beta_{\rm 3D}$ denotes the wavenumber of a waveguide mode, and
$\left(k_x^{\rm L}\right)^2+ \beta_{ \rm 3D}^2= \left(\beta_{\rm 2D}^{\rm L}\right)^2$
and
$\left(k_x^{\rm T}\right)^2+ \beta_{   \rm 3D}^2= \left(\beta_{\rm 2D}^{\rm T}\right)^2$~\footnote{Note that the same notations $k_x^{\rm L, T}$ are adopted in both Eqs.~\eqref{eq:u3D} and Eqs.~\eqref{eq:u-express}, but with different meanings.}.

In Eqs.~\eqref{eq:u3D}, the waveguide modes can be sorted into two decoupled sets associated with $\left\{T_1 , L_1\right\}$ and $\left\{T_2 , L_2\right\}$, respectively, as a result of the reflection symmetry of the plate about the central $x-z$ plane.
The two solutions correspond to the fundamental longitudinal ($L$)- and transverse ($T$)-like modes, respectively, which are highlighted in Fig. 1B in the main text.
Employing the BCs, Eq.~\eqref{eq:Bcs}, the approximated dispersion relations of the $L$- and $T$-modes are derived as
\begin{subequations}
\begin{align}
L\text{-modes:} \quad & \left(\beta_{ \rm 3D}^{\rm L}\right)^2\simeq\frac{\omega^2 \rho}{E}+\frac{\left(\beta_{ \rm 3D}^{\rm L}\right)^2 \left(k_{x}^{\rm L}\right)^2 E}{\omega^2\rho(\mu+1)^2}\left[
\frac{\tan(k_{ x}^{\rm L}w/2)k_{\rm x}^{\rm T}}{\tan(k_{x}^{\rm T}w/2)k_{\rm x}^{\rm L}}-
1
\right],\\
T\text{-modes:} \quad
&\left(\beta_{ \rm 3D}^{\rm T}\right)^2\simeq\frac{\omega^2 \rho}{E}+\frac{\left(\beta_{ \rm 3D}^{\rm T}\right)^2 \left(k_{\rm x}^{\rm L}\right)^2 E}{\omega^2\rho(\mu+1)^2}\left[
\frac{\cot(k_{x}^{\rm L}w/2)k_{\rm x}^{\rm T}}{\cot(k_{ x}^{\rm T}w/2)k_{x}^{\rm L}}-
1
\right].
\end{align}
\label{eq:disp_RWG}
\end{subequations}

Equations~\eqref{eq:disp_RWG} belong to transcendental equations which blur closed-form solutions. In order to derive the closed-form solutions, we again take the long wavelength approximation (assuming $\beta_{ \rm 3D}^{\rm T,L} w\ll 1$) and obtain that
\begin{subequations}
\begin{align}
L\text{-modes:} \quad & \beta_{ \rm 3D}^{\rm L}\simeq \omega\sqrt{\frac{\rho}{E}},\nonumber\\
                                      & \uv_{\rm 3D}^{\rm L} \simeq N e^{i \beta_{\rm 3D}^{\rm L} z} \left[-i \mu  \beta_{\rm 3D}^{\rm L} x\hat x-i \mu  \beta_{\rm 3D}^{\rm L} y\hat y+\hat z \right], \\
T\text{-modes:}   \quad &  \beta_{\rm 3D}^{\rm T} \simeq
\sqrt{\frac{\omega}{w}}
\left(
\frac{12\rho}{E}
\right)^{1/4},\nonumber\\
& \uv_{\rm 3D}^{\rm T} \simeq N e^{i \beta_{\rm 3D}^{\rm T}  z} \left[\hat x+i \beta_{\rm 3D}^{\rm T} x \hat z\right].
\end{align}
\label{eq:disp_RWG2}
Note that the $L$-modes exhibit a linear dispersion under the long wavelength approximation (valid at low frequencies).
\end{subequations}

\begin{figure}[!htp]
\centering
\includegraphics[width=12cm]{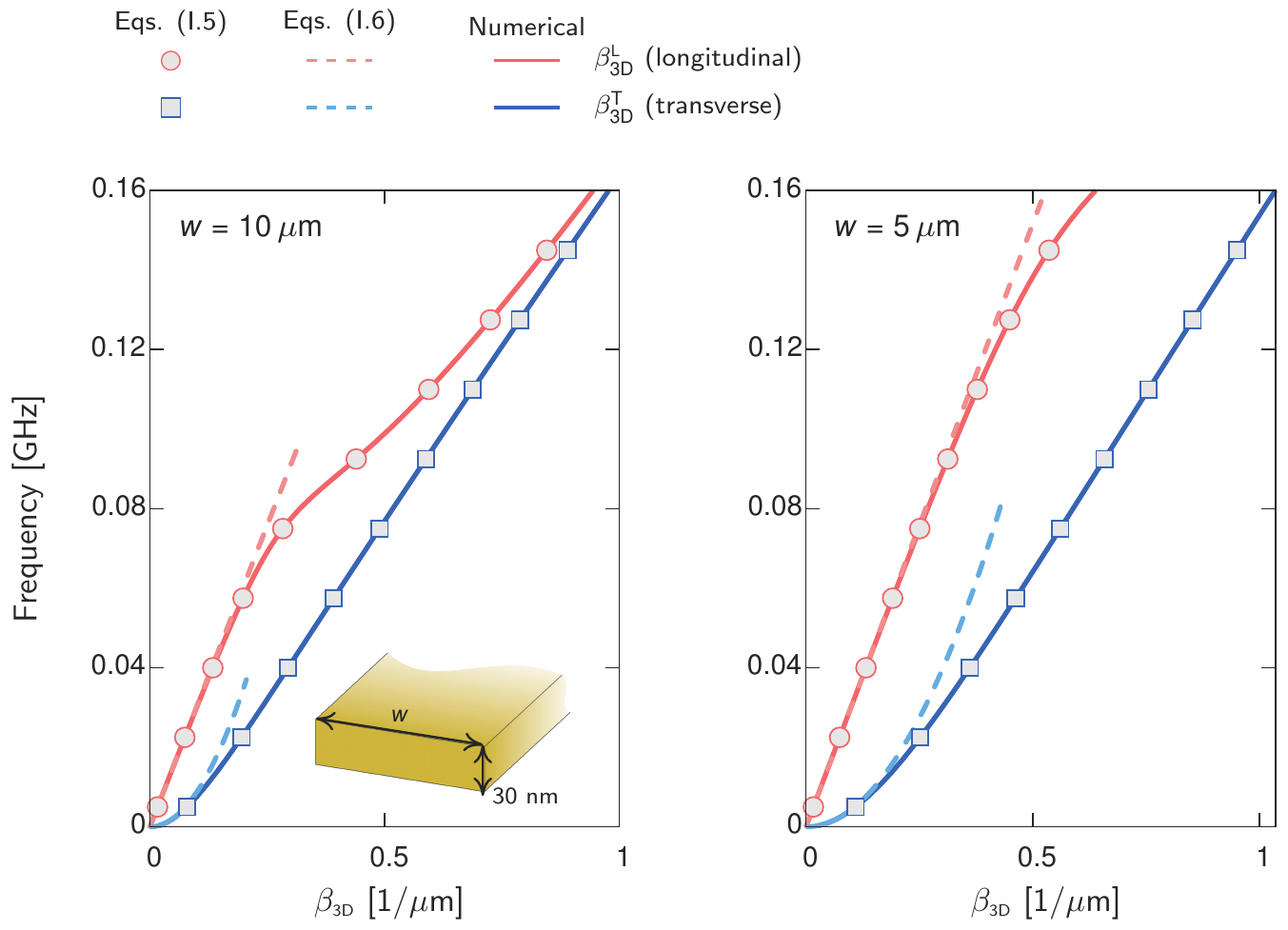}
\caption{{\bf Validation of Eqs.~\eqref{eq:disp_RWG} and \eqref{eq:disp_RWG2} for predicting dispersion relations of elastic waveguide modes of gold rectangular plates}. The studied two plates have the same thickness, 30 nm, but different width, $10\,\mu\rm m$ (left) and $5\,\mu\rm m$ (right).
}
\label{Fig:3D_RWG_modes}
\end{figure}

As shown in Fig.~\ref{Fig:3D_RWG_modes}, we numerically validate the predictive accuracy of Eqs.~\eqref{eq:disp_RWG} and \eqref{eq:disp_RWG2} by examining the dispersion relations of the elastic waveguide modes in a gold rectangular plate with $h=30\,\rm nm$, and $w=10\,\mu\rm m$ (left panel), $5\,\mu\rm m$ (right panel). Here the theoretical results (markers) computed from Eqs.~\eqref{eq:disp_RWG} agree excellently with the fully-numerical results obtained with the COMOSOL Multiphysics (solid lines), while the long-wavelength-approximation results (dashed lines) of Eqs.~\eqref{eq:disp_RWG2} are accurate for small frequencies.

\section{Elastic waves Excited by Temperature Change and Friction Force }

\label{Sec:EWs}

\begin{figure}[!htp]
\centering
\includegraphics[width=14cm]{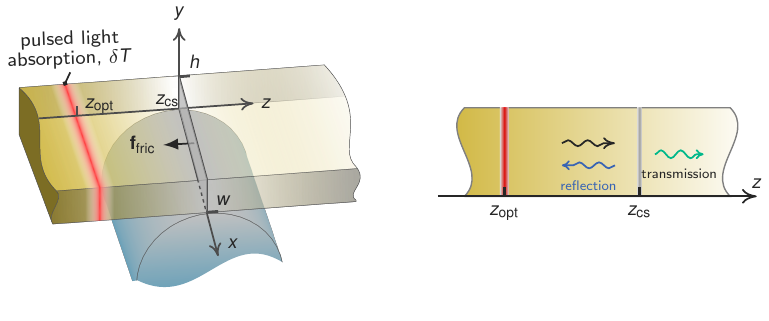}
\caption{
{\bf Sketch of the problem studied in Fig. 1 in the main text}. A gold plate contacts with a curved substrate. An incident light pulse is absorbed by the plate on the left side of the contact surface and induces a temperature change $\delta T$. The temperature change excites elastic modes, which, when passing through the line-shaped contact surface between the plate and the substrate, are resisted by friction force that induces reflected elastic waves. $z_{\rm op}$ and $z_{\rm cs}$ denote the centers of the $z$-coordinates of the optical absorbed power and the contact surface, respectively.
}
\label{Fig:casestudy}
\end{figure}

In this section, we derive the expressions of elastic displacements induced by temperature change (due to optical absorption) and friction force, {\bf supplementing Figs. 1C-D and Eq. (1) in the main text}. For clarification, Fig.~\ref{Fig:casestudy}, duplicating Fig. 1A in the main text, sketches the investigated problem: locomotion of a rectangular gold plate on a curved support driven by pulsed optical absorption.
Here, we focus on the elementary interaction between an incident (thermally-excited) elastic waves and reflected waves induced by the friction force, and, thus, omit the reflected elastic waves due to truncation of the plate in the $z$ axis. For convenience of derivation, we set that the incident elastic wave is induced by a pulsed optical absorption that locates on the left side of the contact surface~\footnote{This setting is different from our experiment, wherein the spatial distribution of the optical absorption overlaps with the contact surface, and generates elastic waves that bounce back and forth inside the plate. Nevertheless, the elementary physical process---interactions between an incident elastic force and friction force---is the same. Consequently, the results derived in this section can be applied to the experiment by taking multiple reflected elastic waves into account (see Sec.~\ref{Sec:Rot_Theory}).}.

We start with the time-dependent linear elastic equation that relates the displacement vector of elastic waves with temperature change and friction force
\begin{align}
&\nablav\times\nablav \times \uv(\rv;t) - \frac{2(1-\mu)}{1-2\mu}\nablav\nablav\cdot\uv(\rv;t) + \frac{2\rho(1+\mu)}{E}\frac{\partial^2\uv(\rv;t)}{\partial t^2}=\nonumber\\
&\quad\quad\quad\quad\quad\quad\quad-\alpha_{\rm th}\frac{2(1+\mu)}{(1-2\mu)}\nablav \delta T(\rv;t) +
\frac{2(1+\mu)}{E}\mathbf f_{\rm fric}(\rv;t),
\label{eq:u-td}
\end{align}
where $\alpha_{\rm th}$ denotes the coefficient of linear thermal expansion of gold, $\delta T\equiv T-T_0$ denotes the temperature change with respect to the ambient temperature $T_0=293\,\text{K}$, and $\mathbf f_{\rm fric}$ denotes the friction force density. We express $\uv$ in terms of $\delta T$ and $\mathbf f_{\rm fric}$ by
\begin{subequations}
\begin{align}
\uv(\rv;t)= \int_{-\infty}^{t}\int d^3\rv' \greenOP(\rv,\rv';t,t')\cdot
\left[-\alpha_{\rm th}\frac{2(1+\mu)}{(1-2\mu)}\nablav'\delta T(\rv';t')+
\frac{2(1+\mu)}{E}\mathbf f_{\rm fric}(\rv';t')
\right],
\end{align}
with elastic Green's tensor $\greenOP(\rv,\rv';t',t)$ defined by
\begin{align}
\left[\nablav\times\nablav \times - \frac{2(1-\mu)}{1-2\mu}\nablav\nablav\cdot+ \frac{\omega^2\rho(1+\mu)}{E}\frac{\partial^2}{\partial t^2}\right] \greenOP (\rv,\rv';t,t') = \delta(\rv-\rv')\delta(t-t')\Iv.
\label{eq:green-td}
\end{align}
\end{subequations}

\subsection{Modal Expansion of Elastic Green's Tensor}

In this subsection, we represent $\greenOP(\rv,\rv';t,t')$ with the elastic waveguide modes of the plate. Firstly, we express $\greenOP(\rv,\rv';t,t')$ with the frequency-domain representation
\begin{subequations}
\begin{align}
\greenOP(\rv,\rv';t,t')=\frac{1}{2\pi}\int d\omega\,  \greenOP(\rv,\rv';\omega) e^{-i\omega(t-t')},
\label{eq:green-fourier-transform}
\end{align}
where $\greenOP (\rv,\rv';\omega)$ defined by
\begin{align}
\left[\nablav\times\nablav \times  - \frac{2(1-\mu)}{1-2\mu}\nablav\nablav\cdot - \frac{2\omega^2\rho(1+\mu)}{E}\right]\greenOP (\rv,\rv';\omega) = \delta(\rv-\rv')\Iv.
\label{eq:green-freq}
\end{align}
\end{subequations}
Then, we denote the wavenumbers and the displacement vectors of the elastic waveguide modes by
\begin{align}
\beta_{\rm 3D}^{\pm m}\;\; \text{and}\;\; \uv_{\rm 3D}^{\pm m}(\rv;\omega)=
\widetilde\uv_{\rm 3D}^{\pm m}(x,y; \omega)e^{ i \beta_{\rm 3D}^{\pm m} z},
\nonumber
\end{align}
where $\pm m=\pm 1, \pm 2,\cdots$ label modes propagating in the $\pm z$ directions, respectively. The $\pm m$ modes relate each other with
\begin{align}
\beta_{\rm 3D}^m=-\beta_{\rm 3D}^{-m} \quad \text{and}\quad
\widetilde \uv_{\rm 3D}^{-m}=\widetilde \uv_{\rm 3D}^{m}\cdot\left[-\hat x\hat x-\hat y\hat y+\hat z \hat z\right].
\nonumber
\end{align}
The modal-expansion formulation of $\greenOP (\rv,\rv';\omega)$ is then given by
\begin{align}
\greenOP (\rv,\rv';\omega)=
    \begin{cases}
    \sum_{m=1}^\infty  \frac{i({v}^{\rm T})^2}{2\omega {v}_{\rm 3D}^{m}}
      \widetilde\uv_{\rm 3D}^{m}(x,y;\omega) \otimes \widetilde\uv_{\rm 3D}^{-m}(x',y';\omega)
e^{i\beta_{\rm 3D}^m |z-z'| }
& \text{if $z>z'$}\\
     \sum_{m=1}^\infty  \frac{i({v}^{\rm T})^2}{2\omega {v}_{ \rm 3D}^{m}}
      \widetilde\uv_{\rm 3D}^{-m}(x,y;\omega) \otimes \widetilde\uv_{\rm 3D}^{m}(x,y;\omega)
e^{i \beta_{\rm 3D}^m |z-z'|}
& \text{if $z<z'$}
    \end{cases}  ,
    \label{eq:Green_0}
\end{align}
where ${v}^{\rm T}=\sqrt{E/2\rho(1+\mu)}$ denotes the velocity of transverse plane waves in bulk gold, and ${v}_{ \rm 3D}^{m}=\partial \omega/\partial \beta_{ \rm 3D}^m$ is the group velocity for the $\rm m^{\rm th}$ waveguide mode. The modal vectors $\widetilde\uv_{\scriptscriptstyle \rm RWG}^{\pm m}(x,y)$ are normalized such that
\begin{align}
\int_{-w/2}^{w/2} dx \int_{-h/2}^{h/2} dy \;  \widetilde\uv_{\rm 3D}^{-m}(x,y) \cdot
\widetilde\uv_{\rm 3D}^{m}(x,y) =1.
\nonumber
\end{align}

\subsection{Approximations}

To derive a closed-form expression for $\greenOP (\rv,\rv';t,t')$ that benefits physics understanding, we propose the following approximations:
\begin{itemize}
  \item only the $L$-modes are retained in the modal-expansion formulation of the Green's tensor,
  \item the long-wavelength approximate expressions for the $L$-modes, Eqs.~\eqref{eq:disp_RWG2}, are used.
\end{itemize}
These approximations physically assume that the low-frequency $L$-modes are dominantly excited. This assumption is reasonable considering that: (i) we use an ns light pulse to drive the plate, so that the frequency of the excited elastic waves is below GHz, satisfying the low-frequency condition; (ii) the absorbed optical power is designed to distribute uniformly in the $x$-direction, so that the $T$-modes cannot be excited as a result of symmetry mismatching.

Taking Eqs.~\eqref{eq:disp_RWG2} into Eqs.~\eqref{eq:Green_0} and then applying Eq.~\eqref{eq:green-fourier-transform}, $\greenOP (\rv,\rv';t,t')$ is derived as
\begin{align}
\greenOP (\rv,\rv';t,t')\simeq
\begin{cases}
\frac{({v}^{\rm T})^2}{2 A {v}_{\rm 3D}^{\rm L}}\widetilde \uv_{\rm 3D}^{+ \rm L} (x, y ) \otimes
                                                \widetilde \uv_{\rm 3D}^{- \rm L} (x', y' )
                                                H\left(t-t'-\frac{|z-z'|}{v_{\rm 3D}^{\rm L}}\right)\quad  & \text{for $z>z'$}\\
\frac{(v^{\rm T})^2}{2 A {v}_{\rm 3D}^{\rm L}} \widetilde \uv_{\rm 3D}^{- \rm L} (x, y ) \otimes
                                                \widetilde \uv_{\rm 3D}^{+ \rm L} (x', y' )
                                                H\left(t-t'-\frac{|z-z'|}{{v}_{ \rm 3D}^{\rm L}}\right)\quad & \text{for $z<z'$}
\end{cases},
\label{eq:green-0-lwa}
\end{align}
with
\begin{align}
\widetilde\uv_{\rm 3D}^{\pm \rm L} (x, y ) \equiv \pm \frac{\mu x\partial}{{v}_{\rm 3D}^{\rm L}\partial t}\hat x \pm
\frac{\mu y\partial}{{v}_{\rm 3D}^{\rm L}\partial t}\hat y + \hat z
\label{eq:u-RWG-LWA},
\end{align}
where $H(x)$ denotes the Heaviside step function with $H(x)=1$ for $x \ge 0$ and $H(x)=0$ otherwise; $v_{\rm 3D}^{\rm L}=\sqrt{E/\rho}$ denotes the velocity of the $L$-modes under the long-wavelength approximation; $A\equiv w\times h$ is the area of the cross section of the plate.

\subsection{Case Study: Fig.1 in the main text}

\label{Sec:Case_Study}
In this subsection, we employ Eqs.~\eqref{eq:green-0-lwa} and ~\eqref{eq:u-RWG-LWA} to study Fig. 1 in the main text. We decompose the induced displacement vector $\uv(\rv;t)$ into
\begin{align}
\uv(\rv;t)\equiv \uv_{\rm th}(\rv; t) + \uv_{\rm fric}(\rv; t),
\end{align}
where $\uv_{\rm th}$ and $\uv_{\rm fric}$ represent the contributions from the temperature change and the friction force, respectively.

\subsubsection{Thermal contribution}

Among three Cartesian components of $\uv_{\rm th}$, its $z$ component (parallel to the propagation direction of the $L$-modes), denoted by $u_{{\rm th};z}$, is dominant (see the modal profile in the inset of Fig. 1B in the main text). $u_{{\rm th};z}$ is derived with the following algebra manipulations:
\begin{align}
u_{{\rm th};z}(\rv;t) & = - \frac{2\alpha_{\rm th}(1+\mu)}{1-2\mu}\int_{-\infty}^t dt' \int d ^3 \rv'  \hat z \cdot \greenOP(\rv,\rv';t,t') \cdot \nablav'\delta T(\rv'; t')\nonumber\\
              & \overset{a}{\simeq} - \frac{\alpha_{\rm th} ({v}^{\rm T})^2 (1+\mu)}{A {v}_{ \rm 3D}^{\rm L}(1-2\mu)}
                \int_{-\infty}^t dt' \int d ^3 \rv'\, \left[{\rm sgn}(z'-z) \left(\frac{\mu x'\partial}{{v}_{\rm 3D}^{\rm L}\partial t}\hat x +
\frac{\mu y'\partial}{{v}_{\rm 3D}^{\rm L}\partial t}\hat y\right) + \hat z\right]
                H\left(t-t'-\frac{|z-z'|}{{v}_{\rm 3D}^{\rm L}}\right) \cdot \nablav'\delta T(\rv'; t')
                \nonumber \\
             & \overset{b}{=} \frac{\alpha_{\rm th} ({v}^{\rm T})^2 (1+\mu)}{A {v}_{\rm 3D}^{\rm L}(1-2\mu)}
                \int_{-\infty}^t dt' \int d ^3 \rv'\,  \delta T(\rv'; t') \nablav'\cdot\left\{
                \left[{\rm sgn}(z'-z) \left(\frac{\mu x'\partial}{{v}_{\rm 3D}^{\rm L}\partial t}\hat x +
\frac{\mu y'\partial}{{v}_{\rm 3D}^{\rm L}\partial t}\hat y\right) + \hat z\right]
                H\left(t-t'-\frac{|z-z'|}{{v}_{\rm 3D}^{\rm L}}\right)\right\}
                \nonumber \\
              &  = \frac{\alpha_{\rm th} E }{2 A \rho ({v}_{\rm 3D}^{\rm L})^2}\int d ^3 \rv'\, {\rm sgn}(z-z') \delta T\left(\rv';t-\frac{|z-z'|}{{v}_{\rm 3D}^{\rm L}}\right)
                \nonumber\\
              & \overset{c}{\simeq} {\rm sgn}(z-z_{\rm opt}) \frac{\alpha_{\rm th}}{2 A}\int d ^3 \rv'
             \delta T\left(\rv';t-\frac{|z-z_{\rm opt}|}{{v}_{\rm 3D}^{\rm L}}\right).
\label{eq:u-z-th_1}
\end{align}
The above derivations involve the following steps:
\begin{enumerate}[label=$\alph*$.\ ]
\item (1) Use $\greenOP$ expressed in Eq.~\eqref{eq:green-0-lwa}; (2) define the sign function ${\rm sgn}(x)$ such that
${\rm sgn}(x)=1$ for $x>0$, ${\rm sgn}(x)=-1$ for $x<0$ and ${\rm sgn}(x)=0$ for $x=0$.
\item Employ the technique of integration by parts.
\item Assume that $\delta T$ localizes in a tiny region centralized at $z=z_{\rm opt}$ and approximate that $\frac{|z-z'|}{{v}_{\rm 3D}^{\rm L}} \simeq \frac{|z-z_{\rm opt}|}{{v}_{\rm 3D}^{\rm L}} $.
\end{enumerate}

Departing from Eq.~\eqref{eq:u-z-th_1}, we further establish a link between $u_{{\rm th};z}$ and the net heat energy absorbed into/leaked from the plate using the heat conduction equation,
\begin{align}
\rho c_{ p} \frac{\partial\, \delta T(\rv; t)}{\partial\, t}-\nablav\cdot K \nablav \delta T(\rv ; t)= Q_{\rm abs}(\rv; t),
\label{eq:heat-eq}
\end{align}
where $c_p$ and $K$ denote the specific heat capacity and thermal conductivity of gold; $Q_{\rm abs}$ denotes the density of the absorbed optical power. With Eq.~\eqref{eq:heat-eq}, we derive that
\begin{align}
\int d^3 \rv \delta T(\rv;t)\overset{a}{=}      &   \int_{-\infty}^t dt' \int d^3\rv  {Q_{\rm abs}(\rv; t')}/(\rho c_p)
+  \int_{0}^t dt' \oint ds    \hat {\mathbf n} \cdot K \nablav \delta T(\rv ; t')/(\rho c_p)\nonumber\\
                           \overset{b}{=}&  \int_{-\infty}^t dt'  \left[P_{\rm abs}(t')-P_{\rm leak}(t')\right]/(\rho c_p)\nonumber\\
                            \overset{c}{\equiv} &  W_{\rm abs}^{\rm eff}(t)/(\rho c_p).
\label{eq:delta_T}
\end{align}
The above derivations involve the following steps:
\begin{enumerate}[label=$\alph*$.\ ]
\item  Apply the integral operation  $\int_{0}^t dt' \int d^3\rv $ to the both sides of Eq.~\eqref{eq:heat-eq} and employ the divergence theorem to transform the volume integral to the surface one.
\item  Define the heat power leaked into the outside environment,
$P_{\rm leak}(t)\equiv -\oint ds \hat {\mathbf n} \cdot K \nablav \delta T(\rv ; t)$.
\item  Define the effective heat energy with
\begin{align}
W_{\rm abs}^{\rm eff}\equiv \int_{-\infty}^t dt'  \left[P_{\rm abs}(t')-P_{\rm leak}(t')\right].\nonumber
\end{align}
\end{enumerate}

Inserting Eq.~\eqref{eq:delta_T} into Eq.~\eqref{eq:u-z-th_1}, we derive that
\begin{align}
u_{{\rm th};z}(\rv;t) & \simeq {\rm sgn}(z-z_{\rm opt}) \frac{\alpha_{\rm th} }{2 A \rho c_{p}}W_{\rm abs}^{\rm eff}\left(t-\frac{|z-z_{\rm opt}|}{{v}_{\rm 3D}^{\rm L}}\right).
\label{eq:u-z-th_3}
\end{align}

\subsubsection{Frictional contribution}

Similar as $\uv_{\rm th}$, the $z$-component of $\uv_{\rm fric}$, denoted by $u_{{\rm fric};z}$, is the dominant one, and is derived as follows
\begin{align}
u_{{\rm fric};z}(\rv;t)
& = \frac{2(1+\mu)}{E}\int_{-\infty}^t dt' \int d^3\rv' \hat z \cdot \greenOP(\rv,\rv';t,t') \cdot {\bf f}_{{\rm fric}}(\rv')
\nonumber\\
& \overset{a} {\simeq } \frac{2(1+\mu)}{E}\int_{-\infty}^t dt' \int d^3\rv' \hat z \cdot \greenOP(\rv,\rv';t,t') \cdot \hat z f_{{\rm fric};z}(\rv')
\nonumber\\
& \overset{b} {\simeq} \frac{({v}^{\rm T})^2(1+\mu)}{A {v}_{\rm 3D}^{\rm L}E}\int_{-\infty}^t dt' \int d^3\rv'
H\left(t-t'-\frac{|z-z'|}{{v}_{\rm 3D}^{\rm L}}\right)f_{{\rm fric};z}(\rv';t')\nonumber\\
& \overset{c} {\simeq} \frac{1}{2A \rho {v}_{\rm 3D}^{\rm L}}\int_{-\infty}^t dt'
H\left(t-t'-\frac{|z-z_{\rm cs}|}{{v}_{\rm 3D}^{\rm L}}\right)F_{{\rm fric};z}(t').
\label{eq:fric-disp}
\end{align}
The above derivations involve the following steps:
\begin{enumerate}[label=$\alph*$.\ ]
\item Application of $\mathbf f_{{\rm firc}}\simeq f_{{\rm firc};z}\hat z$.
\item Use $\greenOP$ expressed in Eq.~\eqref{eq:green-0-lwa}.
\item Note that the friction force exists on the contact surface between the plate and the substrate, which is a line-shaped area centralized at $z=z_{\rm cs}$, so that
    $\int d^3\rv' H\left(t-t'-\frac{|z-z'|}{{v}_{\rm 3D}^{\rm L}}\right)f_{{\rm fric};z}(\rv';t')\simeq
    H\left(t-t'-\frac{|z-z_{\rm cs}|}{{v}_{\rm 3D}^{\rm L}}\right)\int d^3\rv' f_{{\rm fric};z}(\rv';t')\equiv
    H\left(t-t'-\frac{|z-z_{\rm cs}|}{{v}_{\rm 3D}^{\rm L}}\right)F_{{\rm fric};z}(t)$ with $F_{{\rm fric};z}(t) \equiv \int d^3\rv f_{{\rm fric};z}(\rv;t)$.
\end{enumerate}

\subsubsection{Sliding Displacement \& Threshold Power}
\label{eq:SD-STP}

We now employ Eqs.~\eqref{eq:u-z-th_3} and~\eqref{eq:fric-disp} to derive the expression for the sliding displacement of the contact surface, denoted by $u_z^{\rm cs}$. By putting $z=z_{\rm cs}$ (the center of the z-coordinate of the contact surface) in Eqs.~\eqref{eq:u-z-th_3} and~\eqref{eq:fric-disp}, $u_z^{\rm cs}$ is derived as
\begin{align}
u_z^{\rm cs}(t) & \equiv  u_{{\rm th};z}^{\rm cs} + u_{{\rm fric};z}^{\rm cs} \nonumber\\
             &  \simeq {\rm sgn}(z_{\rm cs}-z_{\rm opt}) \frac{\alpha_{\rm th}  }{2 A \rho c_{p}}W_{\rm abs}^{\rm eff}\left(t-t_0\right) + \frac{1}{2A \rho {v}_{\rm 3D}^{\rm L}}\int_{-\infty}^t dt' F_{{\rm fric};z}(t'),
             \label{eq:u-cs}
\end{align}
where $t_0\equiv \frac{|z_{\rm cs}-z_{\rm opt}|}{{v}_{\rm 3D}^{\rm L}}$ represents the time for the $L$-modes traveling from the center of the absorbed optical power to the center of contact surface.

To evaluate $u_z^{\rm cs}$, the friction force needs to be specified. As is explained in the main text, the friction force can be determined by recognizing that
\begin{itemize}
  \item \it{
From a microscopic wave picture, the friction force behaves as a fence resisting the elastic waves from passing through the contact surface.
In the static regime, the static friction force is large enough to totally reflect back the incident elastic waves and thus nullify mechanical oscillations on the contact surface (that is, the incident and reflected elastic waves cancel each other out perfectly), thereby preserving the contact surface in the still state.
On the contrary, in the dynamic regime, the elastic waves are so strong that even the maximum allowable static friction force---the so-called sliding resistance force denoted by $F_{\rm silde}$---cannot nullify them on the contact surface. The plate thus slides.
}
\end{itemize}

\begin{figure}[!htp]
\centering
\includegraphics[width=14cm]{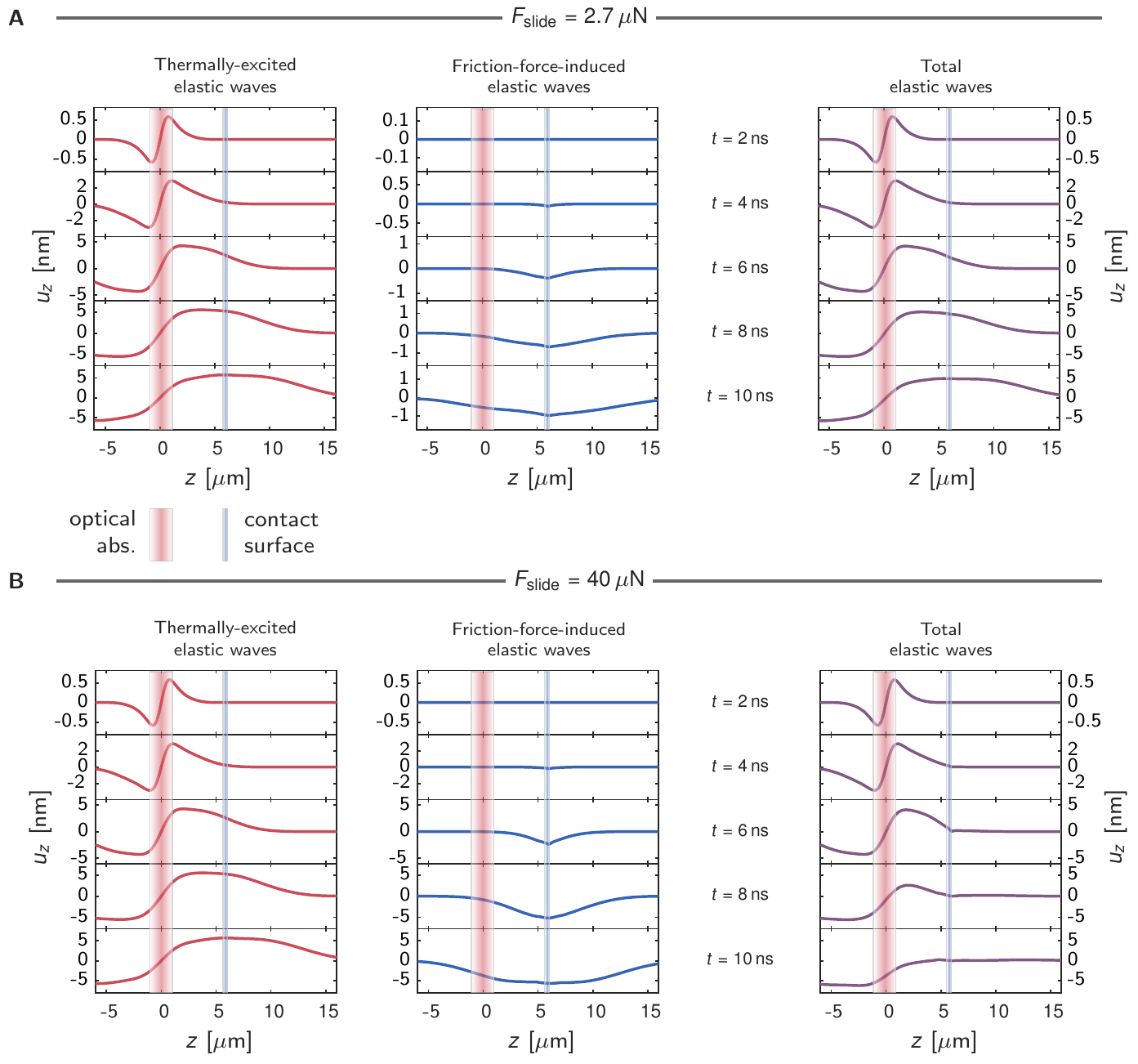}
\caption{{\bf Numerical simulations of excited elastic waves, partitioning contributions from absorbed optical power and friction force}. The simulated system is as the same as studied in Fig. 1C in the main text with $F_{\rm slide}=2.7\,\mu\rm N$ ({\bf A}) and $F_{\rm slide}=40\,\mu\rm N$ ({\bf B}). The total elastic waves (rightmost panels) are the summation of thermally(optical absorption)-excited elastic waves (leftmost panels) and friction-force-induced ones (middle panels).
}
\label{Fig:Cancellation}
\end{figure}

To confirm this wave-cancellation picture, we re-examine Fig. 1C in the main text, and perform additional numerical simulations that partition the total excited elastic waves into the separate contributions from the absorbed optical power and the friction force. As is shown in Fig.~\ref{Fig:Cancellation}A, with $F_{\rm slide}=2.7\,\mu\rm N$, the peak value of the absorbed optical power exceeds $P_{\rm TH}$, so that the amplitude of the thermally-excited elastic waves is larger than that of the friction-force-induced elastic waves. Accordingly, the elastic waves can pass through the contact surface. On the contrary, increasing $F_{\rm slide}$ to $40\,\mu\rm N$, the peak value of the absorbed optical power becomes lower than $P_{\rm TH}$. Consequently, the thermally-excited and friction-force elastic waves cancel out each other, and result in the zero transmission of the elastic waves through the contact surface, as shown in Fig.~\ref{Fig:Cancellation}B.

To concretize this physical picture, we first analyze the static regime wherein the contact surface is motionless. Accordingly, $d u_z^{\rm cs} /dt =0$, which together with Eq.~\eqref{eq:u-cs} gives the $z$-component of the friction force as follows
\begin{subequations}
\begin{align}
F_{{\rm fric};z}(t) \simeq F_{{\rm fric};z}^{\rm static}(t)\quad\text{(static regime)},
\end{align}
with
\begin{align}
F_{{\rm fric};z}^{\rm static}\equiv{\rm sgn}(z_{\rm opt}-z_{\rm cs})\frac{\alpha_{\rm th}{v}_{\rm 3D}^{\rm L}P_{\rm abs}^{\rm eff}(t-t_{0})}{c_p},
\end{align}
where $P_{\rm abs}^{\rm eff}\equiv d W_{\rm abs}^{\rm eff}/dt=P_{\rm abs}-P_{\rm leak} $.
Further, it is known that the magnitude of the static friction force is upper bounded by the sliding resistance $F_{\rm silde}$, i.e., $ |F_{{\rm fric};z}^{\rm static}|<F_{\rm silde}$. This constrainment leads to that
\begin{align}
|P_{\rm abs}^{\rm eff}(t-t_0)|<P_{\rm TH}
\label{eq:PTH1}
\end{align}
\label{eq:static-fric}
\end{subequations}
with
\begin{align}
P_{\rm TH}=\frac{c_pF_{\rm silde}}{\alpha_{\rm th}{v}_{\rm 3D}^{\rm L}}.
\label{eq:PTH}
\end{align}

Apparently, the critical point, at which the plate starts moving, occurs when the constraint condition is just broken, i.e., $|P_{\rm abs}^{\rm eff}(t-t_{0})|=P_{\rm TH}$. This observation renders $P_{\rm TH}$ a clear physical meaning: the minimum of the {\it instantaneous} power needed to overcome the sliding resistance. For this reason, $P_{\rm TH}$ is named as the {\it sliding threshold power}.

As the plate moves, it is subject to the dynamic friction force, whose magnitude generally has the same order as the sliding resistance. Taking this fact into account and for simplicity of analysis, we here assume that the magnitude of the dynamic friction force just equals the sliding resistance, i.e., approximating the dynamic friction force as
\begin{subequations}
\begin{align}
 F_{{\rm fric};z}(t)\simeq{\rm sgn}\left[ F_{{\rm fric};z}^{\rm static}(t)\right]  F_{\rm slide}
  \quad\text{(dynamic regime)},
  \label{eq:dynamic-fric-1}
\end{align}
which is activated under the condition
\begin{align}
|P_{\rm abs}^{\rm eff}(t-t_0)|>P_{\rm TH}.
  \label{eq:dynamic-fric-2}
\end{align}
\label{eq:dynamic-fric}
\end{subequations}

Moreover, exploiting the derived expressions of the friction force, the expression of the sliding displacement of the contact surface, Eq.~\eqref{eq:u-cs}, can be alternatively expressed as \begin{align}
u_z^{\rm cs} (t) \simeq  {\rm sgn}(z_{\rm cs}-z_{\rm opt}) \frac{\alpha_{\rm th}  }{2 A \rho c_{p}} \int_{-\infty}^ t dt'
{\rm sgn}\left[P_{\rm heat}(t'-t_{0})\right]
\left(|P_{\rm heat}(t'-t_{0})|-P_{\rm TH}\right) {H}( |P_{\rm heat}(t'-t_{0})|-P_{\rm TH})
\label{eq:fric-disp-int}
\end{align}
with the term of the Heaviside step function [that is, ${H}( |P_{\rm heat}(t'-t_{0})|-P_{\rm TH})$] explicitly specifying that the sliding occurs when $|P_{\rm heat}(t'-t_{0})|>P_{\rm TH}$.
From Eq.~\eqref{eq:fric-disp-int}, we see that the sliding in principle is possible in both the heating period with $P_{\rm heat}>0$ and the cooling period with $P_{\rm heat}<0$. Two cases contribute to the sliding displacement in opposite directions as indicated by the term ${\rm sgn}\left[P_{\rm heat}(t'-t_{0})\right]$.
Referring to our problem, in the cooling period, the tiny contact surface between the plate and the substrate restricts the cooling efficiency significantly (see Fig.~\ref{Fig:Heat_Cooling}) and results in a negligible cooling power, so that the plate cannot slide. On the contrary, in the heating period, the noticeable absorbed optical power $P_{\rm abs}$ (with $P_{\rm abs}^{\rm eff}\simeq P_{\rm abs}$) can easily exceed $P_{\rm TH}$, thereby driving the sliding of the plate.
In view of this, we can safely replace $P_{\rm heat}$ with $P_{\rm abs}$ in Eq.~\eqref{eq:fric-disp-int} and obtain that
\begin{align}
u_z^{\rm cs} (t) \simeq {\rm sgn}(z_{\rm cs}-z_{\rm opt}) \frac{\alpha_{\rm th}  }{2 A \rho c_{p}} \int_{-\infty}^ t dt'
(P_{\rm abs}(t'-t_{0})-P_{\rm TH}){H}( P_{\rm abs}(t'-t_{0})-P_{\rm TH}).
\label{eq:fric-disp-int2}
\end{align}
Equations~\eqref{eq:u-cs}---\eqref{eq:fric-disp-int2} summarize the main results of this section. Even though a few approximations are made to derive these equations, they show a good predictive accuracy as validated in Fig. 1C in the main text.

\subsubsection{Attenuation effects}

In the above derivations, the elastic attenuation is ignored. However, in reality, the attenuation always exists due to the elastic coupling between the gold nanoplate and the environment, in addition to the intrinsic damping mechanisms, such as lattice defects. To account for the attenuation effects, we here introduce a phenomenological model, which assumes that the elastic waves have frequency-independent life time, denoted by $\tau_{\rm ac}$. This assumption amounts to adding a constant imaginary part $1/(2\tau_{\rm ac} {v}_{\rm 3D}^{\rm L})$ to $\beta_{\rm 3D}^{\rm L}$. In this way, the attenuation effects can be effectively, conveniently incorporated. For instance, the sliding power threshold changes to
\begin{align}
P_{\rm TH}=\frac{c_pF_{\rm silde}}{\alpha_{\rm th}{v}_{\rm 3D}^{\rm L}}e^{t_{0}/(2\tau_{\rm ac})},
\label{eq:PTH_Attenu}
\end{align}
which is multiplied by a factor of $e^{t_{0}/(2\tau_{\rm ac})}$ in comparison with the lossless case. Equation~\eqref{eq:PTH_Attenu} is just Eq. (1) in the main text.


\section{Rotation Displacement of Gold Plates around Micro-fibers: Theoretical Analysis}
\label{Sec:Rot_Theory}
In this section, we theoretically analyze the rotation movement of gold plates around micro-fiber, as has been observed in our experiments. {\bf The results here supplement the discussions attached to Fig. 4C in the main text.}

\begin{figure}[!htp]
\centering
\includegraphics[width=12cm]{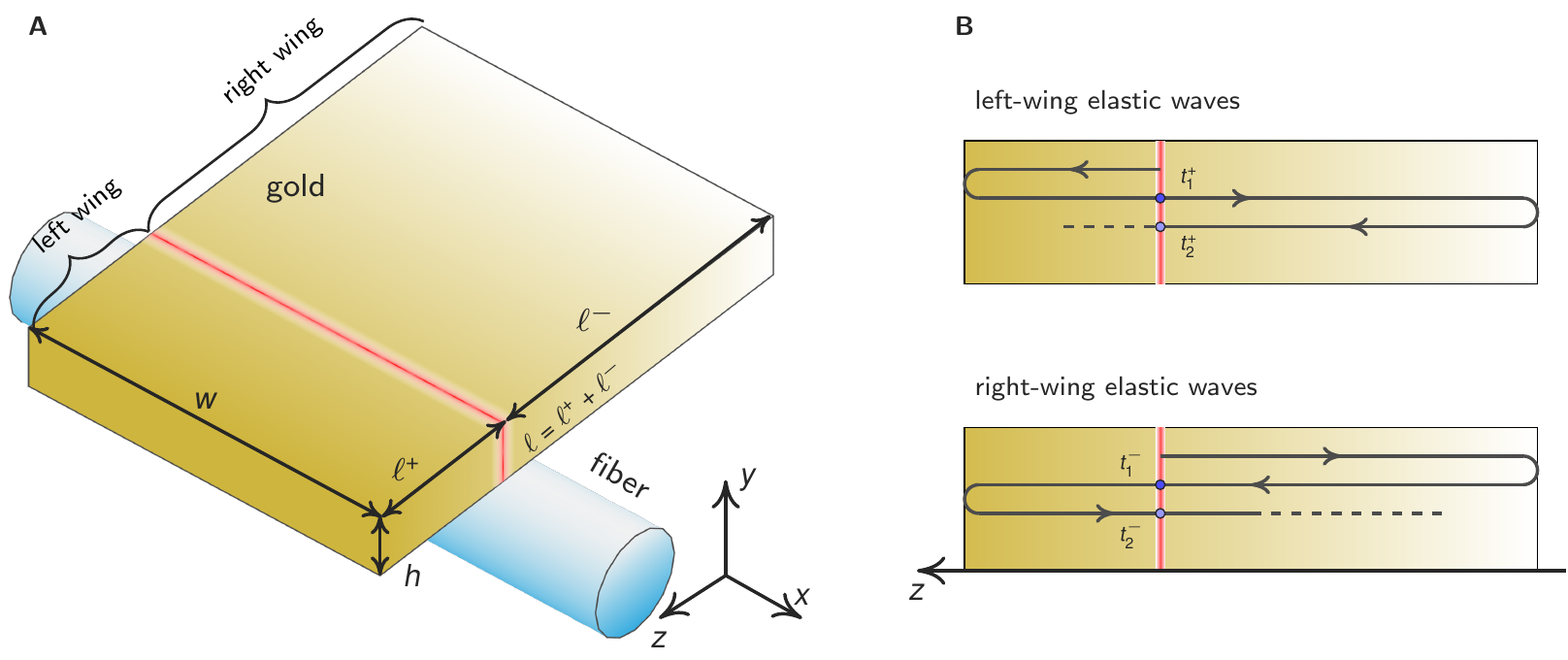}
\caption{{\bf A}. Sketch of the experimental setup: a gold plate is placed on top of a micro-fiber. {\bf B.} Excited elastic waves are classified into left- and right-wing waves, which initially propagate towards the left- and right-wings, respectively, before reflecting at two ends of the plate in the $z-$direction.
}
\label{Fig:Plate_fiber}
\end{figure}

\begin{figure}[!htp]
\centering
\includegraphics[width=8cm]{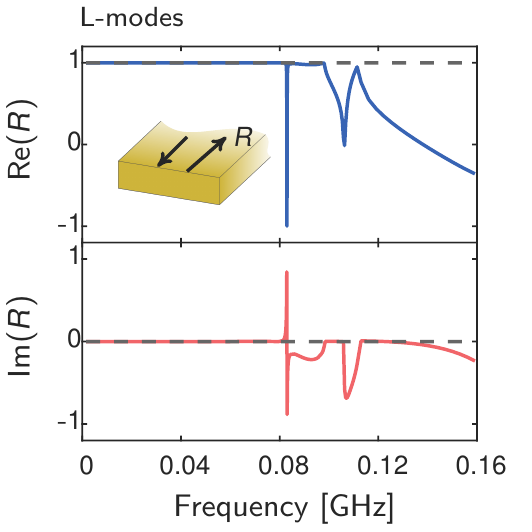}
\caption{
{\bf Reflection spectrum of fundamental $L$-modes}. The cross section of the studied gold plate has a width of 10 $\mu\rm m$ and a thickness of 30 nm. The numerical results (solid lines) are obtained with the COMSOL Multiphysics. The long-wavelength-approximation (LWA) results (dashed lines) represent the limiting values of the reflection coefficients as frequency approaches zero. The reflection coefficient is defined as the ration between the longitudinal component of the reflected $L$-modes and that of the incident ones.
}
\label{Fig:modal_reflection}
\end{figure}

Figure~\ref{Fig:Plate_fiber}A sketches the problem under investigation. The theoretical analysis here below is a direct generalization of the theory developed in Sec.~\ref{Sec:Case_Study} by additionally taking multiple reflected elastic waves into account. The reflection spectrum of the fundamental $L$-modes is computed with COMSOL Multiphysics, as shown in Fig.~\ref{Fig:modal_reflection}. The results show that, at low frequencies, the reflection coefficients approximately equal to $1$.

{\bf Effects of multiple reflections}---The thermally-excited $L-$modes in the plate are classified into the left- and right-wing waves, according to their initial propagation directions towards the left- and right-wing sides, respectively (see Fig.~\ref{Fig:Plate_fiber}B).
The longitudinal components of the elastic displacement vectors carried by left- and right-wing waves are in the opposite directions. Specifically, their initial directions, before they reach the ends of the plate, are parallel to the wave propagation directions, i.e., pointing towards the left- and right-wing sides, respectively, as implied in Eq.~\eqref{eq:u-z-th_3}.
After reflections, the $L$-modes reverse their propagation directions, while the longitudinal components of the elastic displacement vectors do not change the sign because the reflection coefficients equal to $1$ (under the long-wavelength approximation; see Fig.~\ref{Fig:modal_reflection}).
Therefore, the left- and right-wing waves continuously drive the pate to rotate towards the left- and right-wing sides, respectively.
Taking the aforementioned multiple reflections into account and employing Eqs.~\eqref{eq:u-z-th_3} and ~\eqref{eq:fric-disp}, the rotation displacement of the contact surface of the gold plate, denoted by $u_{\rm rot}^{\rm cs}$, is derived as follows
\begin{subequations}
\begin{align}
u_{\rm rot}^{\rm cs}(t) =  \underbrace{u_{{\rm th};z}^{\rm cs}(t)}_{\text{thermal contribution}} +
\underbrace{ u_{{\rm fric};z}^{\rm cs}(t)}_{\text{friction-force contribution}}
\end{align}
with
\begin{align}
u_{{\rm th} ; z}^{\rm cs}(t) \simeq \frac{\alpha_{\rm th}  }{2 A \rho c_{p}}\sum_{n=1}^{\infty} \underbrace{W_{\rm abs}^{\rm eff}\left(t-t_{ n}^{+}\right)e^{-t_{n}^{+}/(2\tau_{\rm ac})}}_{\text{left-wing waves}} - \underbrace{W_{\rm abs}^{\rm eff}\left(t-t_{ n}^{-}\right)e^{-t_{n}^{-}/(2\tau_{\rm ac})}}_{\text{right-wing waves}},
\label{eq:uth_rot}
\end{align}
and
\begin{align}
u_{{\rm fric}; z }(t) \simeq \frac{P_z(t)}{2A \rho {v}_{\rm 3D}^{\rm L}} + \frac{1}{{2A \rho {v}_{\rm 3D}^{\rm L}}}\sum_{n=1}^{\infty} P_z(t-t_{ n}^{+})e^{-t_{ n}^{+}/(2\tau_{\rm ac})}
+
P_z(t-t_{ n}^{-})e^{-t_{ n}^{-}/(2\tau_{\rm ac})}.
\label{eq:uth_fric}
\end{align}
Here $P_z(t)\equiv\int_{-\infty}^t dt' F_{{\rm fric};z}(t')$; $t_{n}^{\pm}= \left[2(n-1)\ell+2\ell^{\pm}\right]/{v}_{\rm 3D}^{\rm L}$ if $n$ is an odd number and $t_{ n}^{\pm}= 2(n-1)\ell/{v}_{\rm 3D}^{\rm L}$ if $n$ is an even number, where $\ell^{\pm}$ denote the lengths of the left and right wings, respectively, and $\ell=\ell^+ + \ell^-$ is the total length of the plate. Physically, $t_{ n}^{\pm}$ represent the propagation time that the left- and right-wing waves take to return to their initial positions after $n$ times reflections (see Fig.~\ref{Fig:Plate_fiber}B). The attenuation effects are approximately incorporated with the damping terms $e^{-t_{ n}^{\pm}/(2\tau_{\rm ac})}$.
\label{eq:disp-Rot}
\end{subequations}

The friction force can be determined by using the same routine as in Sec.~\ref{eq:SD-STP}. First, the static friction force, resulting from $du_z^{\rm cs}(t)/dt=0$, is given by
\begin{subequations}
\begin{align}
F_{{\rm fric}}(t)=F_{{\rm fric}}^{\rm static}(t)
\quad\text{(static regime)},
\end{align}
with
\begin{align}
F_{{\rm fric}}^{\rm static}(t) &\equiv -\frac{\alpha_{\rm th} E }{2 A \rho^2 c_{p}\left({v}_{\rm 3D}^{\rm L}\right)^2}\sum_{n=1}^{\infty} P_{\rm abs}^{\rm eff}\left(t-t_{n}^{+}\right)e^{-t_{\rm tra; n}^{+}/(2\tau_{\rm ac})} - P_{\rm abs}^{\rm eff}\left(t-t_{ n}^{-}\right)e^{-t_{\rm tra; n}^{-}/(2\tau_{\rm ac})}\nonumber\\
&\quad - \frac{1}{{2A \rho {v}_{\rm 3D}^{\rm L}}}\sum_{n=1}^{\infty} F_{{\rm firc};z}(t-t_{\rm tra; n}^{+})e^{-t_{n}^{+}/(2\tau_{\rm ac})},
\label{eq:Fstatic_rot}
\end{align}
and constrained by
\begin{align}
|F_{{\rm fric}}^{\rm static}(t)|<F_{\rm slide}.\nonumber
\end{align}
\end{subequations}
In the dynamic regime, the friction force is given by
\begin{align}
F_{{\rm fric}}(t)={\rm sgn}\left[F_{{\rm fric}}^{\rm static}(t)\right]F_{{\rm slide}}.
\quad\text{(dynamic regime)},
\label{eq:Fdynamic_rot}
\end{align}
with
\begin{align}
|F_{{\rm fric}}^{\rm static}(t)|\ge F_{\rm slide}.
\label{eq:Fdynamic_constr}
\end{align}

{\bf Strong-attenuation approximation---}Equations~\eqref{eq:disp-Rot}--\eqref{eq:Fdynamic_constr} characterize the rotation movement of gold plates. However, they are complicated due to the presence of the infinite series. Therefore, examining them to reveal the physics is not straightforward. To bypass this complication, we simplify Eqs.~\eqref{eq:disp-Rot}--\eqref{eq:Fdynamic_constr} by retaining the leading-order terms in these equations. This amounts to keeping the $n=1$ terms of the series in Eq.~\eqref{eq:uth_rot} and dropping all the series in Eq.~\eqref{eq:uth_fric}. Such an approximation physically corresponds to {\it the strong elastic attenuation case} wherein all high-order reflected waves are sufficiently damped and thus become negligible. Under the strong-attenuation approximation and approximating that $P_{
abs}^{\rm eff}\simeq P_{
abs}$, $u_{\rm rot}^{\rm cs}$ could be arranged in a similar form as Eq.~\eqref{eq:fric-disp-int2} and is expressed as
\begin{align}
u_{\rm rot}^{\rm cs} (t)\simeq &\frac{\alpha_{\rm th}  }{2 A \rho c_{p}} \int_{-\infty}^ t dt'
(P_{\rm abs}(t'-t_{1}^+)-P_{\rm TH}^{+}){H}\left[ P_{\rm abs}(t'-t_{1}^+)-P_{\rm TH}^{+}\right]\nonumber\\
-&\frac{\alpha_{\rm th}  }{2 A \rho c_{p}} \int_{-\infty}^ t dt'
(P_{\rm abs}(t'-t_{1}^-)-P_{\rm TH}^{-}){H}\left[ P_{\rm abs}(t'-t_{1}^-)-P_{\rm TH}^{-}\right],
\end{align}
with
\begin{align}
P_{\rm TH}^{\pm}=\frac{c_pF_{\rm silde}e^{t_{1}^{\pm}/(2\tau_{\rm ac})}}{\alpha_{\rm th}{v}_{\rm 3D}^{\rm L}}
\end{align}
representing the sliding power threshold for the left- and right-wing waves, respectively. Apparently, if $\ell^+<\ell^-$, $P_{\rm TH}^{+}<P_{\rm TH}^{-}$, leading to $u_z^{\rm cs}>0$; on the contrary, if $\ell^+>\ell^-$, we have
$u_z^{\rm cs}<0$. Therefore, the rotation direction points from the long wing to the short wing.

\section{Optical Absorption in Gold-Plate \& Micro-fiber Coupled System}

\begin{figure}[!htp]
\centering
\includegraphics[width=14cm]{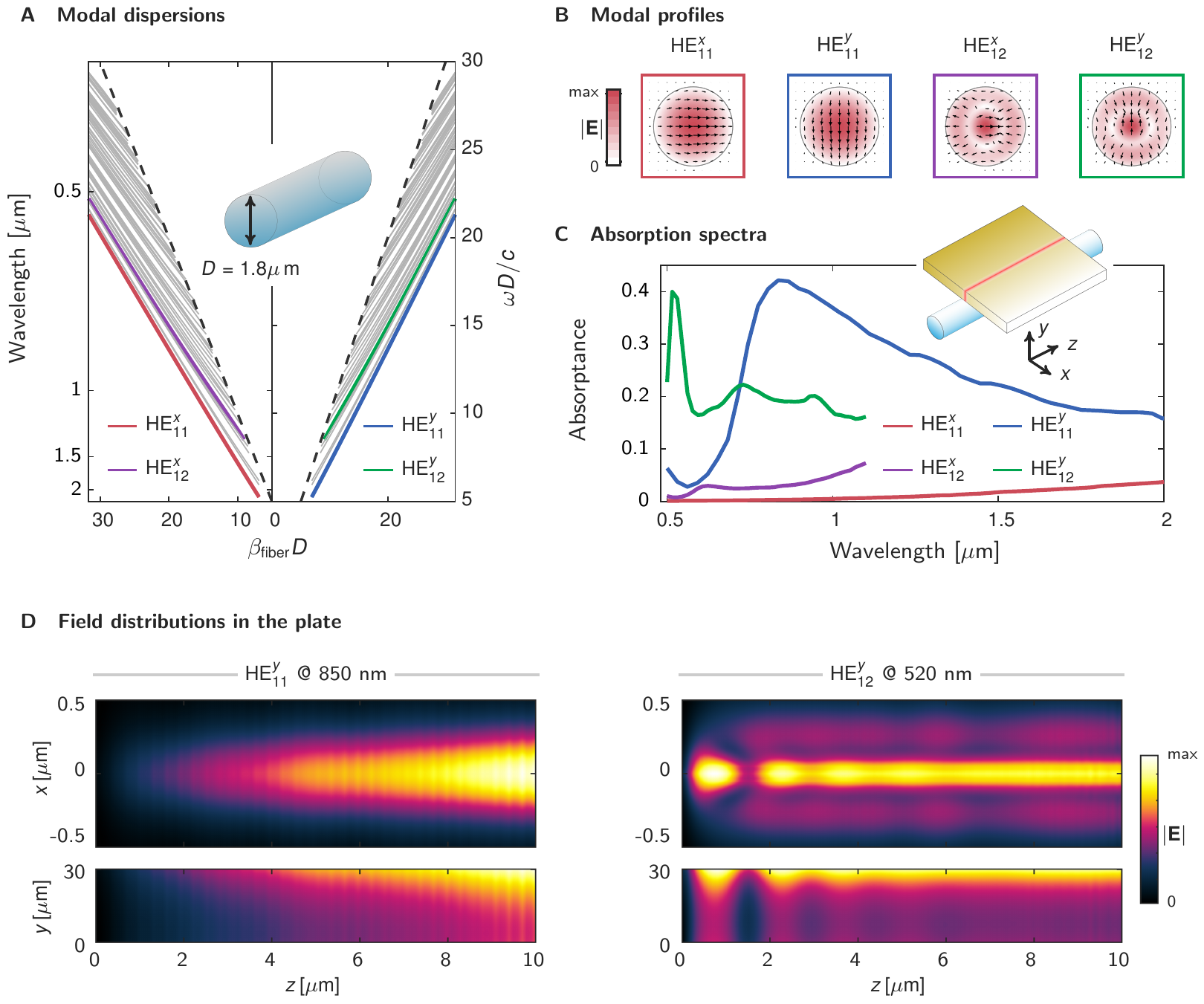}
\caption{ {\bf Optical absorption in gold-plate \& micro-fiber coupled system.} {\bf A.} Dispersion relations of optical waveguide modes of a 1.8 $\mu\rm m$-diameter micro-fiber in air. The refractive index of the micro-fiber is 1.46. {\bf B.} Modal profiles of fundamental ${\rm HE}_{11}^{\rm x,y}$ modes and high-order ${\rm HE}_{12}^{\rm x,y}$ modes. {\bf C.} Absorption spectra of ${\rm HE}_{11}^{\rm x,y}$ and ${\rm HE}_{12}^{\rm x,y}$ modes. The gold plate has a square base shape with side length 10 $\mu\rm m$ and thickness 30 nm. {\bf D.} Distributions of electric field strength in the $x-z$ and $y-z$ planes through the center of the plate for the incidences of the ${\rm HE}_{11}^{\rm y}$ (left) and ${\rm HE}_{12}^{\rm y}$ (right) modes. The wavelengths are at 850 nm and 520 nm for two modes, respectively, which correspond to the spectral positions of their absorption peaks ({\bf B}).
}
\label{Fig:Opt_Abs}
\end{figure}

In our experiments, we investigate gold-plate and micro-fiber coupled system, wherein a gold plate is adhered to a micro-fiber. Micro-fibers with diameter about a few $\mu\rm m$ can support a large number of optical modes in the wavelength range of the employed super-continuum laser (between 450 nm and 2400 nm). Fig.~\ref{Fig:Opt_Abs}A shows the dispersion relations of optical modes in a $1.8\,\mu \rm m$-diameter micro-fiber surrounded by air. Among these modes, we here discuss fundamental ${\rm HE}_{11}^{\rm x,y}$ modes and high-order ${\rm HE}_{12}^{\rm x,y}$ modes, which have the intensity maximum at the center of the fiber core (see Fig.~\ref{Fig:Opt_Abs}B for modal profiles of ${\rm HE}_{11}^{\rm x,y}$ and ${\rm HE}_{12}^{\rm x,y}$ modes) and, thus, can be efficiently excited by external laser pulses whose spatial profiles are generally Gaussian.

Gold plates absorb evanescent electric fields of optical modes and then generate heat. Figure~\ref{Fig:Opt_Abs}C depicts the computed absorption spectra for a square gold plate---with side length 10 $\mu\rm m$ and thickness 30 nm---, which is symmetrically placed on a micro-fiber with diameter 1.8$\mu\rm m$ for the ${\rm HE}_{11}^{\rm x,y}$ and ${\rm HE}_{12}^{\rm x,y}$ modes as incident waves. The results show that the ${\rm HE}_{11}^{\rm y}$ and ${\rm HE}_{12}^{\rm y}$ modes are more efficiently absorbed than the ${\rm HE}_{11}^{\rm x}$ and ${\rm HE}_{12}^{\rm x}$ modes. This is due to that the dominant component of electric fields of the ${\rm HE}_{11}^{\rm y}$ and ${\rm HE}_{12}^{\rm y}$ modes is perpendicular to the surface of the plate, which enables efficient excitations of plasmonic modes in the plate, thereby benefiting optical absorption through field-enhancement effects. While for the ${\rm HE}_{11}^{\rm x}$ and ${\rm HE}_{12}^{\rm x}$ modes, the dominant component of electric fields is instead parallel to the surface of the plate, unfavorable for plasmonic excitations.
Moreover, the plasmonic excitations due to the ${\rm HE}_{11}^{\rm y}$ and ${\rm HE}_{12}^{\rm y}$ incidences result in subwavelength localizations of electric fields in the plate along the $x$ and $y$ directions, as shown in Fig.~\ref{Fig:Opt_Abs}D.

\begin{figure}[!htp]
\centering
\includegraphics[width=10cm]{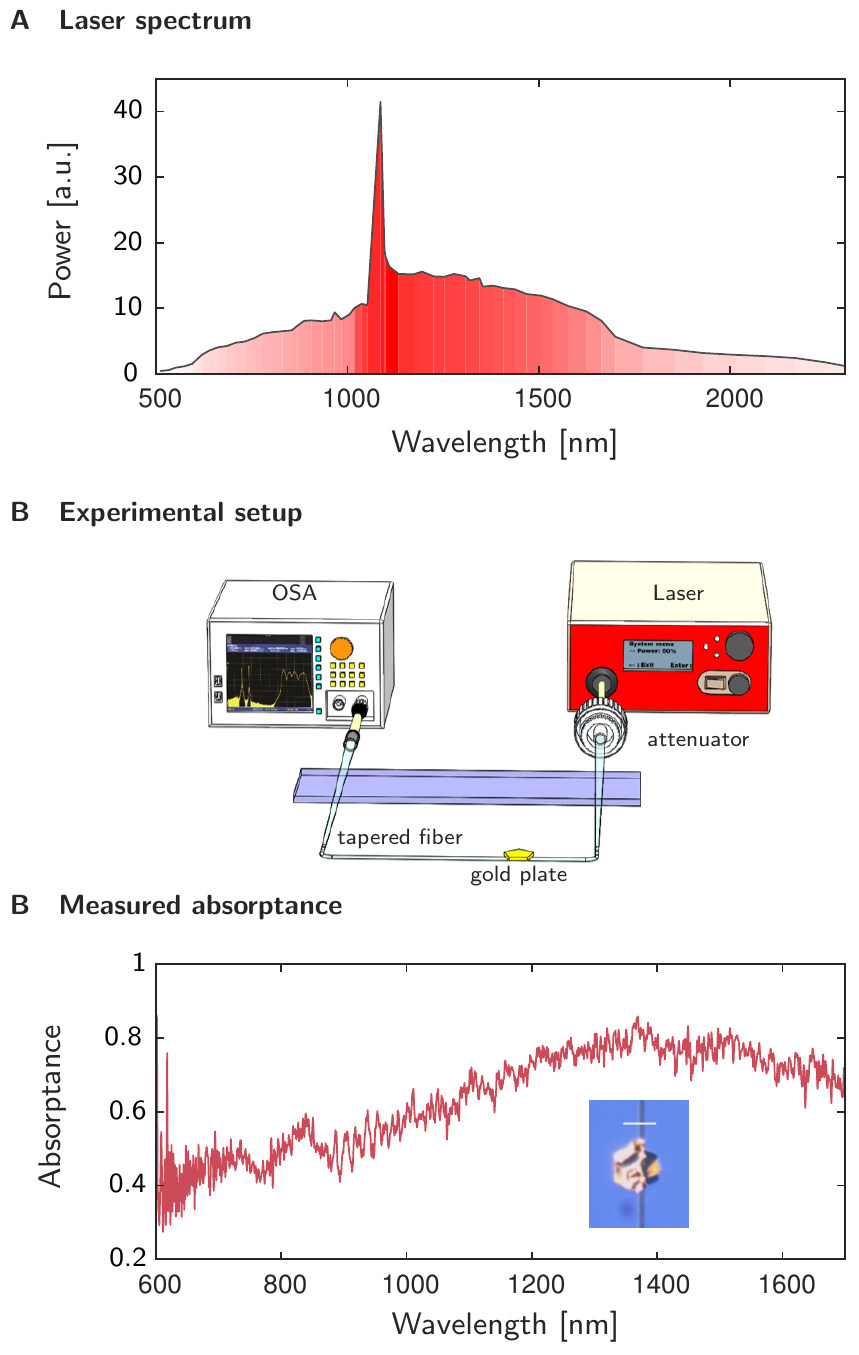}
\caption{ {\bf A.} Power spectrum of the employed supercontinuum laser. {\bf B.} Sketch of experimental setup for measuring absorption spectra.
{\bf C.} Measured absorption spectra of a hexagonal gold-plate \& micro-fiber coupled system. Inset: an optical image of the gold plate on the micro-fiber (scale bar, 15 $\mu\rm m$).
}
\label{Fig:Opt_Abs_Exp}
\end{figure}

We measured the absorption spectra of a gold-plate and micro-fiber coupled system, as shown in Fig.~\ref{Fig:Opt_Abs_Exp}C. Comparing with the simulation results, the measured absorptance is higher. This is attributed to that the size of the measure gold plate is larger than the simulated one. Moreover, it should be noted that the measured spectra are contributed from multi-modes of the micro-fiber and thus cannot be directly, quantitatively mapped to the simulated results of single individual modes in Fig.~\ref{Fig:Opt_Abs_Exp}C.

{\bf Remarks---}Even though we can precisely compute electromagnetic interactions between gold plates and individual optical modes of micro-fibers, it remains challenging to estimate the joint contributions from all modes, because experimentally quantifying coupling efficiencies of external laser pulses into various optical modes in the wide wavelength range of the used super-continuum laser is difficult. In view of this, in the below heat-elastic simulations that demand spatial distributions of absorbed optical power as an indispensable input, we retreat to a simple expression, Eq.~\eqref{eq:Q-abs}, as an approximate representation.

\section {Heating and Cooling Dynamics in Gold-Plate \& Micro-fiber Coupled System}

\begin{figure}[!htp]
\centering
\includegraphics[width=16cm]{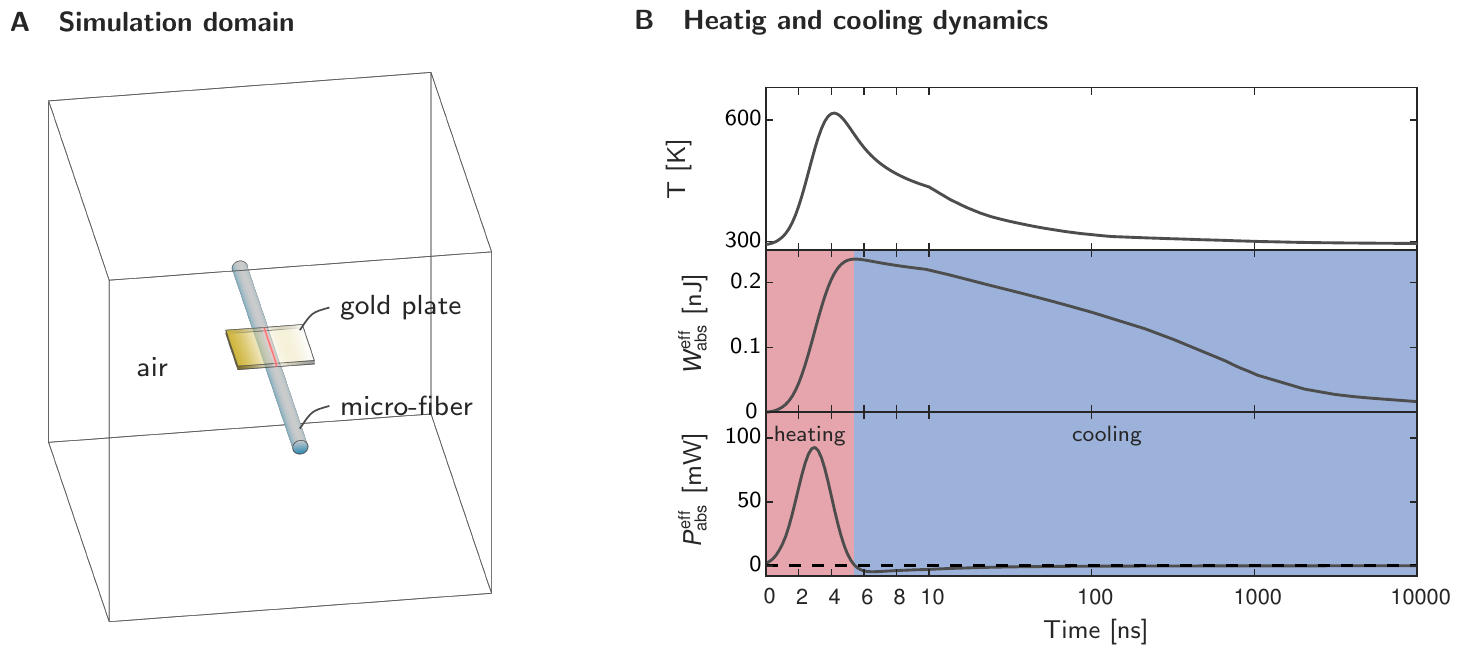}
\caption{{\bf Thermal dynamics in gold-plate \& micro-fiber coupled system driven by a single optical pulse.} {\bf A}. Sketch of the simulation domain. A square gold plate with side length $10\,\mu \rm m$ and thickness 30 nm is placed on top of a micro-fiber with diameter $1.8\,\mu \rm m$ and length $50 \,\mu \rm m$. The two wings of the plate separated by the micro-fiber are of equal length. An air cubic box with side length $50 \,\mu \rm m$ surrounds the plate and the micro-fiber. The driven optical pulse is specified in Eq.~\eqref{eq:Q-abs}. {\bf B.} Temperature at the center of the plate (upper panel), net heat energy $W_{\rm heat}$ (middle panel) and heat power $P_{\rm heat}$ (lower panel) of the plate, as functions of time.
}
\label{Fig:Heat_Cooling}
\end{figure}

In this section, we numerically examine the thermal heating and cooling dynamics in gold-plate $\&$ micro-fiber coupled system.
Figure~\ref{Fig:Heat_Cooling}A sketches the simulation domain comprising of a gold plate, a micro-fiber and air background (their geometrical dimensions are summarized in the caption of the figure).
Both the axial length of the micro-fiber and the background size are set large enough such that further increasing their values does not change simulation results significantly.
The absorbed optical power density $Q_{\rm abs}$ is set as
\begin{align}
Q_{\rm abs}(\rv;t)=\overbrace{P_{\rm peak}}^{\text{peak power}}\underbrace{\frac{1}{\sqrt{\pi}\tau_p}e^{-(t-t_0)^2/\tau_p^2}}_{\text{ temporal distribution}} \underbrace{\frac{1}{wh}\frac{1}{\sqrt{\pi}\ell_p}e^{-(z-z_{\rm cs})^2/\ell_p^2}}_{\text{ spatial distribution}},
\label{eq:Q-abs}
\end{align}
where $t_0=3\,\rm ns$, $\tau_p=1.5\,\rm ns$, $\ell_p=250\,\rm nm$, $z_{\rm cs}=0$ and $P_{\rm peak}=100\,\rm mW$; $w$ and $h$ denote the width and the thickness of the plate, respectively (see the inset in Fig.~\ref{Fig:Heat_Cooling}A).

Figure~\ref{Fig:Heat_Cooling}B shows the temperature at the center of the plate (upper panel), the net heat energy $W_{\rm abs}^{\rm eff}$ stored in the plate (middle panel) and the effective heat power $P_{\rm abs}^{\rm eff}\equiv d W_{\rm abs}^{\rm eff}(t)/dt$
(lower panel), as functions of time. The results show that the plate is rapidly heated within the short period of the optical absorption pulse, and then slowly cooled by transferring heat into the air background and the micro-fiber. The cooling efficiency is low due to (i) low thermal conductivity of air and (ii) tiny contact surface between the plate and the micro-fiber. As a result, $P_{\rm abs}^{\rm eff}$ in the cooling period has a rather small magnitude and its maximum is only about $5$ mW, about $5\%$ of the absorption peak power. The complete cooling takes more than $10\,\mu\rm s$.

\section{Coupled Heat-Elastic Simulations}

In this section, we present more results on modelling locomotion of gold plates around micro-fibers driven by optical pulses, {\bf supplementing Fig. 4 and the associated discussions in the main text}.

The numerical modelling amounts to solving the coupled heat-conduction and elastic equations:
\begin{subequations}
\begin{align}
\rho c_{ p} \frac{\partial\, \delta T(\rv; t)}{\partial\, t}-\nablav\cdot K \nablav \delta T(\rv ; t) = Q_{\rm abs}(\rv; t),
\label{eq:ht}
\end{align}
\begin{align}
\nablav\times\nablav \times \uv(\rv;t) - \frac{2(1-\mu)}{1-2\mu}\nablav\nablav\cdot\uv(\rv;t) + \frac{2\rho(1+\mu)}{E}\frac{\partial^2\uv(\rv;t)}{\partial t^2} =
-\alpha_{\rm th}\frac{2(1+\mu)}{(1-2\mu)}\nablav \delta T(\rv;t) +
\frac{2(1+\mu)}{E}\mathbf f_{\rm fric}(\rv;t).
\end{align}
\end{subequations}
The thermal-mechanical parameters used here are summarized in Table~\ref{Table1}. The expression of absorbed optical power $Q_{\rm abs}$ is given in Eq.~\eqref{eq:Q-abs}. The simulations are performed with COMSOL Multiphysics.

\begin{table}
\begin{tabular}{l*{3}{c}r}
\centering
               & {\small \bf specific heat } &{\small \bf thermal conductivity,}  &{\small \bf coefficient of linear   }  \\
                & {\small \bf capacity}, {\small $c_p$ [J/(kg$\cdot$K)]} &{\small $K$ [W/(m$\cdot$K)]}  &{\small \bf   thermal expansion},{\small  $\alpha_{\rm th}$ [1/K] }  \\
\hline
\hline
{\small gold}         & 120   & 110 & 31.5$\times 10^{-6}$ \\
{\small silica}       & 730   & 1.4 & 0.5$\times 10^{-6}$  \\
{\small air}          & 1000   & 0.03\\
\hline
\hline
\\
\\
\end{tabular}
\begin{tabular}{l*{3}{c}r}
\centering
               & {\small \bf Young's modulus,} & {\small \bf Possion's ratio,} & {\small \bf density,}  \\
               & {\small $E$ [Pa]} & {\small $\mu$} & {\small $\rho$ [kg/$\rm m^3$]} \\
\hline
\hline
{\small gold}          & 70$\times 10^{9}$ & 0.44 & 19300\\
{\small silica}        & 70$\times 10^{9}$ & 0.17  & 2200\\
\hline
\hline
\end{tabular}

\captionsetup{justification=centering}
\caption{\bf Thermal-mechanical parameters used in the numerical simulations.}
\label{Table1}
\end{table}

In experiments, we observe that gold plates move spirally around micro-fibers. The spiral motion has two fundamental degrees of freedom: rotation and translation in the azimuthal and axial directions of micro-fibers, respectively, which will be discuss separately below.

\subsection {Rotation}

\begin{figure}[!htp]
\centering
\includegraphics[width=15cm]{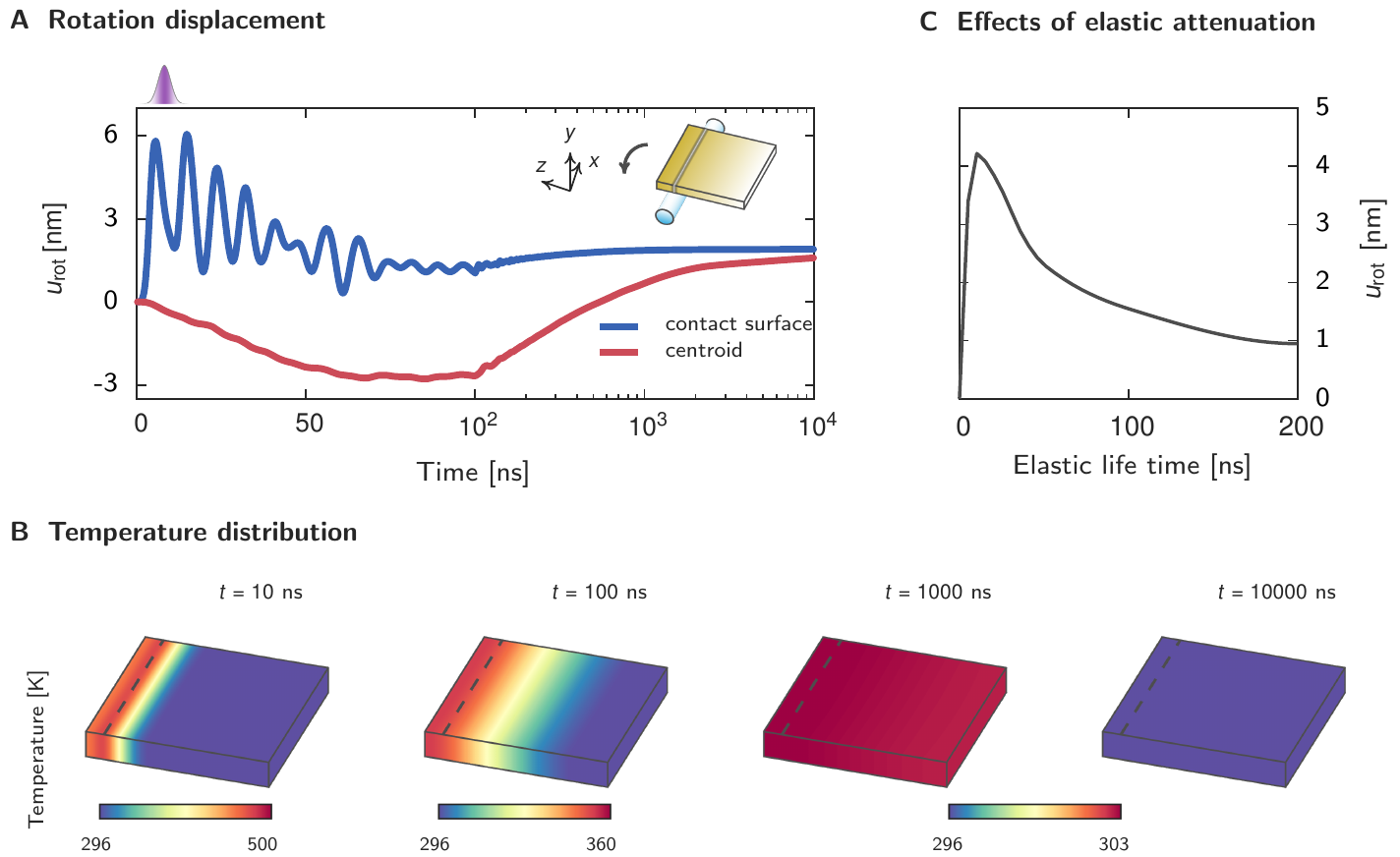}
\caption{{\bf Rotation dynamics in gold-plate \& micro-fiber system, supplemental to Fig. 4C in the main text}. {\bf A.} Temporal evolution of rotation displacements of the contact surface and the centroid of a gold plate driven by a single optical pulse.
{\bf B.} Temperature distribution in the plate at different time.
{\bf C.} Stabilized rotation displacement of the plate driven by an optical pulse as a function of the life time of elastic waves.
The gold square plate with side length $10\,\mu\rm m$ and thickness 30 nm is placed on a micro-fiber with diameter 1.8$\mu\rm m$. The two wings of the plate are of unequal length of $1\,\mu\rm m$ and $9\,\mu\rm m$, respectively. The pulsed optical absorption is specified in Eq.~\eqref{eq:Q-abs}. The line-shaped contact surface between the plate and the micro-fiber is smooth along the axial direction of the micro-fiber and it supports a frictional sliding resistance of $1.5\,\mu\rm N$. The life time of elastic waves is $\tau_{\rm el}=20\,\rm ns$ in $\bf A$-$\bf B$ and is varied in $\bf B$.
}
\label{Fig:Rotate_dynamic}
\end{figure}

We below present the supplementary results for Fig. 4C in the main text.
Figure~\ref{Fig:Rotate_dynamic}A contrasts the rotation displacements of the contact surface and the centroid of the studied gold plate in the rotation direction as functions of time.
Interestingly, it is observed that, at earlier times (smaller than one hundred nanoseconds) when the plate slides noticeably, the centroidal displacement (i.e., the average displacement of the plate) and the contact-surface displacement show opposite signs.
This can be understood by analyzing the friction force (see the middle panel in Fig.~4C in the main text).
In specifics, the friction force points in the opposite direction with the contact-surface displacement, while the centroid of the plate moves in the direction of the friction force due to Newton's second law.
Then, as time increases such that the elastic waves in the plate are significantly damped, the contact surface ceases sliding. At the mean time, the centoridal displacement gradually reverses its sign and finally approaches the value of the contact-surface displacement. This is due to the existence of the static friction force that points in the same direction as the accumulated contact-surface displacement, thereby preventing the contact surface from returning to its initial position and additionally dragging the centroidal displacement towards the direction of the contact-surface displacement.
Figure~\ref{Fig:Rotate_dynamic}B plots the temperature distribution in the plate at different time.


Figure~\ref{Fig:Rotate_dynamic}C depicts the stabilized rotation displacement driven by a single optical pulse as a function of life time of elastic waves. It is observed that, as the elastic life time increases (i.e., the attenuation decreases), the magnitude of the displacement initially increases benefiting from the reduced attenuation, and then decreases as a result of the restored balance between the counter-propagating waves in the plate (see Sec.~\ref{Sec:Rot_Theory}).


\subsection {Translation}

\subsubsection{Supplementary results for Fig. 3D}
\begin{figure}[!htp]
\centering
\includegraphics[width=14cm]{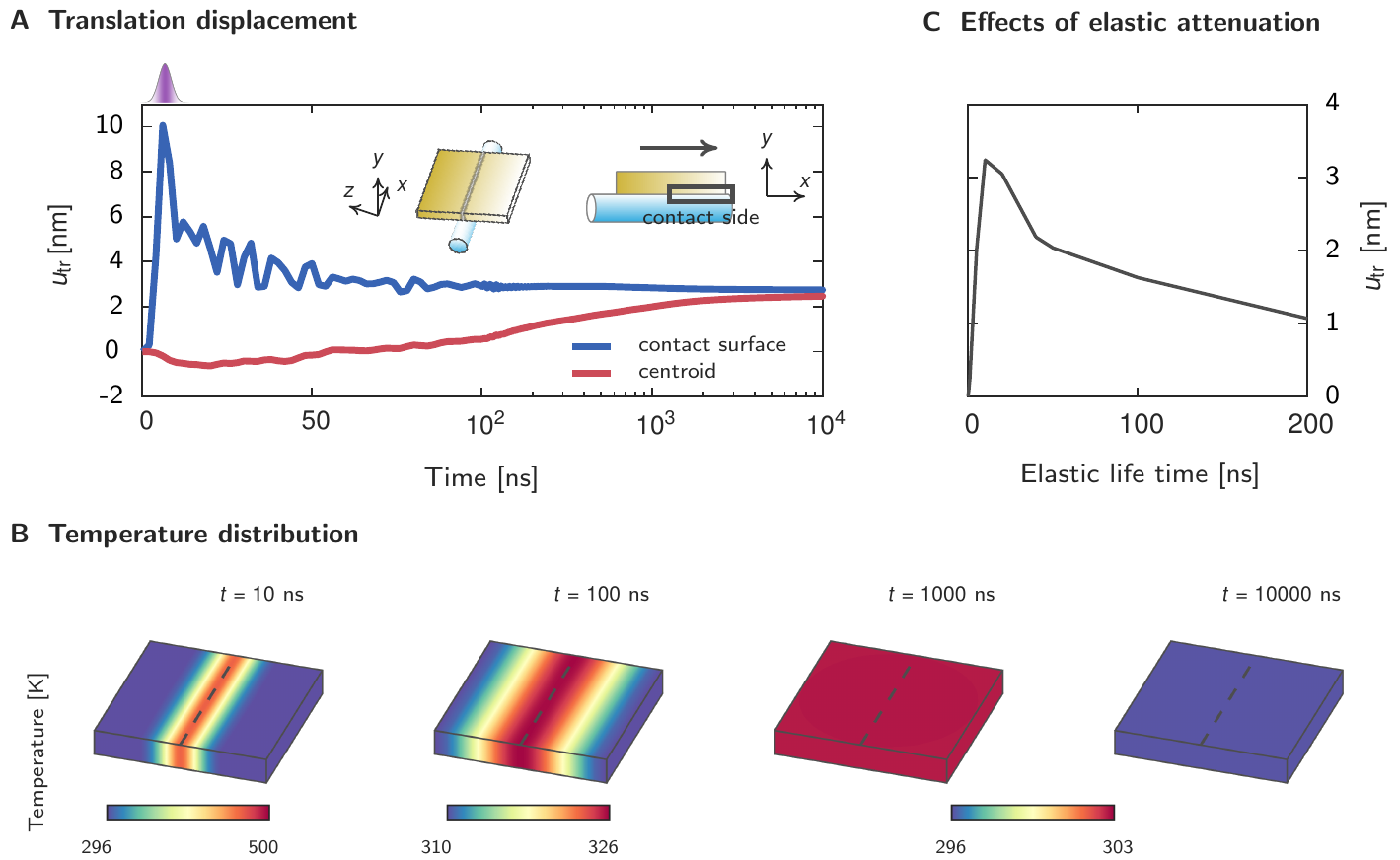}
\caption{
{\bf Translation dynamics in gold-plate \& micro-fiber system, supplemental to Fig. 4D in the main text}. {\bf A.} Temporal evolution of translation displacements of the contact surface and the centroid of a gold plate driven by a single optical pulse.%
{\bf B.} Temperature distribution in the plate at different time.
{\bf C.} Stabilized translation displacement of the plate driven by a single optical pulse as a function of life time of elastic waves.
The gold square plate with side length $10\,\mu\rm m$ and thickness 30 nm is placed on a micro-fiber with diameter $1.8\,\mu\rm m$. The two wings of the plate are of equal length. The pulsed optical absorption is specified in Eq.~\eqref{eq:Q-abs}. The line-shaped contact surface between the plate and the micro-fiber occupies the half part of the ideal smooth touching line, mimicking the contact asymmetry induced by the surface curvature of the plate; the frictional sliding resistance is $1.5\,\mu\rm N$. The life time of elastic waves is $\tau_{\rm el}=20\,\rm ns$ in $\bf A$-$\bf B$ and is varied in $\bf B$.
}
\label{Fig:Translate_dynamic}
\end{figure}

We below present the supplementary results for Fig. 3D in the main text. Figure~\ref{Fig:Translate_dynamic}A contrasts the translation displacements of the contact surface and the centroid of a gold plate in the axial direction of a micro-fiber as functions of time.
The centrodial displacement here shows similar features as in the rotation case in Fig.~\ref{Fig:Rotate_dynamic}, which can be understood by analyzing the friction force (middle panel in Fig.~4D in the main text; following the same reasonings as the rotation case).
Figures~\ref{Fig:Translate_dynamic}B plots the temperature distribution in the plate at different time.
Figures~\ref{Fig:Translate_dynamic}C shows the stabilized translation displacement driven by a single pulse as a function of life time of elastic waves, consistent with the results in the rotation case.

\subsubsection{Effects of asymmetrical distribution of absorbed optical power on translation}

In Fig. 3D  in the main text and Fig.~\ref{Fig:Translate_dynamic}, we set that optical absorption power distributes uniformly along the axial direction of micro-fibers. This setting is derived from the experimental observation that the translation direction is independent of light propagation direction, and the contact asymmetry is identified as the key factor determining the translation direction of the plate (see the associated discussions in the main text).

Detaching our considerations from the experimental facts, one may wonder, if the surface curvature of gold plates is negligible such that the contact surface is “flat" along the axial direction of micro-fibers, how do plates translate? As is discussed in the main text, to enable the translation of plates, the reflection symmetry of the system in the axial direction of micro-fibers needs to be broken. Apparently, in the absence of the contact asymmetry, this requirement can also be achieved by the asymmetrical distribution of the absorbed optical power, which is termed as optical asymmetry in the following discussions.

To model the optical asymmetry, we modify Eq.~\eqref{eq:Q-abs} to
\begin{align}
Q_{\rm abs}(\rv;t)=P_{\rm peak} \frac{1}{\sqrt{\pi}\tau_p}e^{-(t-t_0)^2/\tau_p^2}\frac{1}{wh}\frac{1}{\sqrt{\pi}\ell_p}e^{-(z-z_{\rm cs})^2/\ell_p^2}\underbrace{f_{\rm asy} e^{-(x-x_{\rm asy})^2/w_p^2}}_{\text{asymmetrical distribution}} .
\label{eq:Q-abs2}
\end{align}
Here, the optical asymmetry is introduced by a nonzero value of $x_{\rm asy}$ (the center coordinate of the contact surface in the axial direction of micro-fibers is $x=0$); $f_{\rm asy}$ is the normalization factor such that
$\int_{-w/2}^{w/2} dx f_{\rm asy}  e^{-(x-x_{\rm asy})^2/w_p^2}=1$.

\begin{figure}[!htp]
\centering
\includegraphics[width=16cm]{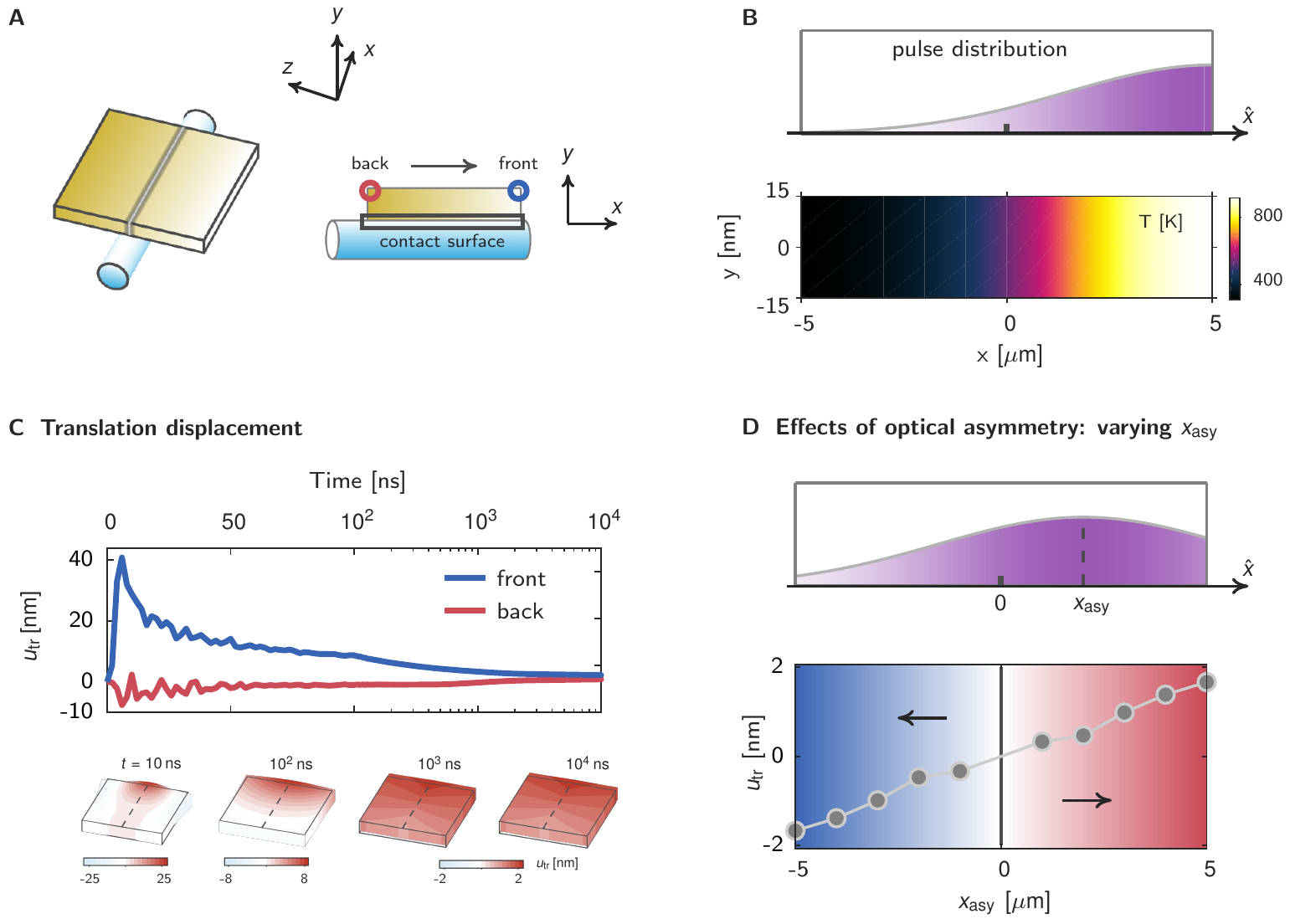}
\caption{ {\bf Effects of asymmetrical spatial distribution of optical absorption power on translation movement of gold plates.} {\bf A.} Sketch of the problem. A gold square plate with side length 10 $\mu\rm m$ and thickness 30 nm is placed symmetrically with two equal-length wings on top of a micro-fiber with a diameter of $1.8\,\mu\rm m$. The plate and the micro-fiber contacts each other smoothly along the axial direction of the micro-fiber and the frictional sliding resistance is $1.5\,\mu\rm N$. The plate is driven by a pulsed optical absorption. The pulse is centered at $3\,\rm ns$ with 3-ns FWHM and its spatial asymmetry along the axial direction of the micro-fiber is parameterized by $x_{\rm asy}$, see Eq.~\eqref{eq:Q-abs2}.
{\bf B.} Distribution profile of optical absorption power with $x_{\rm asy}=5\,\mu\rm m$ through the center of the gold plate in the axial direction of the micro-fiber (upper panel), and simulated temperature distribution at $t=6\,\rm ns$  (lower panel). Note that the temperature at the front edge of the plate is higher than the back edge because the optical absorption power localizes at the front edge.
{\bf C.} Temporal evolutions of translation displacements of two points on the front- and back-edges of the plate, respectively (upper panel), and translation profiles at different times (lower panel). The plate is driven by the pulsed optical absorption as shown in {\bf B}.
{\bf D.} Stabilized translation displacement as a function of $x_{\rm asy}$.
}
\label{Fig:Opt_Asym}
\end{figure}

In the following numerical study, we change the simulation settings of Fig.~\ref{Fig:Translate_dynamic} by (i) employing Eq.~\eqref{eq:Q-abs2} to represent the pulsed optical absorption and setting $x_{\rm asy}=5\,\mu\rm m$ and $w_p=5\,\mu\rm m$, and (ii) removing the contact asymmetry. To examine the translation dynamics, we select two points locating on the front and back edges of the plate in the $x$-direction (see Fig.~\ref{Fig:Opt_Asym}A). Note that the magnitude of absorbed optical power decreases monotonically from the front edge of the plate to the back edge (see the upper panel in Fig.~\ref{Fig:Opt_Asym}B).

As shown in Fig.~\ref{Fig:Opt_Asym}C, the selected front and back points translate in the opposite directions initially, and the magnitude of the translation displacement of the front point is noticeably larger than the back point. The latter observation is due to the temperature at the front edge is higher than the back one (see the lower panel in Fig.~\ref{Fig:Opt_Asym}B), resulting in larger thermal expansion. As time increases, the difference of the translation displacements of two points decreases as a result of thermal contraction, and, finally, converge to the same value. Notably, in this process, the translation direction of the back point reverses from the negative $x$-direction to the positive $x$-direction, while that of the front point retains the positive $x$-direction unchanged. Therefore, the net translation direction points from the back point to the front point, (i.e., towards the side localizing more optical absorption power). Moreover, the lower panel of Figs.~\ref{Fig:Opt_Asym}C shows the translation profiles of the plate at different times.

Figure~\ref{Fig:Opt_Asym}D plots the stabilized translation displacement of the plate as a function of the parameter of the optical asymmetry, $x_{\rm asy}$. It is seen that, as $x_{\rm asy}$ varies from $-5\,\mu\rm m$ to $5\,\mu\rm m$, the direction of the translation displacement reverses its sign.

In summary, the above results evidence that, in the absence of the contact asymmetry, the optical asymmetry can also enable the translation. In the presence of both the contact and optical asymmetries, the translation direction shall be determined by the dominant one. The investigations in this direction shall be discussed in a future paper by us.

\section{Fabricated gold plates}

\begin{figure}[!htp]
\centering
\includegraphics[width=14cm]{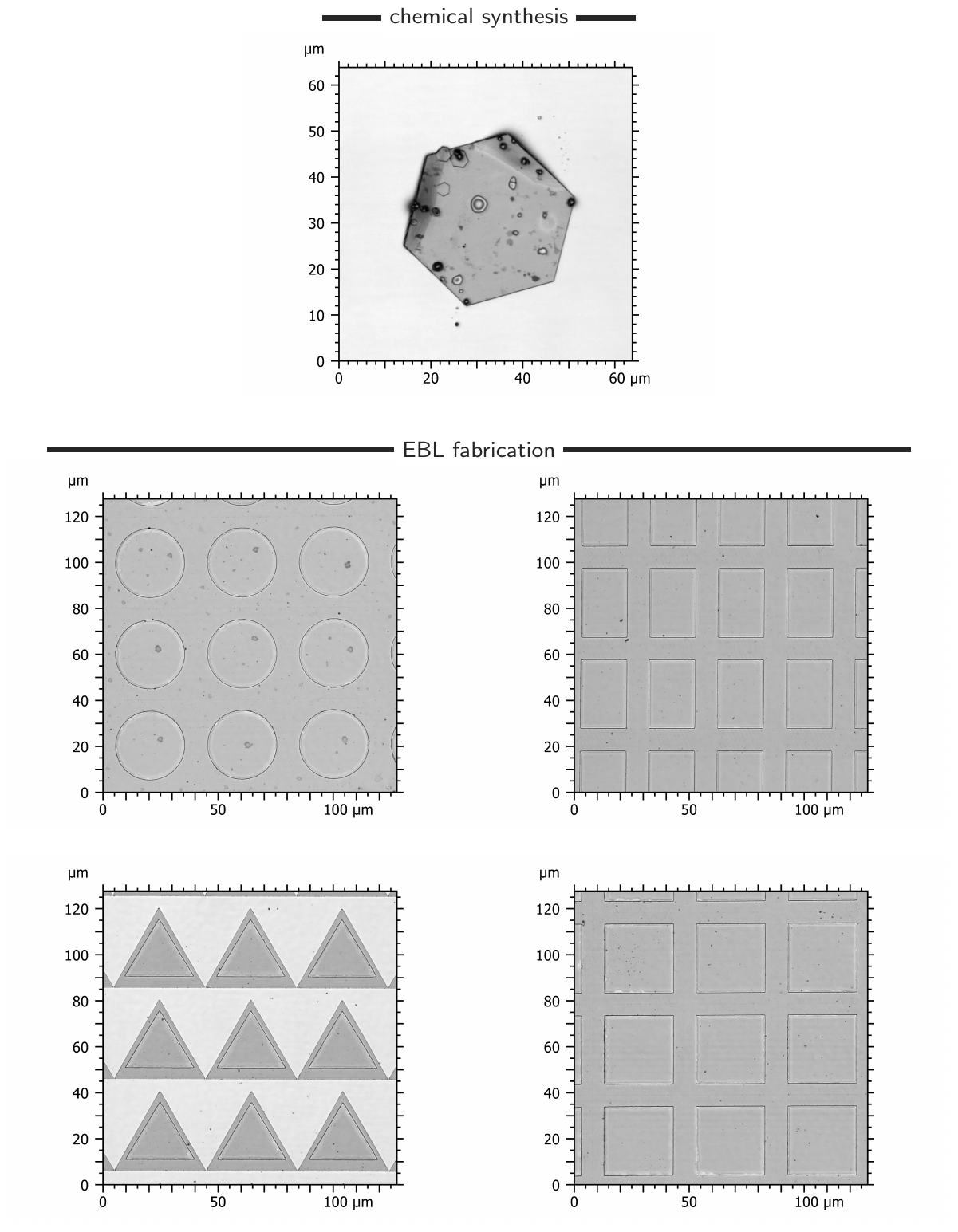}
\captionsetup{justification=centering}
\caption{{\bf Confocal laser scanning microscopy images of fabricated gold plates}.
}
\label{Fig:LSM_GPlates}
\end{figure}

Figure ~\ref{Fig:LSM_GPlates} plots confocal laser scanning microscopy images of fabricated gold plates of different shapes.

\section{Experimental characterizations of spiral motion of gold plates around micro-fibers}

\subsection  {Optical Images of Spiral Motion}

\begin{figure}[!htp]
\centering
\includegraphics[width=14cm]{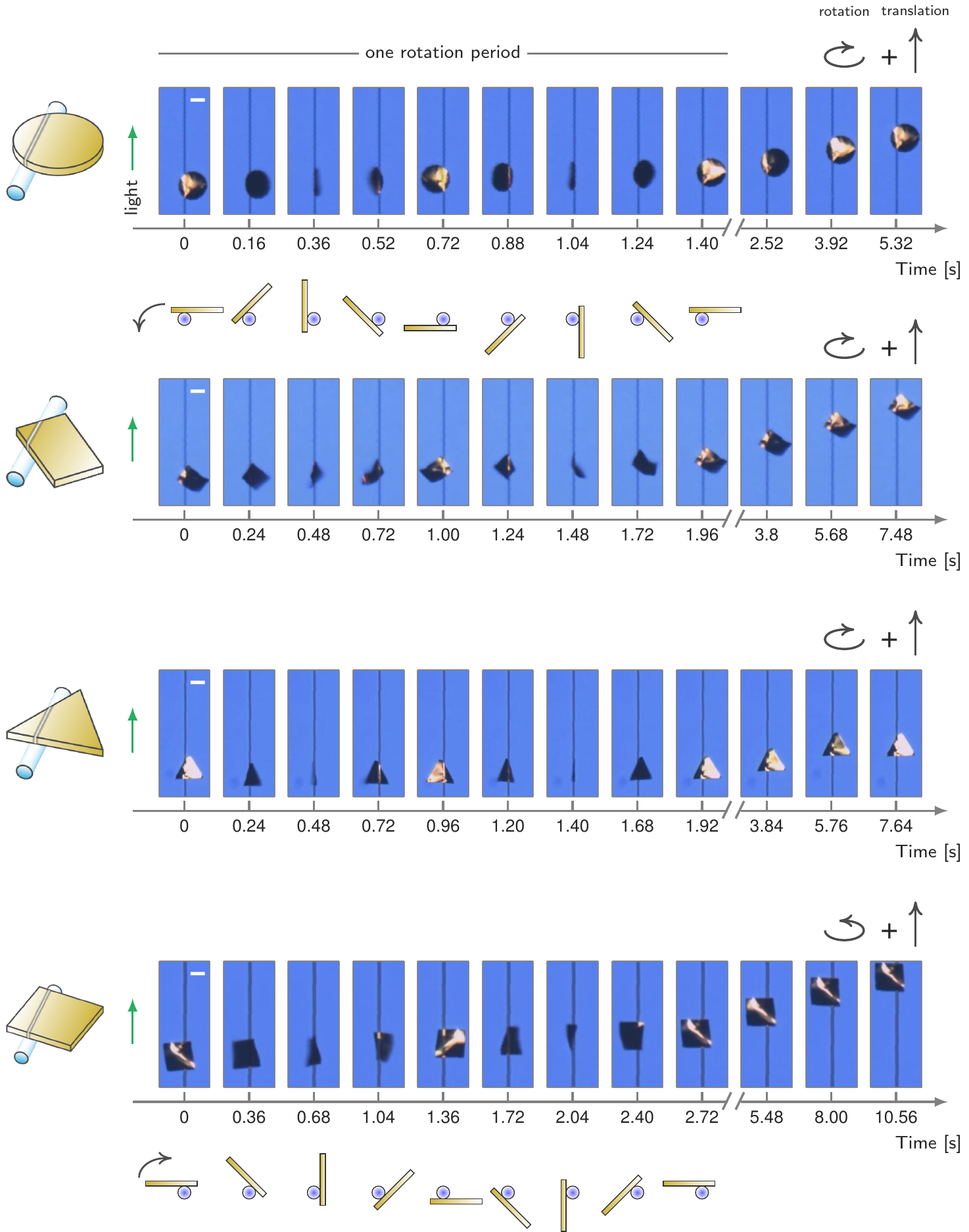}
\caption{{\bf Sequential optical images of gold plates of various shapes spirally moving around micro-fibers.} All scalar bars represent $15\,\mu\rm m$. The parameters of the laser source are as the same as in Fig. 2 in the main text.
}
\label{Fig:Spiral_Motion}
\end{figure}

Figure~\ref{Fig:Spiral_Motion} demonstrates the recorded temporal sequential optical images of gold plates moving around micro-fibers spirally. The demonstrated gold plates here have different base shapes, including circle, rectangle, triangle and square (from up to down).
\\
\\

\begin{figure}[!htp]
\centering
\includegraphics[width=14cm]{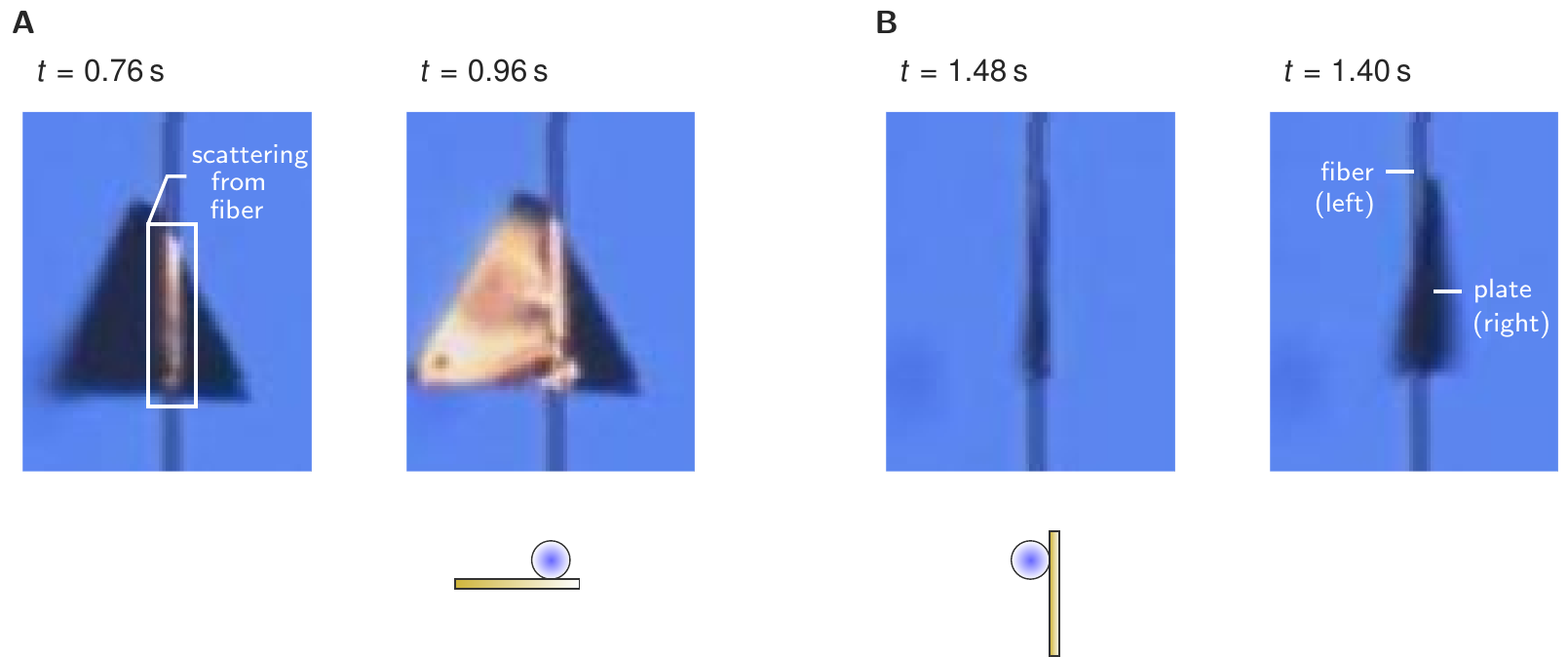}
\caption{{\bf Identifying relative positions between a gold plate and a micro-fiber by examining optical images.}
}
\label{Fig:Examine_RelPos}
\end{figure}

{\bf Technical remarks on determining rotation direction---}The identification of the rotation direction needs to know relative positions between gold plates and micro-fibers at different times.
This task however is not straightforward, because, in optical images, optical fibers with low refractive index are almost invisible when they overlap with non-transparent gold plates (see Fig.~\ref{Fig:Spiral_Motion}).
Nevertheless, some fine features from the optical images can be exploited to specify the relative positions. Taking the triangle plate in Fig.~\ref{Fig:Spiral_Motion} as an example, the optical image at $t=0.72\,\rm s$, as zoomed in Fig.~\ref{Fig:Examine_RelPos} A, show a clear line-shaped scattering pattern along the axial direction of the micro-fiber. This regular pattern is from light scattered by the gold plate. Further, we deduce that the gold plate locates below the micro-fiber. If not this scattering will be blocked by the plate and cannot be seen. Then, examining the optical image at $t=0.96\,\rm s$, wherein the area of the plate is at the maximum and which is temporally close to the image at $t=0.72\,\rm s$, we can conclude that the plate is just perpendicularly below the micro-fiber. Next, we examine the optical image at $t=1.48\,\rm s$, as shown in Fig.~\ref{Fig:Examine_RelPos}B. The image clearly demonstrates the micro-fiber is on the left side of the plate. Consequently, for the optical image at $t=1.40\,\rm s$ (temporally close to optical image at $t=1.48\,\rm s$), wherein the area of the plate is at the minimum, we deduce that the plate just touches the rightmost of the micro-fiber (see Fig.~\ref{Fig:Examine_RelPos}B). Knowing two specified positions, the rotation direction is determined unambiguously.

\begin{figure}[!htp]
\centering
\includegraphics[width=14cm]{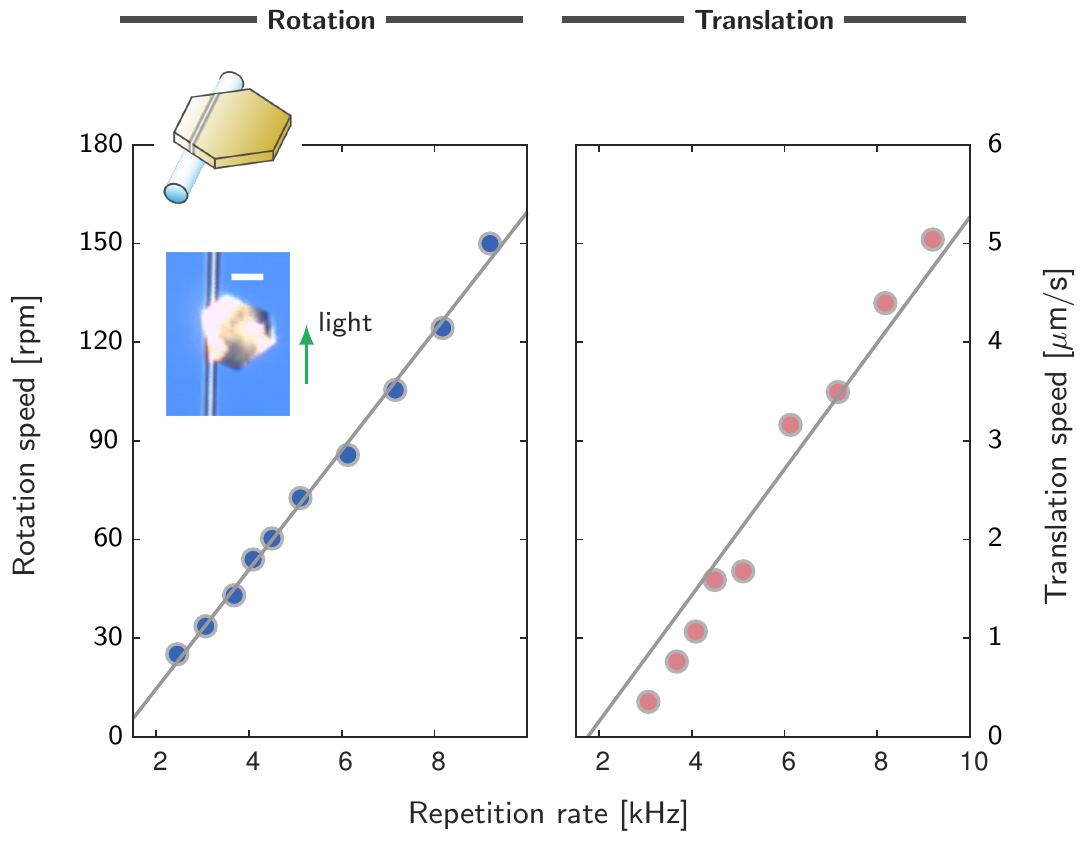}
\caption{{\bf Rotation and translation speeds of a hexagonal gold plate spirally} moving around a micro-fiber as a function of repetition rate of laser pulses. Inset: optical image of the gold plate (scalar bar, $15\,\mu\rm m$). The used super-continuum laser pulses here have 6.8--mW average power.
}
\label{Fig:Speed_Rep}
\end{figure}

\subsection {Controlling Motion Speed by Varying Repetition Rate of Laser Pulses }

We measured the dependence of motion speeds of gold plates on the repetition rate of the used laser pulses. As shown in Fig.~\ref{Fig:Speed_Rep}, both the rotation and translation speeds of a hexagonal plate increase (approximately) linearly as the repetition rate increases, evidencing that the spiral motion of the plate is driven by single individual pulse in a stepwise manner.




\subsection {Manipulating Spiral Motion by Adjusting Relative Positions between Gold Plates and Micro-fibers}

\begin{figure}[!htp]
\centering
\includegraphics[width=16cm]{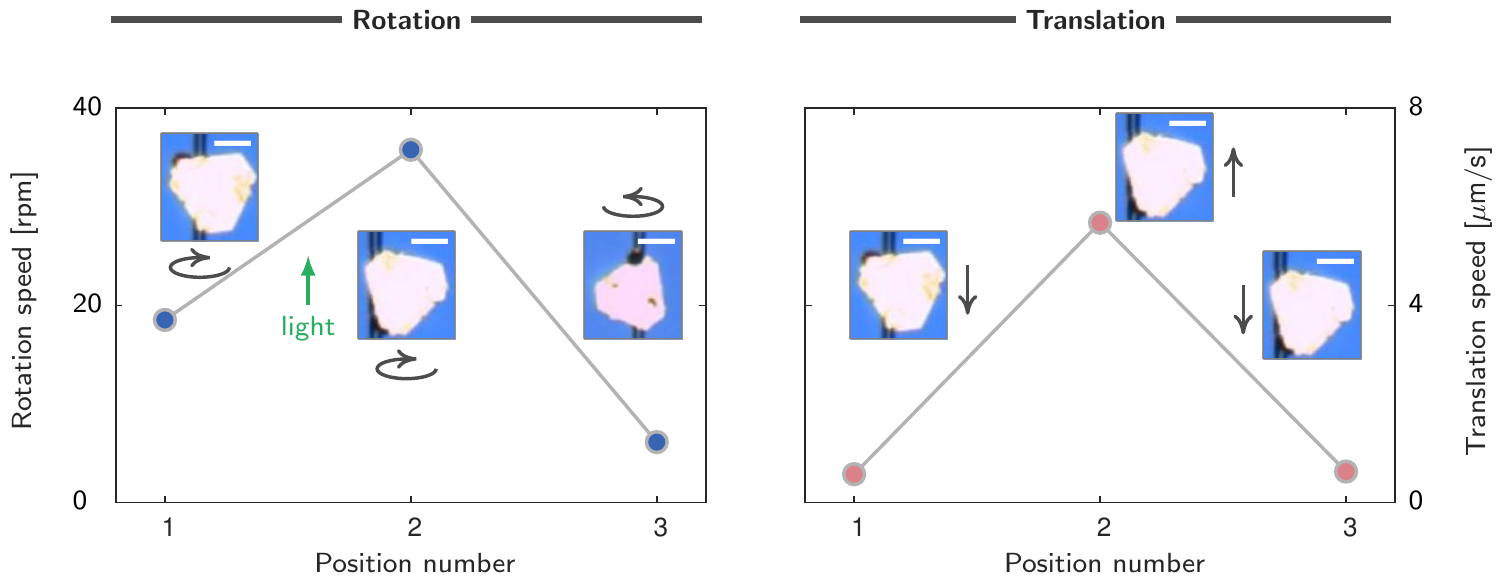}
\caption{{\bf Varying spiral rotation of a gold plate by adjusting its relative position with respect to the contacted micro-fiber.} The parameters of the laser source are as the same as in Fig. 2 in the main text. All scale bars represent $15\,\mu\rm m$.
}
\label{Fig:Adjust}
\end{figure}

As has been discussed in the main text, the translation and rotation directions of gold plates depend on the asymmetries of their two wings and contact surfaces, respectively. Apparently, the wings' asymmetry can be conveniently controlled by adjusting the relative positions between plates and micro-fibers. However, for the contact asymmetry induced by the surface curvature of gold plates, the control is not straightforward, because the curvature features of plates are almost invisible under optical microscope, with which we adjust the positions of plates. Nevertheless, it is still possible to accidentally alter the contact asymmetry by randomly moving plates on micro-fibers. Figure~\ref{Fig:Adjust} shows that by adjusting the positions of a gold plate on a micro-fiber, its motion directions  as well as motion speeds can be adjusted . Notably, the rotation speeds decrease significantly as two wings become more asymmetric, consistent with Fig. 4E in the main text.
\\
\\
\\

\subsection  {Spiral Motion in a Vacuum Chamber}

The experiments demonstrated in the main text and above in the Supplementary Information are performed in air. As is known, air molecules can exert two types of forces on their surrounded object: photophoretic force and drag force (i.e., air resistance). The photophoretic force is proportional to the air pressure and the temperature gradient on the object, with the magnitude estimated to be in the order of pN~\cite{Lu:2017}. Differently, the drag force relates with the air density, the area and the velocity of the object~\cite{Batchelor:1967}, i.e., $F_{\rm drag}=C_D\rho A v^2/2$ with drag coefficient $C_D \sim 1$, air density $\rho$, object area $A$, and velocity $v$. Its magnitude in our case is in the order of 100 pN, considering the instantaneous velocity of the plate is $\sim$1 m/s. Consequently, both the photophoretic and drag forces are significantly smaller than the friction force ($\mu N$), thereby negligibly contributing to the spiral motion.

\begin{figure}[!htp]
\centering
\includegraphics[width=16cm]{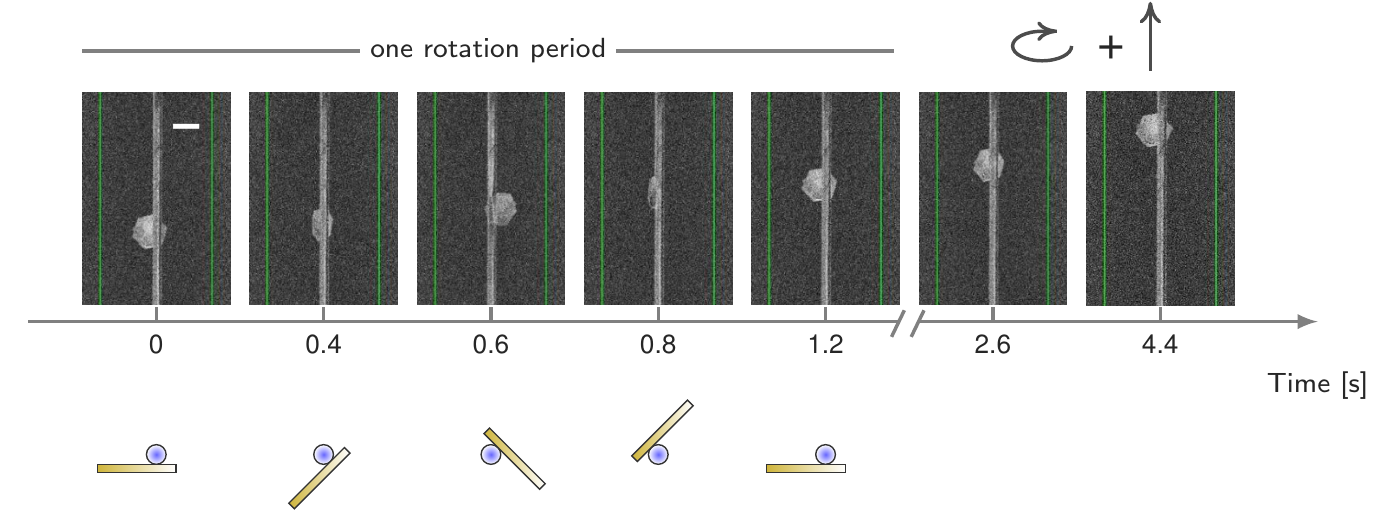}
\caption{{\bf Sequential scanning electron microscopy images recording the spiral motion of a gold plate in vacuum}.
Scale bar: 10 $\mu$m.
}
\label{Fig:Spiral_vacuum}
\end{figure}

To confirm that the surrounding air is inessential to the observed spiral motion, we repeat experiments in a vacuum chamber. As expected, we observe that gold plates can also spirally move around microfibers in vacuum, as is shown in Fig.~\ref{Fig:Spiral_vacuum}.

\section{Surface Topography of Gold Plates}

\begin{figure}[!htp]
\centering
\includegraphics[width=14cm]{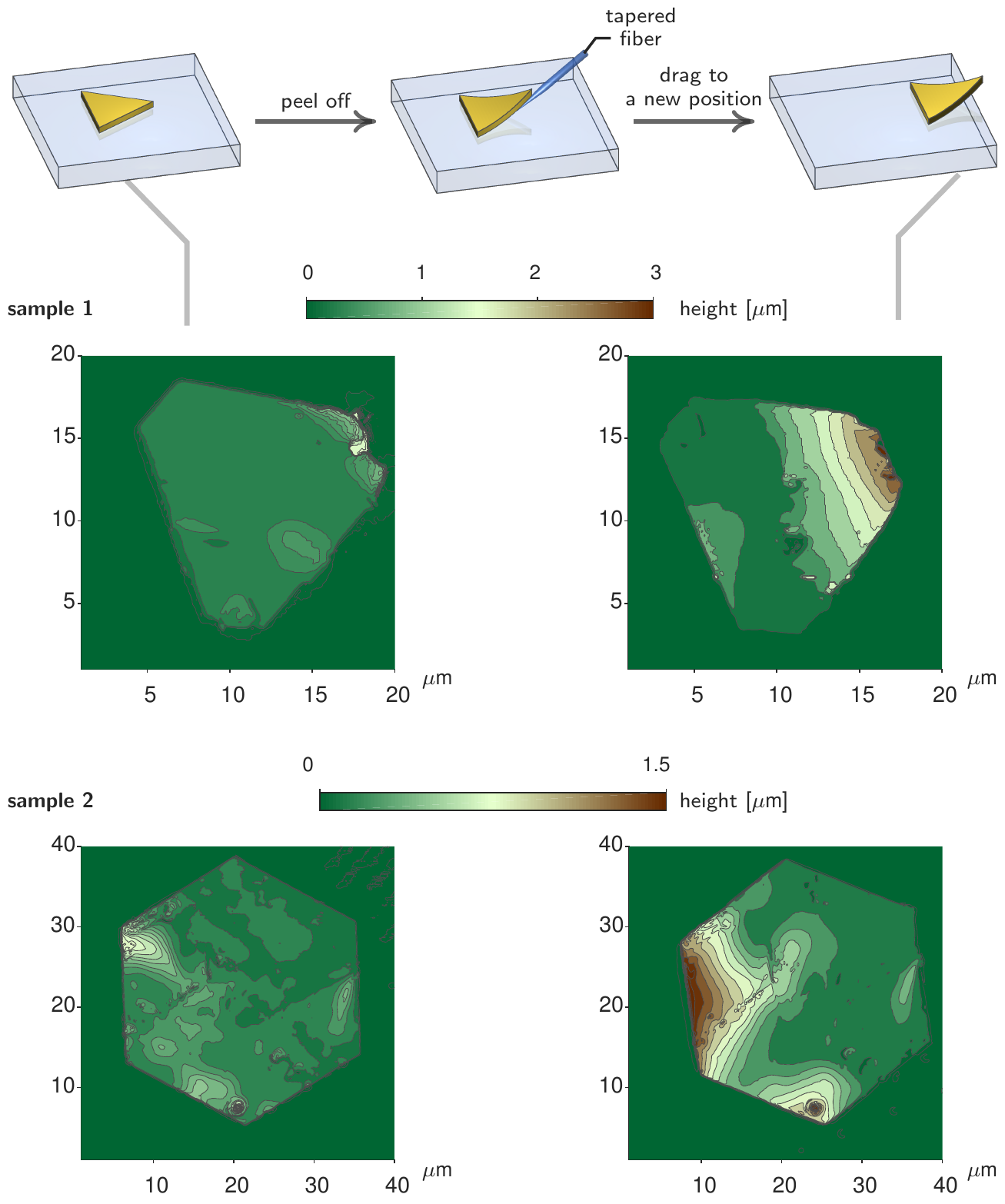}
\caption{{\bf Confocal laser scanning microscopy images of two exemplified gold plates} before and after that they are peeled off from glass substrates by tapered fibers and then dragged to new positions.
}
\label{Fig:LSM_Curve}
\end{figure}

Figure ~\ref{Fig:LSM_Curve} gives CLSM images of  two golds plate before and after that they are peeled off from a glass substrate by the tapered fibers and then are dragged to new positions on the substrates. The images show that the plates inevitably becomes more curved after the peeling-off process, a necessary step of transferring gold plates from glass substrates onto micro-fibers.

\begin{figure}[!htp]
\centering
\includegraphics[width=8cm]{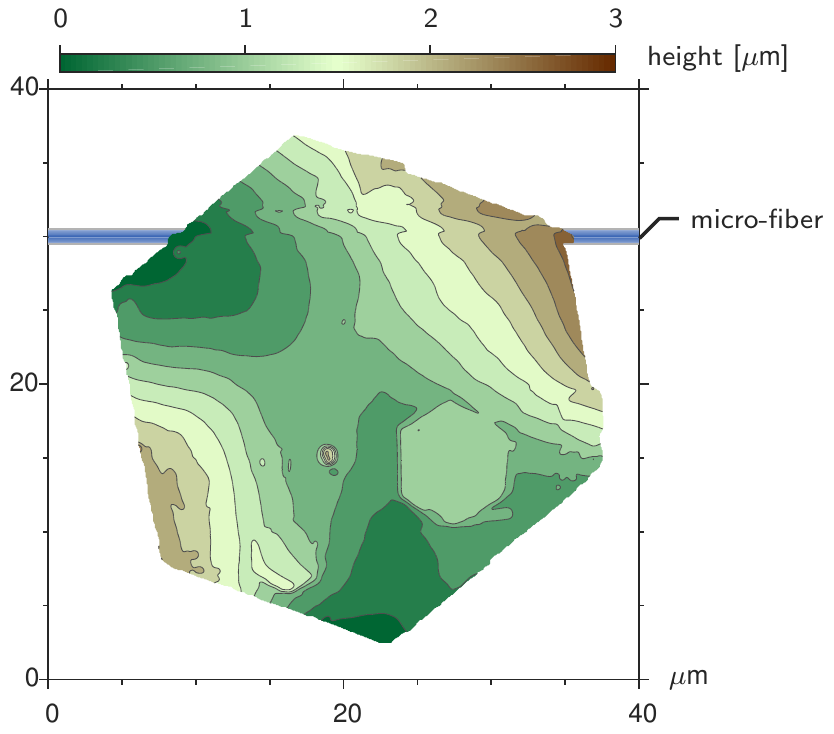}
\caption{{\bf Confocal laser scanning microscopy image of a hexagonal gold plate on top of a micro-fiber}.
}
\label{Fig:LSM_SI}
\end{figure}

Figure~\ref{Fig:LSM_SI} shows a confocal laser-scanning-microscopy image (CLSM) of a hexagonal gold plate on a micro-fiber, {\bf supplementing Fig. 4F in the main text}. The image shows that the surface of the plate is apparently curved, thereby resulting in that the contact surface between the plate and micro-fiber only occupies part of the ideal one when the plate is flat.

\bibliographystyle{apsrev4-1}
\bibliography{supplement}